\documentclass[floats,floatfix,showpacs,amssymb,prd,twocolumn,superscriptaddress,nofootinbib,reprint]{revtex4-2}

\usepackage{amssymb,amsmath,verbatim,mathtools,needspace,enumitem,etoolbox,graphicx,physics,microtype,afterpage,xspace,tabularx,lmodern,multirow}
\usepackage[dvipsnames]{xcolor}
\definecolor{linkcolor}{rgb}{0.57, 0.05, 0.15}
\usepackage[unicode, colorlinks=true, linkcolor=linkcolor, citecolor=linkcolor, filecolor=linkcolor, urlcolor=linkcolor, linktocpage, breaklinks]{hyperref}
\usepackage[all]{hypcap}
\usepackage[T1]{fontenc}
\usepackage[utf8]{inputenc}

\usepackage{tensind}
\tensordelimiter{?}
\DeclareGraphicsExtensions{.bmp,.png,.jpg,.pdf}
\usepackage{verbatim}
\usepackage[normalem]{ulem}
\usepackage{orcidlink}
\usepackage{soul}
\usepackage{aas_macros}

\usepackage[flushleft]{threeparttable}
\usepackage{booktabs}

\newcommand{\V}{\mathcal{V}}
\newcommand{\D}{\mathcal{D}}

\newcommand{\II}{{\rm II}}
\newcommand{\III}{{\rm III}}

\begin{document}
\title{Searching for Population III stars with line intensity mapping cross-correlations}

\author{Sahil Hegde\orcidlink{0000-0002-9370-8061}$^\dagger$}
\email{sahil@astro.ucla.edu}
\affiliation{Department of Physics \& Astronomy, University of California, Los Angeles, 475 Portola Plaza, Los Angeles, CA 90095, USA}
\altaffiliation{NSF Graduate Research \& NASA FINESST Fellow}

\author{Guochao Sun\orcidlink{0000-0003-4070-497X}}
\affiliation{CIERA and Department of Physics and Astronomy, Northwestern University, 1800 Sherman Ave, Evanston, IL 60201, USA}
\affiliation{NSF-Simons AI Institute for the Sky (SkAI), 172 E. Chestnut St., Chicago, IL 60611, USA}

\author{Julian B. Mu\~{n}oz \orcidlink{0000-0002-8984-0465}}
\affiliation{Department of Astronomy, The University of Texas at Austin,
2515 Speedway, Stop C1400, Austin, TX 78712, USA}
\affiliation{Cosmic Frontier Center, The University of Texas at Austin, Austin, TX 78712, USA}

\author{Steven R. Furlanetto\orcidlink{0000-0002-0658-1243}}
\affiliation{Department of Physics \& Astronomy, University of California, Los Angeles, 475 Portola Plaza, Los Angeles, CA 90095, USA}

\begin{abstract}
    Decades of searches for Population~III stars in individual galaxies have yielded a few potential candidates, but a statistically robust characterization of the demographics of the first stars in the lowest-mass systems remains elusive. Line intensity mapping (LIM), an observational technique that measures fluctuations in the aggregate emission from the entire galaxy population --- including the faintest sources --- offers an alternative strategy that is especially well-suited for the Pop~III era. With the recent launch of SPHEREx and the rapid development of a number of complementary LIM studies, we are poised to place some of the first LIM constraints on sites of star formation at high-redshift. In this work, we expand an analytical model for LIM power spectra, \texttt{Zeus21/oLIMpus}, to include Pop~III stars and the emission lines identified as diagnostic signatures of star formation with a low-metallicity, top-heavy IMF, such as H$\alpha$ and HeII. We introduce a flexible framework to estimate the measurement uncertainties associated with instrument and survey configurations, and apply these to study LIM signatures of the first stars in mock surveys carried out with SPHEREx and potential next-generation instruments. We quantify the sensitivity of the LIM signal to variations in Pop~II and III parameters and forecast joint limits that can be placed on the Pop~III star formation efficiency (SFE) and IMF shape with SPHEREx. We find that next-generation instruments will be necessary to comprehensively survey the Pop~III theoretical landscape --- both with respect to `classical' and `exotic' Pop~III models --- and identify design improvements that will enable such studies. Finally, we carry out a Fisher analysis to characterize synergies between SPHEREx and JWST, suggesting that joint constraints on the Pop~III SFE and extensions to conventional Pop~III models may be within reach.
\end{abstract}

\maketitle

\section{Introduction}\label{sec:intro}
The canonical picture of hierarchical structure formation predicts that the first generation of stars  --- termed Population III, or Pop III --- formed from small clouds of pristine gas in the first minihalos ($m_h \sim 10^6 M_\odot$; for recent reviews see Refs.~\cite{bromm_formation_2013, loeb_first_2013, klessen_first_2023}). With star formation efficiencies of $10^{-3}-10^{-2}$, these clouds gave rise to $10^2-10^4 M_\odot$ clumps of the first luminous sources in the universe \citep{mckee_formation_2008, chon_impact_2024, hegde_efficient_2025}. In turn, by virtue of their pristine nurseries, it is believed that Pop~III stars formed with a more top heavy IMF than observed locally, with typical masses on the order of $10-100 M_\odot$ \citep{abel_formation_2002, bromm_formation_2002, stacy_constraining_2013, hirano_one_2014, hirano_primordial_2015, jaura_trapping_2022, prole_fragmentation_2022, sharda_interplay_2023, sharda_magnetic_2025,  sharda_population_2025, lake_stellar_2025}. Such massive stars would quickly exhaust their nuclear fuel and live short, few Myr lives ending in luminous core collapse or pair-instability supernovae (SNe; \cite{barkat_dynamics_1967, fryer_pair_2001, heger_nucleosynthetic_2002, heger_how_2003, heger_nucleosynthesis_2010, kasen_pair_2011, hummel_source_2012, whalen_finding_2013, tolstov_multicolor_2016, lazar_probing_2022, jeon_hunting_2026}). Thus, though such Pop~III-hosting minihalos would be ubiquitous across the universe, the conjunction of these fleeting lifetimes with their small total masses makes a direct detection of individual Pop~III star forming clumps an exceedingly challenging endeavor.

Indeed, this observational quest has stymied astronomers for decades. Wide and deep searches alike with generations of increasingly sophisticated instruments have been largely fruitless, though the advent of the James Webb Space Telescope (JWST) has yielded some potential candidates \citep{maiolino_jwst-jades_2024, wang_strong_2024, vanzella_pristine_2025, fujimoto_glimpse_2025, morishita_pristine_2025, cai_metal_2025, fujimoto_glimpsed_2025, trussler_out_2026, reumert_hot_2026, maiolino_search_2026}. Deep photometric surveys carried out in the first few years of JWST operation have identified a number of candidate Pop~III systems and spectroscopic follow-up has bolstered these claims, but number statistics are still small. These endeavors often attempt to identify sources that display the hallmarks of top-heavy, Pop~III star formation (no metal lines, strong nebular emission features, and signatures of a moderately hard ionizing spectrum) and leverage pencil beam observations of lensed fields in their search for the faintest galaxies in the universe \citep{schaerer_properties_2002, zackrisson_spectral_2011, katz_challenges_2023, zackrisson_detection_2024, trussler_on_2023, schaerer_observable_2025}. Ultimately, these probes rely on serendipity. In kind, so-called `archaeological' searches for peculiar abundance patterns and relics of Pop~III star formation in the local universe have identified extremely metal-poor, second generation stars, but no truly metal-free stars have yet been detected \citep{chiti_detailed_2023, chiti_enrichment_2024, chiti_enrichment_2026, bonifacio_most_2025, ji_nearly_2026, frebel_metal_2026}.

Line intensity mapping (LIM) is an observational strategy that takes the opposite tack. Rather than attempting to resolve individual sources in small regions of the sky, LIM surveys seek to measure the aggregate emission from sources of a particular emission line over wide swaths of the sky with low spatial resolution but exquisite spectral sensitivity (for some recent reviews see e.g., \cite{kovetz_line-intensity_2017, bernal_line_2022, chang_line_2026}). By measuring every photon of the line of interest at a range of frequencies, LIM experiments yield tomographic information along the line of sight, statistically mapping out the \textit{entire} source population across the sky, \textit{including the faintest sources}. LIM provides a powerful --- and complementary --- probe to conventional galaxy surveys, which are most sensitive to the bright end of the population \citep{furlanetto_cross_2007, lidz_probing_2009, visbal_measuring_2010, visbal_cross_2023, sun_limfast_2023, sun_LIMFAST_2026}. In principle, LIM can be applied to any detectable line and offers a statistical lens into the astrophysical and cosmological drivers of that emission \citep{furlanetto_cosmology_2006, munoz_ethos_2021, munoz_impact_2022, cruz_effective_2025, sun_limfastIII_2025, sun_LIMFAST_2026, libanore_effective_2025, libanore_new_2025, kovetz_when_2026, lazare_when_2026, cruz_rise_2026}. As such, it has been applied in a number of contexts, ranging from star formation indicators such as $[{\rm CII}]$158$\mu$m and CO \citep{lidz_intensity_2011, yang_multitracer_2021, schaan_multi_2021, schaan_astrophysics_2021, kannan_THESAN_2022, sato-polito_multitracer_2023, liu_effects_2024, bracks_forecasting_2026} to tracers of the intergalactic medium (IGM) structure, such as the 21-cm and Ly$\alpha$ lines \citep{zaldarriaga_21_2004, furlanetto_global_2006, mesinger_21cmfast_2011, sun_limfast_2023, cruz_effective_2025}. 

Because intensity mapping surveys are sensitive to \textit{all emission}, they are especially well suited to shed light on the faintest galaxy populations, including Pop~III hosts \citep{sun_revealing_2021}. With this in mind, Ref.~\cite{visbal_looking_2015} proposed applying LIM techniques to search for Pop~III star formation via the HeII 1640\AA\ recombination line (hereafter HeII), a typical diagnostic of a hard ionizing spectrum owing to its 54.4 eV ionization potential \citep{oh_he_2001, schaerer_properties_2002, schaerer_transition_2003}. Pop~I and II galaxies (with more metal-enriched, and thus lower-mass stellar populations) are unlikely to produce HeII emission, so a detection of the emission signature is regarded as a smoking gun for the presence of very massive stars or accreting black holes (BHs, \cite{cassata_HeII_2013, mas-ribas_boosting_2016}). Building from the proof-of-concept first introduced in Ref.~\cite{visbal_looking_2015} --- which relied on simple analytical estimates of Pop~III and II star formation --- Ref.~\cite{parsons_probing_2022} explored the power of HeII LIM to study Pop~III star formation more quantitatively using the \texttt{LIMFAST} semi-numerical simulation. Using the ratio of HeII to H$\alpha$ emission detected in a mock LIM survey, Ref.~\cite{parsons_probing_2022} demonstrate how such complementary measurements can ground the variation in the signal due to the overall magnitude of star formation in the universe and thus isolate the sensitivity to the Pop~III IMF. However, though faster than a conventional simulation, due to its reliance on a simulated box and detailed tracking of the relevant radiation backgrounds, a semi-numerical simulation such as \texttt{LIMFAST} is still limited in its flexibility and efficiency of exploring the parameter space. 

Here we leverage the power of the lightweight and flexible intensity mapping framework \texttt{Zeus21}/\texttt{oLIMpus} to explore our ability to constrain the demographics of Pop~III star formation with H$\alpha$ and HeII in a truly statistical manner \citep{munoz_effective_2023, cruz_effective_2025, libanore_effective_2025}. \texttt{Zeus21}\footnote{https://github.com/ZeusCosmo/Zeus21}/\texttt{oLIMpus}\footnote{https://github.com/slibanore/oLIMpus} analytically models the spatial fluctuations of a particular line by capturing the nonlinearity and nonlocality of the high-redshift Pop~II star formation rate density (SFRD) with a lognormal random field, and Ref.~\cite{cruz_effective_2025} extend this early work to include Pop~III stars and their associated (spatially varying) feedback. This \textit{effective} model for star formation has been shown to produce predictions of the 21-cm and LIM signals in excellent agreement with state-of-the-art semi-numerical simulations such as \texttt{21cmFAST} and \texttt{LIMFAST} in a fraction of the time.

Our paper is organized as follows. In this work, we expand the \texttt{oLIMpus} framework --- which only traces emission due to Pop~II stars --- to include the Pop~III modeling introduced in Ref.~\cite{cruz_effective_2025} (Section~\ref{sec:oLIMpus}). We then introduce new modules to compute Pop~III line luminosities for a flexibly-defined Pop~III IMF (Section~\ref{sec:line_lum}) and to estimate measurement uncertainties for a particular LIM observing configuration (Section~\ref{sec:estimating_SNR}). In Section~\ref{sec:parameter_constraints}, we leverage the efficiency of this framework to statistically forecast parameter constraints with a Fisher analysis. We first explore the power of forthcoming LIM surveys alone --- both in terms of `classical' Pop~III models (Section~\ref{sec:classical_popIII}) and extensions to those models (Section~\ref{sec:model_extensions}) --- and then expand our analysis to identify synergies between galaxy surveys carried out with JWST and existing LIM experiments to home in on the demographics of Pop~III stars (Section~\ref{sec:popIII_UVLF}). We discuss some limitations and caveats to our modeling in Section~
\ref{sec:caveats} and we conclude in Section~\ref{sec:conclusion}.

\section{Power spectra with \texttt{Zeus21+oLIMpus}}\label{sec:oLIMpus}
We begin by briefly summarizing the methodology underpinning the \texttt{Zeus21}/\texttt{oLIMpus} framework \citep{munoz_effective_2023, libanore_effective_2025} --- which we employ to efficiently generate LIM power spectra --- and our extensions to it to incorporate the contributions of Pop III stars \citep{cruz_effective_2025}.

\subsection{SFR or line luminosity densities for Pop II stars}\label{sec:density_bckgd}
Building from the formalism first introduced in the \texttt{Zeus21} framework to analytically model fluctuations in the 21-cm brightness temperature during cosmic dawn --- which reflect the evolution of radiation fields on \textit{large} scales --- \texttt{oLIMpus} refines that effective model to make predictions for LIM science \citep{munoz_effective_2023, cruz_effective_2025, libanore_effective_2025}. \texttt{Zeus21} and \texttt{oLIMpus} both rely on the crucial observation that, to first order, the local star formation rate (SFR) or line luminosity density can be \textit{ effectively} modeled as a biased tracer of the density field --- which itself is lognormally distributed \citep{coles_lognormal_1991}. In other words, the mean SFR or line luminosity density smoothed on a scale $R$ is simply $\langle \rho_{\rm SFR, L}\rangle_R\propto e^{\gamma_R^{\rm SFR, L}\delta_R}$, where the $\gamma_R^{\rm (i)}$ represent the non-linear biases scaling the density dependence of the SFR or luminosity density integrals and $\delta_R$ represents the local overdensity smoothed on a scale $R$. Thus, correlation functions of the SFRD, luminosity density, or 21-cm brightness temperature simply reduce to correlation functions of this lognormal variable summed over the relevant scales, which can be computed analytically. 

At small scales ($R\lesssim 3\ {\rm Mpc}$) and later times ($z\lesssim 10$), overdensities $\delta$ grow and the first-order lognormal approximation begins to break down. As such, the \texttt{oLIMpus} framework extends the approximations to second order --- namely 
\begin{equation}\label{eq:effective_density}
    \rho_{\rm i}(z|\delta_R) =  \overline{\rho}(z) \frac{\exp\Big(\gamma_R^{\rm i}\delta_R + \gamma_R^{\rm NL, i}\delta_R^2\Big)}{\mathcal{N}_R(z)},
\end{equation}
where the previous proportionality can be cast as an equality by introducing the average (Eulerian) density $\overline{\rho}(z)$ and the factor $\mathcal{N}_R(z)$ ensures that the distribution is properly normalized; i.e., 
\begin{equation}
    \Bigg\langle\frac{\rho(z|\delta_R)}{\overline{\rho}(z)}\Bigg\rangle_{\mathcal{N}_R} = 1,
\end{equation}
where we have now omitted the subscript on $\rho$, which is understood to generically represent the luminosity- or SFR-density as discussed below. This results in a normalization factor that can be expressed in terms of the effective biases and the variance of the matter fluctuations on a scale $R$, $\sigma_R^2$:
\begin{equation}
    \label{eq:norm}
    \mathcal{N}_R(z) = \frac{1}{\sqrt{1-2\gamma_R^{\rm NL}\sigma_R^2}}\exp\Bigg(\frac{\gamma_R^2\sigma_R^2}{2-4\gamma_R^{\rm NL}\sigma_R^2}\Bigg).
\end{equation}
All that remains to set up the basic building blocks of the SFRD and luminosity density power spectra are then the effective biases $\gamma$ and $\gamma^{\rm NL}$. 

Inspection of Eq.~(\ref{eq:effective_density}) indicates that the effective biases can be computed as first and second derivatives of the (log) of the SFR or luminosity density. To specify these derivatives, \texttt{oLIMpus} and \texttt{Zeus21} leverage the extended Press-Schechter (EPS) formalism and the Sheth-Mo-Tormen (SMT) description to characterize the density-dependent halo mass function (HMF) and its integrals \citep{press_formation_1974, bond_excursion_1991, sheth_ellipsoidal_2001}. That is, the global density of star formation or line luminosity is
\begin{equation}
    \label{eq:avg_Lag_dens}
    \Big\langle \rho^{\rm Lag}(z)\Big\rangle = \int dm_h \frac{dn}{dm_h}(m_h, z)f(m_h, z),
\end{equation}
where this is understood to be the mean in Lagrangian space, denoted by the superscript `Lag' (because of the calibration of the SMT HMF\footnote{We note that we have omitted the full expression for the SMT HMF here for brevity --- the full expression is presented in all of the preceding works \citep{munoz_effective_2023, cruz_effective_2025, libanore_effective_2025}.}) and $f(m_h,z)$ associates halos with star formation and/or line luminosity (e.g., for the SFRD $f(m_h,z) = f_bf_{\rm duty}f_\star\dot{m}_h$, and the latter three terms in that expression all depend on $m_h$ and $z$ --- see Sections~IIB and III in Ref.~\cite{cruz_effective_2025} for the full expressions).

The local density in a region of comoving radius $R$ can be expressed similarly: 
\begin{align}
    \label{eq:density_integral}
    \begin{split}
        \rho(z|\delta_R) &=  (1+\delta_R)\rho^{\rm Lag}(z|\delta_R)\\
    &= (1+\delta_R)\int dm_h \frac{dn}{dm_h}(z|\delta_R)f(m_h, z),
    \end{split}
\end{align}
where the $(1+\delta_R)$ factor converts the integral from Lagrangian to Eulerian space. Following the EPS formalism, the overdensity-modulated HMF is a rescaled version of the mean HMF,
\begin{equation}
    \frac{dn}{dm_h}(z|\delta_R) = \frac{dn}{dm_h}(z)\frac{dn^{\rm PS}/dm_h(\delta_R)}{\langle dn^{\rm PS}/dm_h(\delta_R)\rangle},
\end{equation}
and 
\begin{equation}
    \frac{dn^{\rm PS}/dm_h(\delta_R)}{\langle dn^{\rm PS}/dm_h(\delta_R)\rangle} = \mathcal{C}\frac{\tilde{\nu}}{\nu_0}\frac{\sigma_{m_h}^2}{\tilde{\sigma}^2}e^{a_{\rm EPS}(\tilde{\nu}^2-\nu_0^2)/2}
\end{equation}
with 
\begin{equation}
    \tilde{\nu} = \frac{\delta_c-\delta_R}{\tilde{\sigma}}, \nu_0=\frac{\delta_c}{\sigma_{m_h}}, \tilde{\sigma}^2 = \sigma_{m_h}^2 - \sigma_R^2,
\end{equation}
$a_{\rm EPS} = a_{\rm ST} = 0.707$, and $\mathcal{C} = 0.3222$, the latter of which are calibrated with numerical simulations \citep{sheth_large_1999}. 

This expression also specifies the mean density in Eulerian space, which differs because of the additional $(1+\delta_R)$ factor
\begin{equation}
    \label{eq:eulerian_dens}
    \overline{\rho}(z) = \frac{\Big\langle (1+\delta_R)\rho^{\rm Lag}(z|\delta_R)\Big\rangle}{\Big\langle\rho^{\rm Lag}(z)\Big\rangle}\Big\langle\rho^{\rm Lag}(z)\Big\rangle = \phi_R^{\rm LtoE}\Big\langle\rho^{\rm Lag}(z)\Big\rangle,
\end{equation}
where 
\begin{equation}
    \label{eq:LtoE_correction}
    \phi_R^{\rm LtoE}(z) = \frac{1+\big(\gamma_{R,\rm Lag} - 2\gamma_{R,\rm Lag}^{\rm NL}\big)\sigma_R^2}{1-2\gamma_{R,\rm Lag}^{\rm NL}\sigma_R^2}.
\end{equation}
Derivatives of the (log) of the Eulerian or Lagrangian versions of Eq.~(\ref{eq:density_integral}) then yield the effective biases for the quantity of interest.

\subsection{From density to power spectrum}\label{sec:power_spectrum_bckgd}
The lognormal parameterization of the density dependence of the SFR- and line luminosity densities enables analytical computation of the associated correlation function. \texttt{oLIMpus} extends the computation of this quantity from the original \texttt{Zeus21} result by computing the full correlation function to second order in the lognormal density variables. That is, for $a=e^{\gamma_1\delta_{R_1}+\gamma_1^{\rm NL}\delta_{R_1}^2}/\mathcal{N}_1$ and $b=e^{\gamma_2\delta_{R_2}+\gamma_2^{\rm NL}\delta_{R_2}^2}/\mathcal{N}_2$ smoothed on scales $R_1$ and $R_2$ (with the appropriate normalizations $\mathcal{N}_{1,2}$), the correlation function is \citep{libanore_effective_2025}
\begin{equation}\label{eq:full_CF}
    \langle ab\rangle =\exp\left[\dfrac{N^{R_1R_2}}{D^{R_1R_2}}-\log C^{R_1R_2}\right]{-1}, 
\end{equation}
where 
\begin{equation}
\label{eq:CF_buildingBlocks}
\begin{aligned}
     N^{R_1R_2} &= \gamma_1\gamma_2\xi^{R_1R_2}+\\
     &\quad+\gamma_1^2\sigma_{R_1}^2\left[\frac{1}{2}-\gamma_2^{\rm NL}\sigma_{R_2}^2\left(1-\frac{\big(\xi^{R_1R_2}\big)^2}{(\sigma_{R_1}\sigma_{R_2})^2}\right)\right]+\\
     &\quad+\gamma_2^2\sigma_{R_2}^2\left[\frac{1}{2}-\gamma_1^{\rm NL}\sigma_{R_1}^2\left(1-\frac{\big(\xi^{R_1R_2}\big)^2}{(\sigma_{R_1}\sigma_{R_2})^2}\right)\right],\\
     D^{R_1R_2} &=1-2\gamma_2^{\rm NL}\sigma_{R_2}^2-2\gamma_1^{\rm NL}\sigma_{R_1}^2+\\
     &\quad+4\gamma_1^{\rm NL}\sigma_{R_1}^2\gamma_2^{\rm NL}\sigma_{R_2}^2\left(1-\frac{\big(\xi^{R_1R_2}\big)^2}{\sigma_{R_1}^2\sigma_{R_2}^2}\right),\\
    C^{R_1R_2}&=\sqrt{D^{R_1R_2}}\mathcal{N}_1\mathcal{N}_2.
\end{aligned}
\end{equation}
In these expressions $\xi^{R_1 R_2}(r) = G^2(z)\xi_m^{R_1R_2}(r,z=0)$ is the matter correlation function scaled by the growth function $G(z)$ and smoothed over scales $R_1, R_2$. Altogether, this complicated expression reduces to a rescaled version of that matter correlation function at large ($r\gtrsim 7\ {\rm Mpc}$) scales, with the local deviations from the matter power specified by the combination of the linear and nonlinear effective biases $\gamma$. Thus, the Fourier transform of Eq.~(\ref{eq:full_CF}) scaled by the mean SFR- or luminosity density (which depends on smoothing scale $R$) along with any relevant unit conversion factors yields the power spectrum of that quantity of interest:
\begin{equation}
    \label{eq:power_spectrum}
    \begin{split}
        P(k,z) = {\rm FT}\Bigg[A^2(z,R)\Bigg(\exp\bigg(\frac{N^{R}}{D^{R}}-\log C^{R}\bigg)-1\Bigg)\Bigg],
    \end{split}
\end{equation}
where we have set $R_1=R_2=R$ because the SFRD and luminosity density are local quantities, and we have absorbed the mean density, normalization, and any conversion factors into a redshift- and scale-dependent term $A(z,R)$.\footnote{The expression presented in Eq.~(\ref{eq:power_spectrum}) is appropriate for an auto-power spectrum of interest, where $a$ and $b$ in Eq.~(\ref{eq:full_CF}) represent the same observable. In general, the formalism can be applied to cross-correlate two (local) signals as well, so long as the effective biases $\gamma_i$ and $\gamma_i^{\rm NL}$ are appropriately defined. In addition, the prefactor in front of the correlation function then becomes $A(z,R)B(z,R)$.}   In practice, the value for $R$ is set by the observational resolution or the minimum scale for which the second order lognormal approximation is valid ($R_{\rm min}\sim 1\ {\rm Mpc}$, see Ref.~\cite{libanore_effective_2025}). In the case of the SFRD power spectrum, for example, $A(z,R) = \overline{\rho}_{\rm SFR}(z, R) = \phi_R^{\rm LtoE}(z)\Big\langle \rho_{\rm SFR}^{\rm Lag}(z)\Big\rangle$ (the corresponding coefficients for line luminosity density are summarized in Section IIIB of Ref.~\cite{libanore_effective_2025}).

\subsection{Redshift-space distortions} \label{sec:RSD}
In addition to the aforementioned contributions to the power spectrum, we additionally account for perturbations to the observable positions of emitting sources in redshift-space due to their peculiar velocities imprinted by the matter overdensity field, an effect known as \textit{redshift-space distortions} (RSD). This introduces an artificial enhancement to overdensities and thus the variations in the clustering signal, which manifest as an overall boost to the observed power spectrum. 

We quantify the effects of RSD as in Ref.~\cite{libanore_effective_2025} (see Appendix C therein, which follows Refs.~\cite{kaiser_clustering_1987, peebles_principles_1993, dodelson_modern_2003}), and summarize a few salient pieces here for completeness.\footnote{We note that the derivation outlined in Ref.~\cite{libanore_effective_2025} is only appropriate at very low redshift, where the Hubble parameter $H(z)\approx H_0$ and the scale factor $a\to 1$. Here, we write out the expressions for the position in redshift space $\vec{x}_s$ and the line-of-sight velocity $\vec{v}$ (which satisfies $v \equiv |\vec{v}| = ifaH\delta/k$, which both involve factors of the Hubble parameter and scale factor, rather than $H_0$ \citep{dodelson_modern_2003}. Ultimately these extra factors cancel and the result is the same, but we reproduce the full expressions here for completeness.} The positions in redshift space (where source distances are associated with their redshifts; denoted by $\vec{x}_s$) are related to the real space positions (denoted by $\vec{x}$) through $x_s = z/(aH(z))$ and $z = aH(z)x + \vec{v}\cdot \hat{x}$. Because the local number of emitting sources is conserved under this coordinate transformation, the densities in redshift- and real-space can be related by the Jacobian
\begin{align}\label{eq:RS_density}
\begin{split}
    \rho_s(\vec{x}_s) &= J^{-1}\rho(\vec{x})\simeq \bigg(1 - \frac{\partial}{\partial x}\frac{\vec{v}\cdot\hat{x}}{aH}\bigg)\rho(\vec{x}) \\
    &\approx \bigg(e^{\gamma_R\delta_R(\vec{x}) + \gamma_R^{\rm NL}\delta_R^2(\vec{x})} - \frac{\partial}{\partial x}\frac{\vec{v}\cdot\hat{x}}{aH}\bigg)\frac{\overline{\rho}(z)}{\mathcal{N}_R(z)},
\end{split}
\end{align}
where in the first line we have applied the Kaiser approximation, and in the second we have introduced the second order lognormal approximation for the density (Eq.~(\ref{eq:effective_density})), keeping terms only to first order in both lines.

With these expressions, the correlation function of the redshift-space densities is then 
\begin{align}
    \label{eq:RSD_CF}
    \begin{split}
    \Big\langle &\rho^A_s(\vec{x}_{s,1})\rho^B_s(\vec{x}_{s,2})\Big\rangle = \Bigg\langle \bigg(e^{\gamma_R\delta_R(\vec{x}) + \gamma_R^{\rm NL}\delta_R^2(\vec{x})} - \frac{\partial}{\partial x}\frac{\vec{v}\cdot\hat{x}}{aH}\bigg)  \\ 
    &\qquad \times \bigg(e^{\gamma_R\delta_R(\vec{x}) + \gamma_R^{\rm NL}\delta_R^2(\vec{x})} - \frac{\partial}{\partial x}\frac{\vec{v}\cdot\hat{x}}{aH}\bigg)\Bigg\rangle A(z,R)B(z,R),
    \end{split}
\end{align}
where the $A$ and $B$ factors make explicit that this expression can apply to the cross-correlation of a local signal as well. As before, the Fourier transform of this quantity yields the power spectrum. 

For the cross-correlation of two lines with rest frequencies $\nu_1$ and $\nu_2$ and intensities $I_i$, for example, this expression becomes\footnote{We note that this expression is subtly different from that in Ref.~\cite{libanore_effective_2025}, in which the last two terms are incorrectly labeled with the same subscripts $\overline{I}_{\nu_i}P_{\nu_i m}$. This would break the symmetry of the cross power and artificially enhance the inferred signal. We correct that procedure here and in the \texttt{oLIMpus} code.}
\begin{align}\label{eq:cross_wRSD}
    \begin{split}
        P_{\nu_1\nu_2}^{\rm RSD}(k,z) = &P_{\nu_1\nu_2}(k,z) + \overline{I}_{\nu_1}(z)\overline{I}_{\nu_2}(z)f^2(z)\mu^4 P_m(k,z) \\
        &+ f(z)\mu^2\overline{I}_{\nu_1}(z)P_{\nu_2 m}(k,z) \\
        &+ f(z)\mu^2\overline{I}_{\nu_2}(z)P_{\nu_1 m}(k,z),
    \end{split}
\end{align}
where $\mu$ is the cosine of the angle between $\vec{k}$ and the line of sight and $f(z) = d\ln G(z)/d\ln a$ is the differential growth rate. The first term is the real space power spectrum (Eq.~(\ref{eq:power_spectrum})), the second term is the power spectrum of the two velocity fields (which trace the matter field), and the final two terms are the cross-correlation of one emission field with the density field (traced by the other line). These intensity-matter cross-power spectra $P_{\nu_i m}$ are similar to before (see Appendix C of Ref.~\cite{libanore_effective_2025} for the full expression), and include one factor of the mean intensity of the line of interest. 

The remainder of our treatment of RSD is identical to that of Ref.~\cite{libanore_effective_2025}, so we direct the interested reader to that work for any further detail and discussion.

\subsection{Shot noise}\label{sec:shot_noise}
Because the line emission is sourced by a discrete population of galaxies embedded within a clustered field of DM halos, there is an additional contribution to the power spectrum --- the so-called `shot' noise effect. \texttt{oLIMpus} employs the standard parameterization of shot noise. That is, the shot power is proportional to the second moment of the source luminosity function:
\begin{equation}
    \label{eq:shot_noise}
    P_{\rm shot}(z)  \propto \int d m_h  \frac{dn}{dm_h} \big\langle L^2(m_h,z)\big\rangle,
\end{equation}
where the $\langle\cdot\rangle$ brackets make explicit that this is computed relative to the mean line luminosity inferred from the distribution of luminosities as a function of halo mass and the proportionality can be made an equality by accounting for the relevant factors converting line luminosity to mean intensity (and which we omit here for brevity). 

As in Ref.~\cite{libanore_effective_2025}, we set the line luminosity (described in detail in Section~\ref{sec:line_lum}) to be the median of a lognormal relation between halo mass and luminosity:
\begin{equation}\label{eq:stochastic_L}
    \tilde{L}_\nu(m_h,z) = \int \frac{dL'}{\sqrt{2\pi\sigma_L^2}}\exp\Bigg[-\frac{\big(\log L' - \log L_\nu(m_h, z)\big)^2}{2\sigma_L^2}\Bigg],
\end{equation}
where the median of the distribution is given by the (log of the) deterministic line luminosity relation given in Eq.~(\ref{eq:line_lum_from_sfr}) and the variance is a free parameter $\sigma_L^2$, represented in units of dex (in principle this can be independently fit through a comparison of different star-formation tracers as in Ref.~\cite{munoz_relatively_2026}).\footnote{The variance of the $L_\nu-m_h$ distribution can also vary with halo mass (see e.g., \cite{furlanetto_bursty_2022, gelli_impact_2024, munoz_relatively_2026, lazare_when_2026}), though here we take it to be a constant for simplicity.} Therefore, introducing stochasticity amounts to increasing the average luminosity density as it allows smaller halos to scatter into brighter luminosity bins. 

Shot noise introduces a scale independent contribution to the power spectrum. At large scales, this effect is subdominant relative to the clustering signal, but at the smallest scales, this becomes increasingly important and can enhance power and thus signal detectability \citep{kovetz_when_2026, lazare_when_2026}.

\subsection{\texttt{oLIMpus} with Pop III stars}\label{sec:popIII_bckgd}
The preceding sections outline the procedure for computing power spectra sourced by a single stellar population whose SFRs or luminosities vary only with density. Here, mirroring the procedure introduced in Ref.~\cite{cruz_effective_2025}, we extend the \texttt{oLIMpus} framework to include Pop~III stars, whose abundance is spatially modulated by the relative DM-baryon velocity and inhomogeneous Lyman-Werner (LW) radiation backgrounds in addition to the local overdensity (see e.g., \cite{tseliakhovich_relative_2010, mcquinn_impact_2012, naoz_simulations_2012, naoz_simulations_2013, Munoz:2019rhi, lake_supersonic_2021, Munoz:2021psm, lake_early_2024, williams_supersonic_2023, williams_lighting_2024, hegde_self-consistent_2023, nebrin_starbursts_2023, kulkarni_critical_2021, klessen_first_2023} for some recent background on these feedback effects).

As in Ref.~\cite{cruz_effective_2025}, we model Pop~III stars by modifying the star formation efficiency and duty cycle in Eqs.~(\ref{eq:avg_Lag_dens}-\ref{eq:density_integral}) and we adopt the same fiducial model described therein (see their Fig. 2). Briefly, in this classical Pop~III framework, Pop~III stars form in minihalos of $m_h\sim 10^6-10^8 M_\odot$ (up to the atomic cooling threshold) while Pop~II stars dominate in more massive systems, from the atomic cooling threshold up through a peak halo mass of $m_{\rm peak}\sim 10^{11} M_\odot$. In the Pop~III, H$_2$-cooling dominated regime, star formation is regulated by feedback affecting the buildup of H$_2$ (e.g., \cite{hegde_self-consistent_2023, nebrin_starbursts_2023}) and star formation proceeds less efficiently, yielding efficiencies nearly an order of magnitude lower than in metal-enriched environments (leading to the formation of a few, more massive stars in each star-forming cloud; \cite{bromm_formation_2013, loeb_first_2013, klessen_first_2023}). However, given the uncertainties in all of these parameters, in this work we explore possible deviations and the broadly unconstrained parameter space that may result.

Ref.~\cite{cruz_effective_2025} demonstrated that, in the presence of these feedback and environmental processes, to a good approximation, the density building block introduced above (Section~\ref{sec:density_bckgd}) can be separated as follows
\begin{equation}\label{eq:density_popIII}
    \rho^{\rm III}(z|\delta_R, v_{\rm bc}^2) = \overline{\rho}(z)\V(z|v_{{\rm bc}, R}^2)\D(z|\delta_R),
\end{equation}
where the (over)density and velocity factors are defined as
\begin{equation}
    \D\big(z|v_{{\rm bc}, R}^2\big) = \frac{\rho\big(z|\delta_R, v_{\rm bc, avg}^2\big)}{\overline{\rho}(z)}
\end{equation}
and
\begin{equation}
    \V\big(z|v_{{\rm bc}, R}^2\big) = \frac{\rho\big(z|\delta_R, v_{{\rm bc}, R}^2\big)}{\rho\big(z|\delta_R, v_{\rm bc, avg}^2\big)}.
\end{equation}
Note that here, the numerator in the overdensity building block is the same as the generic expression introduced in Eq.~(\ref{eq:effective_density}). In addition to carrying out the density integrals summarized in Eq.~(\ref{eq:density_integral}), Ref.~\cite{cruz_effective_2025} also incorporate the effects of LW feedback by modifying the linear biases $\gamma_R\to \gamma_R + \Delta\gamma_R\Big(J_{\rm LW}(z|\delta_R)\Big)$ based on a calculation of the evolving LW radiation background, which itself depends on the past SFRD history (see their Section VA for more details on this procedure).

Ref.~\cite{cruz_effective_2025} demonstrated that the velocity term---which only affects the Pop III hosting ``minihalos''---can be written as a log-$\chi^2$ variable, or
\begin{equation}\label{eq:vel_mod}
    \V(z|\delta_R, v_{{\rm bc}, R}^2)\approx \frac{\Lambda_Re^{-\lambda_R\tilde{\eta}_R} + \Omega_Re^{-\omega_R\tilde{\eta}_R}}{\Lambda_R(1+2\lambda_R)^{-3/2} + \Omega_R(1+2\omega_R)^{-3/2}},
\end{equation}
where $\tilde{\eta}_R \equiv 3v_{{\rm bc}, R}^2/\sigma_{\rm bc}^2$ provides a $\chi^2$ representation of the relative velocity field with three degrees of freedom (as it is the sum of 3 Gaussian random variables), $\Lambda_R$ and $\Omega_R$ are normalizations, and $\lambda_R$ and $\omega_R$ are the effective velocity biases.

\begin{figure}
    \centering
    \includegraphics[width=\linewidth]{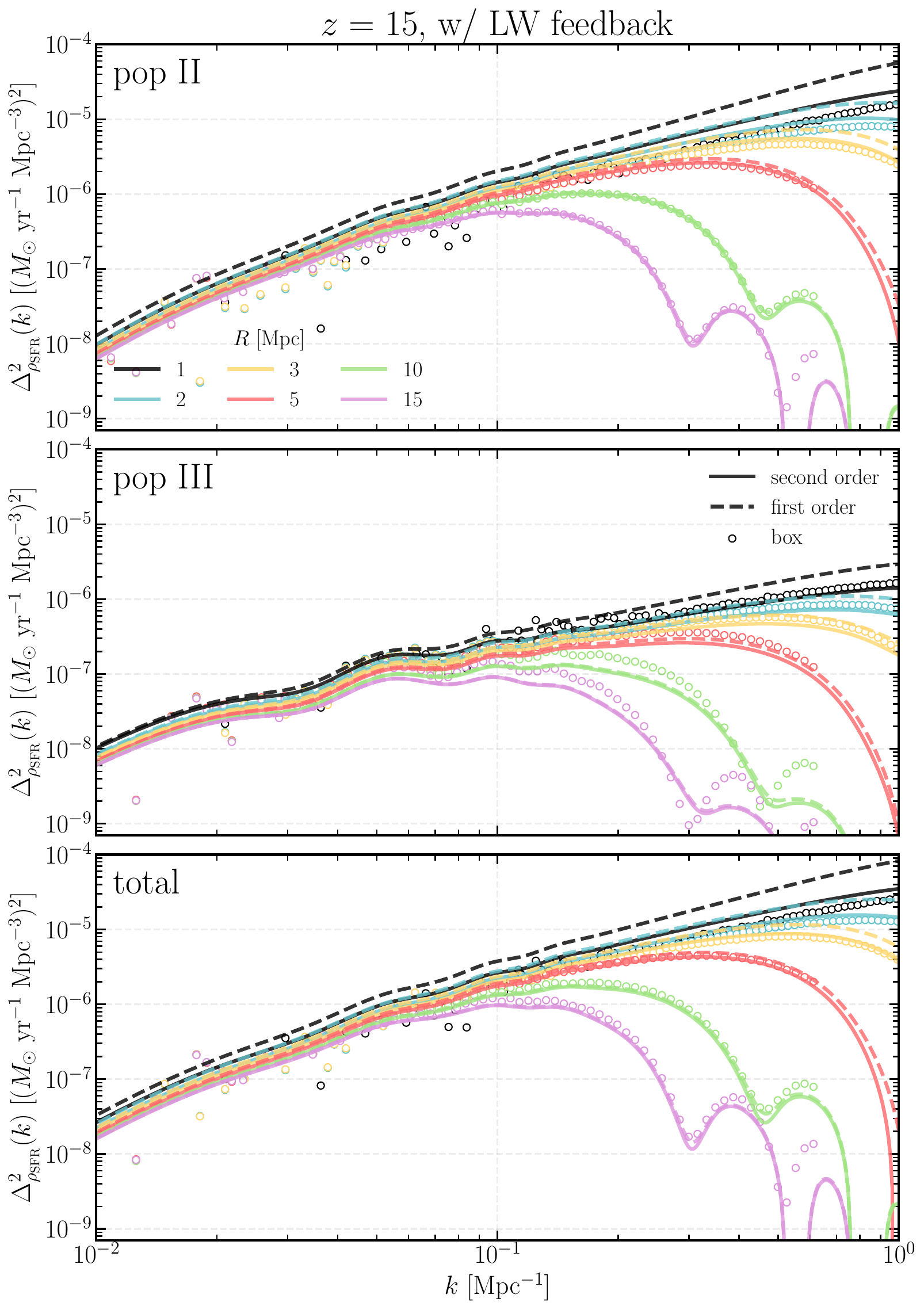}
    \caption{\textbf{\texttt{oLIMpus}/\texttt{Zeus21} agree well with a numerical calculation of the SFRD auto power spectrum.} Comparison between the SFRD power spectrum computed with a numerical simulation (points) and the analytic approximation from \texttt{oLIMpus}/\texttt{Zeus21} at $z=15$ for a range of smoothing radii (different colors). In order to resolve the large scales while maintaining computational efficiency we use boxes of different sizes and pixel scales for each of the specified smoothing scales --- in particular, $(L_{\rm box}/{\rm Mpc}, N_{\rm box})= (300,300), (600,300), (900, 300), (1000, 200)$ for $R=1,2,3,$ and 5+ Mpc, respectively. The large-scale difference in shape in the Pop III case is due to the lack of radiation backgrounds in the numerical calculation. We note that in order to maintain the appropriate scale-dependent conversion between a Lagrangian density field and the Eulerian SFR- or luminosity density as defined in \texttt{Zeus21}, the numerical density field is smoothed \textit{before} computing Eq.~(\ref{eq:density_integral}). Though we only show the agreement at one redshift here, we note that similar agreement is seen down to $z\sim 5$ and $R\sim 1\ {\rm Mpc}$.}
    \label{fig:SFRD_PS_ex}
\end{figure}

As before, expressing the density and velocity terms with these approximate forms enables a fully analytical expression of their correlation functions, albeit with somewhat more bookkeeping due to the two multiplicative factors. Ultimately, we need the total density correlation function or power spectrum, so we must calculate the correlation function of the following expression
\begin{align}\label{eq:full_density_IIandIII}
\begin{split}
    \rho(z|\delta_R, v_{\rm bc}^2) &= \overline{\rho}^\II(z)\D^\II(z|\delta_R) \\ &+ \overline{\rho}^\III(z)\D^\III(z|\delta_R)\V\big(z|v_{{\rm bc}, R}^2\big)
\end{split}
\end{align}
The correlation function of this quantity is then
\begin{align}
\begin{split}
    \xi_{\rho}(r) &= \langle \rho_{R_1}(z_1)\rho_{R_2}(z_2)\rangle \\ &- \langle \rho_{R_1}(z_1)\rangle\langle\rho_{R_2}(z_2)\rangle,
\end{split}
\end{align}
which, after substitution of the terms from Eq.~(\ref{eq:full_density_IIandIII}) can be expressed in terms of correlation functions of the density and velocity factors:
\begin{equation}
\label{eq:CF_popIIIandII}
    \begin{split}
        \xi_{\rho}(r) &= \overline{\rho}^\II_1(z)\overline{\rho}_2^\II(z)\xi_\D^\II(r) \\
        &+ \overline{\rho}^\II_1(z)\overline{\rho}^\III_2(z)\xi_\D^{\rm II\times III}(r) \\
        &+\overline{\rho}^\III_1(z)\overline{\rho}^\II_2(z)\xi_\D^{\rm III\times II}(r) \\
        &+ \overline{\rho}_1^\III(z)\overline{\rho}_2^\III(z) \big[\xi_\V(r) \xi_\D^\III(r) + \xi_\V(r) + \xi_\D^\III(r)\big],
    \end{split}
\end{equation}
where the density correlation functions\footnote{The cross terms $\xi^{\rm II\times III}$ can be found using Eq.~(\ref{eq:full_CF}) as well, taking $\gamma_1 = \gamma_{R_1}^{\rm II}$ and $\gamma_2 = \gamma_{R_2}^{\rm III}$, for example.} are those outlined in Eq.~(\ref{eq:full_CF}). We note that in the case of SFRD or line auto power, the mean densities are the same, $\rho^{\rm (i)}_1 = \rho^{\rm (i)}_2$, but make explicit that these can in principle differ if, for example, we are computing the cross power for lines $\nu_1$ and $\nu_2$. We have omitted the full expressions of the velocity correlation function here for brevity but note that it is defined in Appendix A of Ref.~\cite{cruz_effective_2025}. As in Eq.~(\ref{eq:power_spectrum}), the Fourier transform of this correlation function yields the power spectrum of SFR- or luminosity density fluctuations, which are shown as the solid curves in Fig.~\ref{fig:SFRD_PS_ex}.

\subsection{Numerical validation}\label{sec:maps}
To validate the analytical calculations outlined in Sections~\ref{sec:density_bckgd}-\ref{sec:popIII_bckgd}, we carry out a numerical calculation of the same power spectrum with a simulated box, following a similar procedure to that described in Ref.~\cite{libanore_effective_2025} (see also Refs.~\cite{munoz_robust_2019, munoz_impact_2022}). Briefly, given a box size and pixel resolution, we generate a Gaussian random field of densities and three Gaussian fields of velocities (corresponding to the three components of the relative velocity field), the latter of which we subsequently square and sum to compute the squared magnitude of the velocity field. Following the procedure outlined above (Sec.~\ref{sec:density_bckgd}), we then smooth these fields with a 3D top-hat window function of size $R$ and compute the SFR- or luminosity density for each pixel with mean overdensity $\delta_R$ and velocity $v_{{\rm bc}, R}$ with Eq.~(\ref{eq:eulerian_dens}), where the dependence on density is assumed to manifest through the halo mass function and the velocity through the star formation duty cycle (absorbed into $f(m_h,z)$) in the Pop III case. Finally, we use the \texttt{powerbox} code \citep{murray_powerbox_2018} to compute the power spectrum, which is then shown as the individual points in Fig.~\ref{fig:SFRD_PS_ex}.\footnote{There is a subtle difference in this procedure compared to that presented in Ref.~\cite{libanore_effective_2025}, where the box is smoothed \textit{ after} the SFR- or luminosity density field is generated. This, however, incorrectly maps the field from Lagrangian to Eulerian space, which we remedy here.}

Taking the fiducial parameters given in Table I of Ref.~\cite{cruz_effective_2025}, we compare these two approaches to computing the SFRD power spectrum in Fig.~\ref{fig:SFRD_PS_ex}. From this, we find excellent agreement between the analytical and numerical power spectra to within a factor of 2, even down to the smallest spatial and smoothing scales. In the Pop III case, we note that the difference in shape of the analytical and numerical estimates at large smoothing scales is due to the lack of self-consistently computed LW radiation backgrounds in the numerical calculation.

\begin{figure*}
    \centering
    \includegraphics[width=\linewidth]{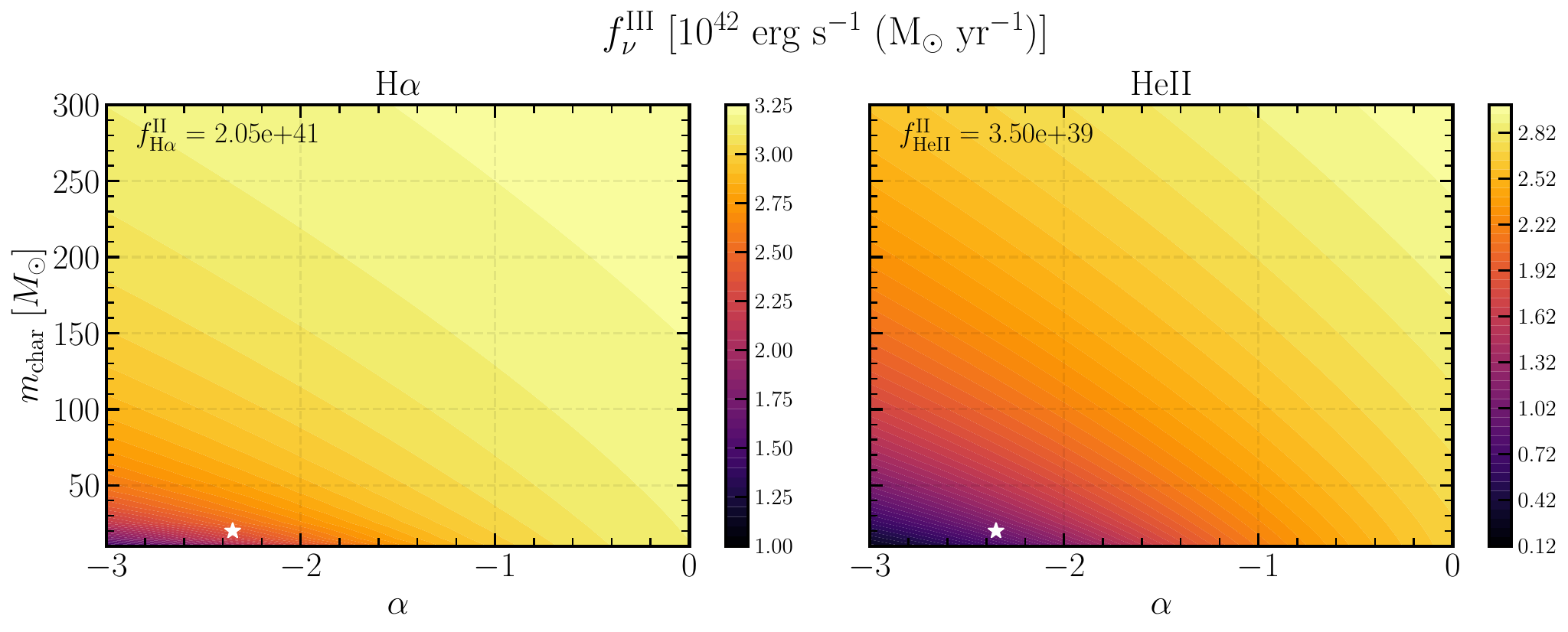}
    \caption{\textbf{HeII is a sensitive tracer of the presence of massive stars}. Contours of the Pop~III line coefficients mapping SFR to H$\alpha$ (left) or HeII (right) luminosity for a range of IMF parameters (see Eq.~(\ref{eq:chabrier_imf})), computed using Eq.~(\ref{eq:f_line}) with the tables compiled in Ref.~\cite{mas-ribas_boosting_2016}. We list the associated coefficients for Pop~II stars in the upper left corner of each panel and highlight our fiducial choice for Pop~III IMF parameters with a white star. We note that for all the IMFs explored, at fixed SFR, Pop~III stars are more effective producers of both H$\alpha$ and HeII ionizing photons.}
    \label{fig:line_coeffs}
\end{figure*}

\section{Pop III line luminosities}\label{sec:line_lum}
We next describe our procedure for connecting a galaxy's SFR to the particular emission lines that might be of interest for a Pop~III intensity mapping measurement. While \texttt{oLIMpus} has built-in models for a few commonly used lines, these generally correspond to higher metallicity stellar populations, so here we expand the framework to include estimates based on theoretical models for Pop III stars (e.g., \cite{marigo_zero_2001, schaerer_properties_2002, kubat_spherically_2012, mas-ribas_boosting_2016}), allowing for flexible variation of the Pop~III IMF. 

For a particular recombination line with rest frequency $\nu$, the line luminosity associated with a single galaxy is
\begin{equation}\label{eq:line_lum_from_sfr}
    L_\nu(m_h, z) = (1-f_{\rm esc, \nu})\big[\big\langle f_\nu^\II\big\rangle\dot{m}_\star^\II + \big\langle f_\nu^\III\big\rangle\dot{m}_\star^\III\big]
\end{equation}
where $f_{\rm esc, \nu}$ is the escape fraction of ionizing photons (for that line) and the angle brackets $\langle\cdot \rangle$ represent an average over a standard Salpeter IMF for Pop~II stars and a flexible IMF for Pop~III as described below (note that these terms represent the luminosity per unit star formation). 

The IMF-averaged line fluxes are computed using the compilation of Pop III stellar tables from Ref.~\cite{mas-ribas_boosting_2016} as follows. For a given IMF $\phi(m) = dN/dm$, the mean cumulative number of ionizing photons produced per unit stellar mass in a burst is given by
\begin{equation}
    \label{eq:ionizing_photon_counts}
    \bigg\langle\frac{N_{i}}{m_\star}\bigg\rangle = \frac{\int_{m_{\rm min}}^{m_{\rm max}}Q_i(m)t_{\rm life}(m)\phi(m)dm}{\int_{m_{\rm min}}^{m_{\rm max}}m\phi(m)dm}
\end{equation}
for a species $i = {\rm H,\ HeI,\ HeII}$, where the $Q_i$ are the ionizing photon production rates and $t_{\rm life}(m)$ is the stellar lifetime. For each such ionization, a fraction $f_{\rm rec, \nu}$ will result in recombinations that produce a line photon. The product of this quantity with the energy of a single line photon yields a constant that maps each ionizing photon produced to a total energy output in recombination line luminosity. Put together, the total luminosity per unit star formation associated with a given line can then be computed through
\begin{equation}\label{eq:f_line}
    \langle f_{\nu} \rangle = f_{\rm rec, \nu}E_\nu \bigg\langle\frac{N_{i}}{m_\star}\bigg\rangle.
\end{equation}
In this work, unless otherwise specified, we will describe Pop~III stars with a Chabrier-like IMF of the form
\begin{equation}\label{eq:chabrier_imf}
    \frac{dN}{dm} \propto m^{\alpha}\exp\Bigg[-\bigg(\frac{m_{\rm char}}{m}\bigg)^\beta\Bigg],
\end{equation}
with $\beta = 1.6$, and the characteristic mass $m_{\rm char}$ and IMF slope $\alpha$ varied, but fiducially set to $20\ M_\odot$ and $-2.35$, respectively, consistent with Refs.~\cite{hegde_self-consistent_2023, hegde_efficient_2025}, though we note that shape of the IMF is highly uncertain and could be substantially more top-heavy than that chosen here \citep{stacy_building_2016, he_simulating_2019, klessen_first_2023, chon_impact_2024, lake_stellar_2025, sharda_role_2019, sharda_importance_2020, sharda_interplay_2023, sharda_magnetic_2025, sharda_population_2025}. While not exactly identical to the IMFs discussed in any one of those works, we choose the aforementioned fiducial parameters to straddle the extremes of the parameter space --- namely producing more massive stars than a typical Salpeter or Chabrier IMF while still maintaining the Salpeter-like slope at high masses. In this sense, our chosen IMF represents a conservative --- though nevertheless top-heavy --- descriptor of the pristine population. We relax the assumed values for $\alpha$ and $m_{\rm char}$ and explore variations to the IMF in subsequent sections below.

For Pop~II stars, we compute the associated ${\rm H}\alpha$ scaling as $f_{\rm H\alpha}^\II = \xi_{\rm ion}E_{\rm H\alpha}f_{\rm rec, H\alpha}/\kappa_{\rm UV}$, with $\xi_{\rm ion} = 10^{25.29}\ {\rm Hz\ erg^{-1}}$ from JWST spectra of galaxies at $z\sim 1-6.7$ \citep{pahl_spectroscopic_2025}\footnote{We note that the H$\alpha$ luminosity is highly sensitive to the age of the stellar population (and thus the burstiness). Therefore, in detail the instantaneous luminosity in a given burst may exceed the lifetime-averaged value we adopt here for simplicity.} and $\kappa_{\rm UV} = 1.15\times 10^{-28}\ M_\odot {\rm \ yr^{-1}\ (erg\ s^{-1}\ Hz^{-1})^{-1}}$ is taken to be a standard SFR-UV luminosity conversion factor for Pop II stars modeled with a Salpeter IMF \citep{madau_cosmic_2014}. The HeII scaling is taken from Ref.~\cite{cassata_HeII_2013} (inferred from observations of Wolf-Rayet stars at $z\sim 2-4.6$), and is listed for comparison in the right panel of Fig.~\ref{fig:line_coeffs}.

Fig.~\ref{fig:line_coeffs} shows the sensitivity of these coefficients to the Pop~III IMF. While the H$\alpha$ luminosity varies by a factor of 3 over the wide range of IMF parameters, the HeII 1640 $\AA$ luminosity varies by a factor of nearly 30 over that same range, highlighting the utility of the HeII line luminosity as a diagnostic of the presence of massive stars. Because Pop~III stars dominate the HeII signal in typical models, the relative correlation between the H$\alpha$ and HeII signals is inherited from the host halo populations of Pop~II and III stars, respectively (see Section~\ref{sec:popIII_bckgd} and Fig.~2 in Ref.~\cite{cruz_effective_2025}). 

Because of the direct scaling given by Eq.~(\ref{eq:line_lum_from_sfr})), and using the definition that $\dot{m}^{\rm (i)}_\star(m_h, z) \equiv \dot{m}_h(m_h, z) f_\star^{\rm (i)}(m_h)f_b f_{\rm duty}^{\rm (i)}(v_{{\rm bc}})$, Eq.~(\ref{eq:density_integral}) can be straightforwardly recast as the comoving line emissivity (or line luminosity density) in a region of radius $R$:
\begin{equation}\label{eq:line_emissivity}
\begin{split}
    \rho_{\nu}(z) &= \frac{1+\delta_R}{\nu}\int \frac{dn}{dm_h}L_{\nu}(m_h)dm_h \\
    &= \frac{1-f_{\rm esc, \nu}}{\nu}\bigg[f_\nu^\II\rho_{\rm SFR}^\II + \langle f_\nu^\III\rangle \rho_{\rm SFR}^\III\bigg].
\end{split}
\end{equation}
The mean line intensity is then just a linear combination of the local SFRDs,
\begin{equation}\label{eq:mean_intensity}
\begin{split}
    J_{\nu}(z|\delta_R, &v_{{\rm bc}, R}) = \frac{c}{4\pi}\frac{\rho_{\nu}(z)}{H(z)} \\
    &= \frac{c(1-f_{\rm esc, \nu})}{4\pi\nu_{0}H(z)}\bigg[f_\nu^\II\rho_{\rm SFR}^\II + \langle f_\nu^\III\rangle \rho_{\rm SFR}^\III\bigg],
\end{split}
\end{equation}
where we have omitted the dependence of the SFRD on density and velocity for brevity. Throughout this work we will assume that $f_{\rm esc, \nu} \sim 0$ for simplicity.

Thus, the same formalism presented in Section~\ref{sec:power_spectrum_bckgd} (specifically Eqs.~(\ref{eq:effective_density}), (\ref{eq:CF_popIIIandII}), and (\ref{eq:power_spectrum})) can be applied to the luminosity density to compute the correlation function and line intensity power spectra for a Pop III IMF of choice.

\section{Observational forecasts}\label{sec:obs_forecast}
We next leverage our framework to forecast the optimal observing configurations needed to probe the Pop~III IMF with intensity mapping and suggest complementary probes that can help to constrain the signal.

\newcolumntype{C}{>{$}c<{$}}
\begin{table*}
\caption{\label{tab:survey_specs} Specifications for the representative intensity mapping surveys referenced in this work}
\begin{threeparttable}
\begin{ruledtabular}
\begin{tabular}{C|CCC}
  {\rm parameters} & {\rm SPHEREx-deep}\tnote{a}  & {\rm CDIM}\tnote{b} & {\rm CDIM+}\tnote{b} \\
  \midrule
  R &  100 & 300 & 500\\
  \sigma_{\rm noise}\ [{\rm erg\ s^{-1}\ cm^{-2}\ sr^{-1}\ Hz^{-1}}]& 3\times 10^{-20} & 10^{-19} & 10^{-20} \\
  {\rm pixel\ size\ ['']}  & 6.2 & 1 & 1\\
  \Omega_{\rm survey}\ [{\rm deg}^2] & 200 & 31 & 31 
\end{tabular}
\end{ruledtabular}
\begin{tablenotes}
    \item[a] The instrument sensitivity and spectral resolution are computed from the post-flight measurements described in Ref.~\cite{bock_spherex_2026}. Depending on the redshift of interest, however, $R$ can differ by a factor of 3 (i.e., from $R=41$ to $R=110$ for the different bands), so we approximate this as a constant $R\sim100$ for the simple forecasts described here.
  \item[b] These are motivated by proposed specifications for the Cosmic Dawn Intensity Mapper (CDIM) and an improved version (higher spectral resolution and sensitivity), as specified in Ref.~\cite{parsons_probing_2022} (see also Ref.~\cite{heneka_optimal_2021}).
\end{tablenotes}
\end{threeparttable}
\end{table*}

\begin{figure*}
    \centering
    \includegraphics[width=\linewidth]{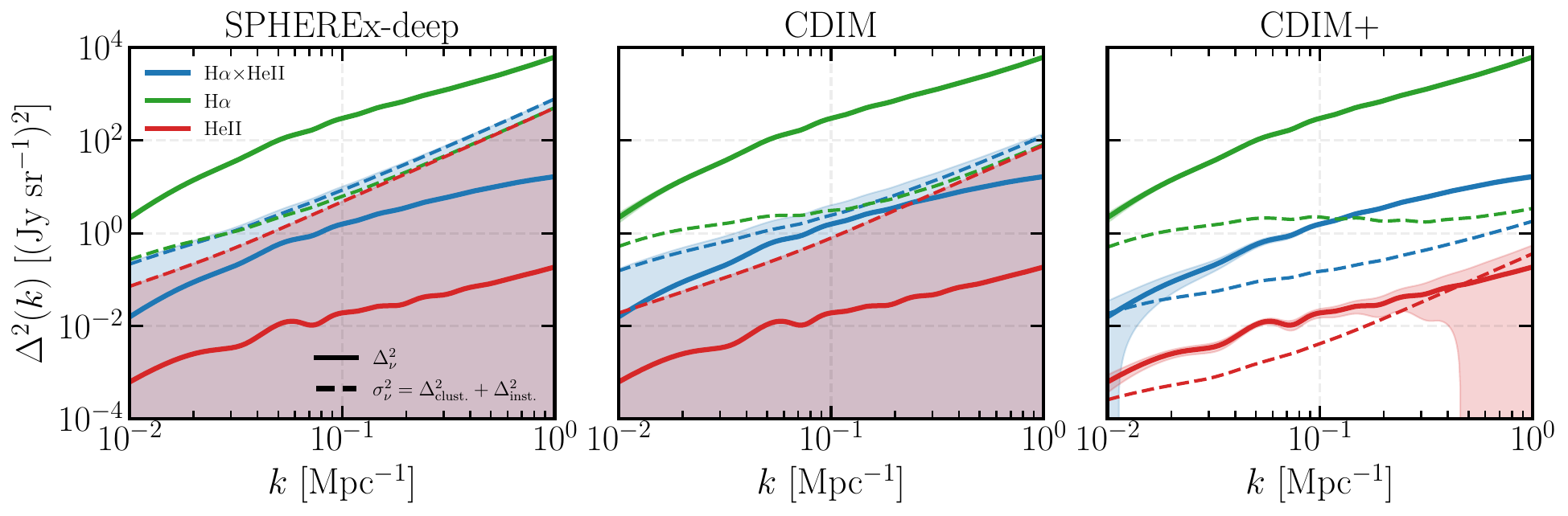}
    \caption{\textbf{The HeII signal is only detectable in cross-correlation in the optimistic survey case.} Example auto- and cross-power spectra with noise forecast for the three surveys summarized in Table~\ref{tab:survey_specs} at $z=5$ following the procedure described in Section~\ref{sec:estimating_SNR}. The different line colors represent the different lines, with the shaded region and dashed line representing the measurement uncertainty given by Eqs.~(\ref{eq:auto_power_uncertainty}) and (\ref{eq:cross_power_uncertainty}).}
    \label{fig:line_PS_noise_ex}
\end{figure*}

\subsection{Observational effects}\label{sec:estimating_SNR}
We first summarize our procedure for estimating the measurement uncertainties that will affect our ability to constrain the underlying physics from a measurement of the LIM signal. Here we characterize two major observational effects which will modify our signal. First, the instrumental noise will introduce a scale-independent contribution to the measured power. Next, the finite resolution of our survey and the instrument can both spatially and spectrally attenuate the measured power. We briefly summarize the key components of these estimates here and direct the interested reader to Refs.~\cite{lidz_intensity_2011, sun_LIMFAST_2026} for more details. Note that the following assumes a spherically-averaged power spectrum with a directionally-independent on-sky sensitivity $\sigma_{\rm noise}$. 

The variance of an individual Fourier mode in the observed auto power spectrum for a line with a rest frequency $\nu$ is the sum of a sample variance and thermal noise term:
\begin{equation}
    \label{eq:auto_power_uncertainty_permode}
    \sigma^2_{\nu\nu}(k,\mu) = \bigg[P_{\nu\nu} + \frac{P_{\nu\nu}^{\rm noise}}{W^2(k,\mu)}\bigg]^2 = P_{\nu\nu, \rm obs}^2.
\end{equation}
The average thermal noise power spectrum is scale-independent and depends on the beam size and spectral resolution:
\begin{equation}
    \label{eq:P_noise}
    P_{\nu \nu }^{\rm noise} = \sigma_{\rm noise}^2V_{\rm vox},
\end{equation}
with $V_{\rm vox} = \chi^2(z)\Omega_{\rm beam}\delta\nu d\chi/d\nu$ for a beam of size $\Omega_{\rm beam}$, where $\chi(z)$ is the comoving distance to $z$, $d\chi/d\nu$ maps from frequency to distance as $d\chi/d\nu = c(1+z)/(\nu H(z))$, and the frequency resolution is $\delta\nu = \nu/R(1+z)$. The window function accounts for the finite spatial and spectral resolution of the instrument as
\begin{equation}
    W^2(k,\mu) = e^{-k^2\sigma_\perp^2-k^2(\sigma^2_\parallel-\sigma_\perp^2)\mu^2},
\end{equation}
where $\mu=\cos\theta$ is the angle between $\vec{k}$ and the line of sight, $\sigma_\parallel = d\chi/d\nu \delta\nu$, and $\sigma_\perp = \chi\sqrt{\Omega_{\rm beam}}$. The number of independent modes sampled is
\begin{equation}
    \label{eq:N_modes}
    N_{\rm mode}(k) = \ln (10)k^3\Delta\log_{10} k \frac{V_s}{4\pi^2},
\end{equation}
for a survey of volume 
\begin{equation}
    V_s = \chi^2(z)\Omega_s \nu \frac{d\chi}{d\nu}\bigg(\frac{1}{1+z-dz} - \frac{1}{1+z+dz}\bigg),
\end{equation}
where $\Omega_s$ represents the sky coverage of the survey and $dz=1$ is the half-width in redshift space of a given slice. 

Reducing the variance by the number of observed modes and averaging over the line-of-sight angle $\mu$, we finally have
\begin{equation}
    \label{eq:auto_power_uncertainty}
    \frac{1}{\sigma^2_{\nu\nu}(k)} = \int_0^1 \frac{N_{\rm mode}(k)}{\sigma^2_{\nu\nu}(k,\mu)}d\mu.
\end{equation}
The cross-power uncertainty is similar, with:
\begin{equation}
    \label{eq:cross_power_uncertainty}
    \frac{1}{\sigma^2_{\nu\nu'}(k)} = \int_0^1 \frac{2N_{\rm mode}(k)}{P_{\nu\nu'}^2(k) + \sigma_{\nu\nu}(k,\mu)\sigma_{\nu'\nu'}(k,\mu)}d\mu.
\end{equation}
We reproduce the full derivation of this expression in Appendix~\ref{app:cross_power_noise} for completeness.

For a given $k$-range, the signal-to-noise ratio of an observation is
\begin{equation}\label{eq:SNR}
    {\rm S/N} = \sqrt{\sum_{k \rm \ bins}\Bigg(\frac{P_{\nu\nu}(k)}{\sigma_{\nu\nu}(k)}\Bigg)^2},
\end{equation}
and likewise for the cross-power SNR.

\subsection{Observing Pop~III stars in the fiducial model}
In Table~\ref{tab:survey_specs}, we summarize the relevant quantities for a few existing and potential intensity mapping surveys and in Fig.~\ref{fig:line_PS_noise_ex}, we apply these specifications to the fiducial IMF and star formation model (shown in Fig.~\ref{fig:line_coeffs}) to forecast the possible observational sensitivity at $z\sim 5$ (the highest redshift for which the H$\alpha$ line is captured within the SPHEREx observing window).  

From these estimates, we find that the contribution of Pop~III stars is most pronounced in the HeII auto-power spectrum (as is expected given the relative coefficients in Fig.~\ref{fig:line_coeffs}), evidenced by the clear velocity acoustic oscillations present even at such late times. However, due to the overall low amplitude of the signal and the instrumental specifications, it is unlikely that we will be able to detect the HeII auto power with a current generation instrument, such as SPHEREx. Indeed, it will be challenging to make a confident detection of this auto-power signal in even the most optimistic proposed modification to the Cosmic Dawn Intensity Mapper (CDIM; \cite{cooray_CDIM_2019}), designated the `improved' case in Ref.~\cite{parsons_probing_2022} (though a more top-heavy IMF than the one we fiducially assume here could improve the detectability). 

Instead, as is argued in Ref.~\cite{parsons_probing_2022}, it is more promising to leverage the higher SNR offered by the stronger H$\alpha$ signal --- which includes a significant contribution from Pop~II stars --- and search for the HeII signal in cross correlation. Despite the boost in SNR offered by cross-correlating the signals, it is still unlikely that the HeII contribution will be detectable over the redshift range for which SPHEREx is sensitive to the H$\alpha$ line ($z\lesssim 6$). Indeed, even given the present configuration of the proposed CDIM instruments, only the H$\alpha$ autocorrelation signal is detectable at high significance. If this instrument were modified to the `improved' configuration suggested in (\cite{parsons_probing_2022}; i.e., reducing the instrumental noise by an order of magnitude and increasing the spectral resolution, here labeled `CDIM+'), then the cross-correlation signal is in principle detectable at $z\sim 5$.\footnote{Diversity in star formation histories \citep{caplar_stochastic_2019, iyer_diversity_2020, tacchella_stochastic_2020, sun_seen_2023, hegde_efficient_2025}, metallicity evolution \citep{hegde_efficient_2025}, and temporal evolution of line luminosities \citep{munoz_relatively_2026}, for example, can introduce significant scatter in the line luminosity-halo mass relation and de-correlate the signal between two lines, even on clustering-dominated, linear scales \citep{yang_empirical_2022, liu_effects_2024, sun_LIMFAST_2026}. Incorporating such effects is, however, beyond the scope of the relatively simple star formation model explored in this work, especially in the context of the exploratory forecasts of interest here, so we defer such investigation to future work.} 

\begin{figure}
    \centering
    \includegraphics[width=\linewidth]{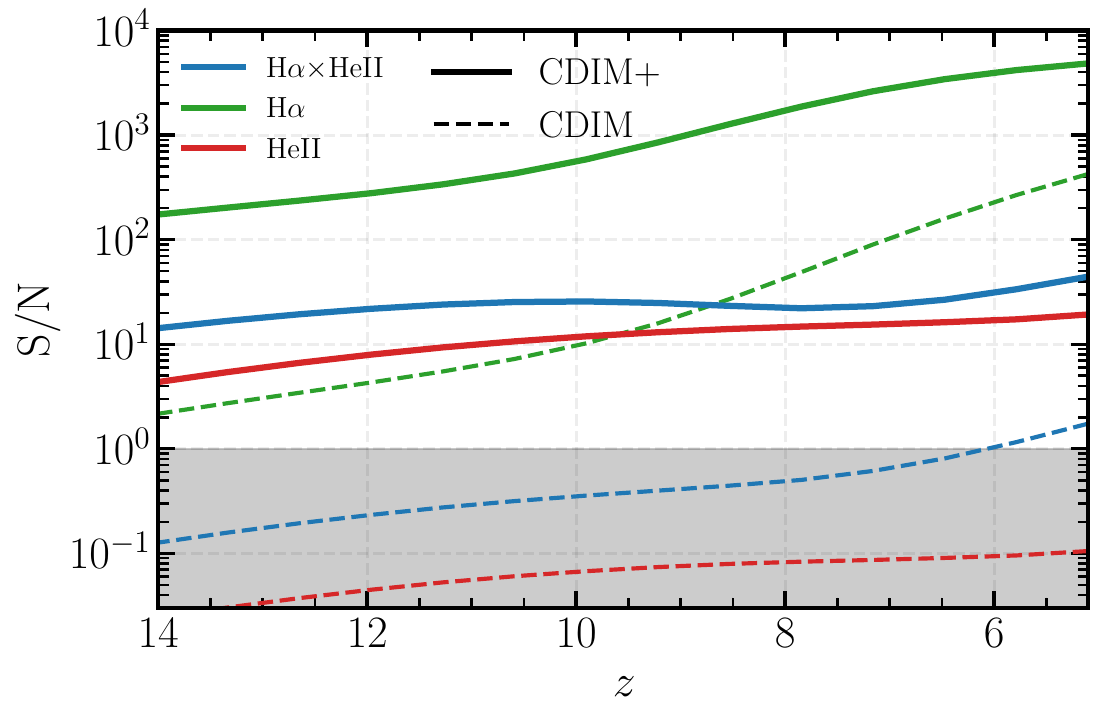}
    \caption{\textbf{The fiducial H$\alpha\times$HeII signal is only detectable at S/N$\gtrsim 5$ at the latest times in the most optimistic  configurations.} S/N estimates over a range of redshifts for the survey specifications summarized in Table~\ref{tab:survey_specs} for auto- and cross-correlation power spectra of the H$\alpha$ and HeII lines. We have grayed out the region below S/N$=1$ to guide the eye but include the curves that fall below to highlight the low significance in this range.}
    \label{fig:SNR_forecast_fid}
\end{figure}

\begin{figure*}
    \centering
    \includegraphics[width=\linewidth]{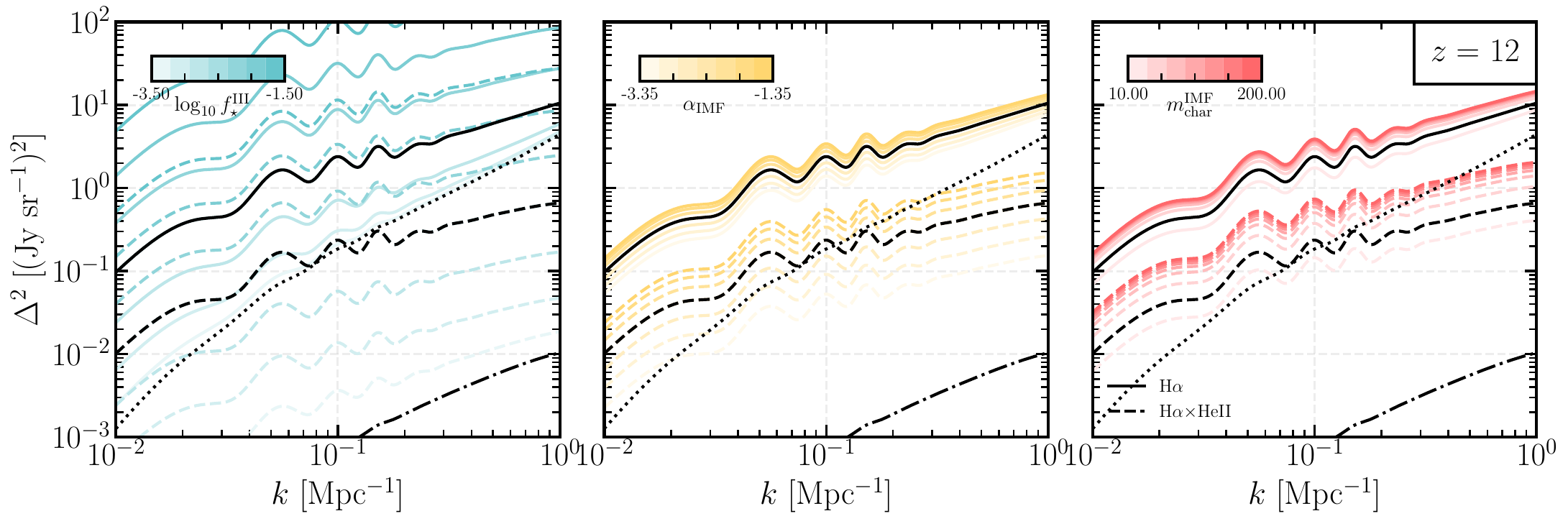}
    \caption{\textbf{At early times, variations in the Pop~III physics sensitively modulate the LIM signal.} The H$\alpha$ auto- and ${\rm H\alpha\times HeII}$ cross power at $z=12$ subject to variations in the parameters characterizing the underlying Pop~III model, with a darkening color indicating a larger value of that parameter (see colorbar). From left to right, we vary the normalization of the Pop~III SFE, the slope of the Pop~III IMF, and the characteristic mass of the IMF. We show the signal with no Pop~III stars as a dotted (H$\alpha$) and dot-dashed (H$\alpha\times$HeII) line in each panel to highlight the significant contribution of Pop~III stars to the shape and normalization of the signal at these early epochs.}
    \label{fig:model_var_classicalIII}
\end{figure*}

\begin{figure}
    \centering
    \includegraphics[width=\linewidth]{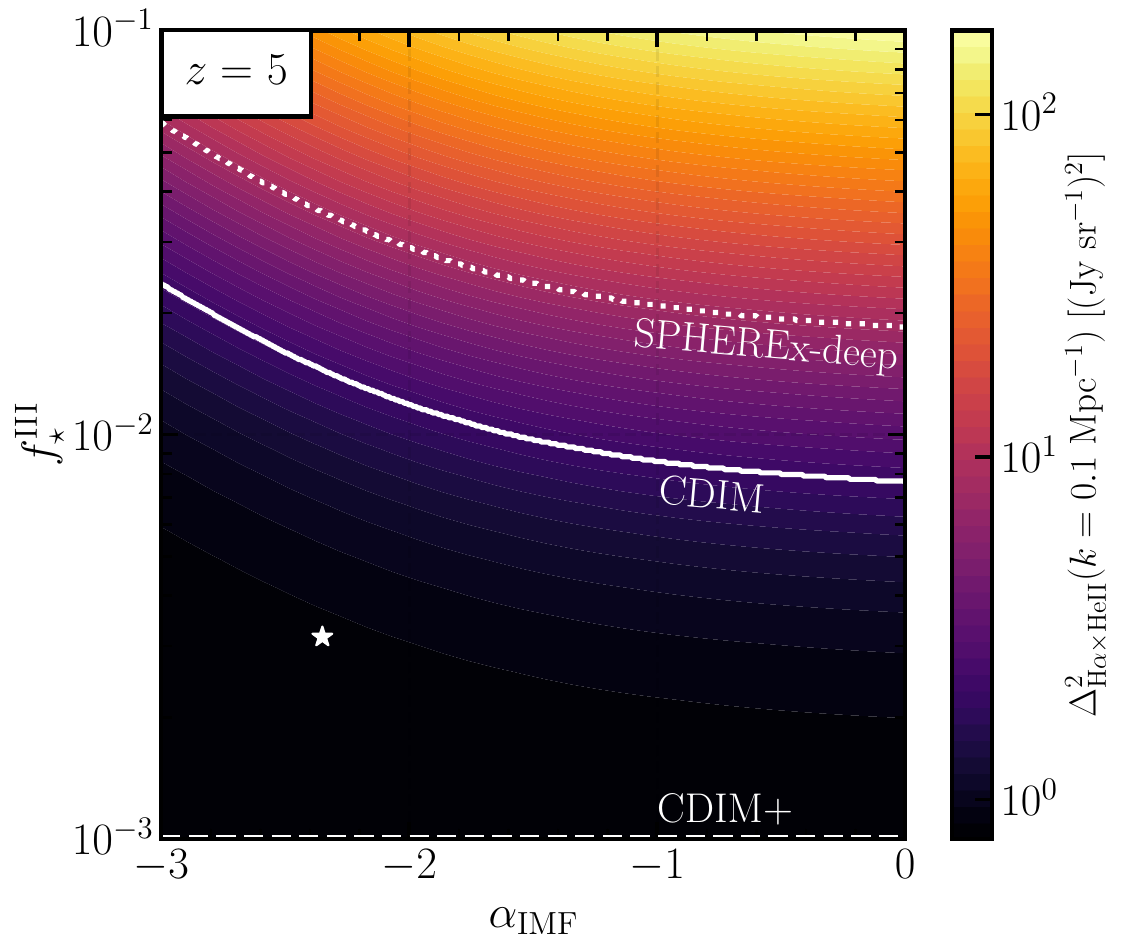}
    \caption{\textbf{SPHEREx and CDIM will be able to place \textit{joint} limits on the Pop~III SFE and IMF slope.} The forecast H$\alpha\times$HeII cross power at $k=0.1\ {\rm Mpc^{-1}}$ at $z=5$ for a range of Pop~III IMF slopes, $\alpha_{\rm IMF}$ and star formation efficiencies $f_\star^{\rm III}$. The \textit{minimum} $(\alpha_{\rm IMF},f_\star^{\III}$) combination necessary to detect the signal in this $k$-bin with $S/N=1$ is indicated with a white line for each of the three sample observing configurations summarized in Table~\ref{tab:survey_specs} (dotted for SPHEREx-deep, solid for CDIM, and dashed for CDIM+). We select a S/N threshold of 1 here for demonstration, but note that the total S/N summed over the $k$-range of interest (30 logarithmic bins from $k=0.01-1\ {\rm Mpc^{-1}}$) is typically a factor of 10 larger than that of the individual $k$-bin used here (of width $\Delta\log_{10}k = 0.1$). We indicate the fiducial parameter set with a white star, reinforcing the previous observation that SPHEREx and CDIM are unlikely to be able to measure the cross power in a classical Pop~III model. The CDIM+ line is shown as a flat line along the lower axis because the $S/N=1$ threshold for that survey configuration falls below the parameter range explored here; CDIM+ would thus be sensitive to the entire Pop~III parameter space shown.}
    \label{fig:parameter_var_classicalIII}
\end{figure}

In Fig.~\ref{fig:SNR_forecast_fid}, we summarize these qualitative conclusions by estimating the total S/N for each of the auto- and cross-power forecasts over a range of redshifts and observing configurations.\footnote{Note that we omit SPHEREx from the discussion in this Fig. because it is only sensitive to H$\alpha$ at $z\lesssim 6$ due to its spectral sensitivity range.} Here, to maintain consistency with Ref.~\cite{parsons_probing_2022}, we sum over 30 $k$-bins from $k_{\rm min} = 0.1\ h/{\rm Mpc}$ to $k_{\rm max}=3\ h/{\rm Mpc}$ in Eq.~(\ref{eq:SNR}). We note, however, that there is significant constraining power below $k_{\rm min}=0.1\ h/{\rm Mpc}$ (see e.g., Fig.~\ref{fig:line_PS_noise_ex}), so we focus our analysis in subsequent sections down to $k_{\rm min}=0.01\ h/{\rm Mpc}$. As is discussed in detail in Appendix~\ref{app:parsons_comparison}, the more detailed treatment of line emission, Pop~III star formation, and feedback included in this work result in a lower amplitude of Pop~III star formation and HeII emission (by nearly 2 dex) compared with past work, and \textit{thus signals with a HeII contribution are only potentially detectable in the most optimistic observing configurations}. Though the star formation model introduced in Ref.~\cite{parsons_probing_2022} maintains a comparable global SFRD to that used in this work, the differing demographics of host halos, coupled with the feedback effects and line luminosity scaling, suppress our fiducial signal by nearly two orders of magnitude and suppress the SNR as a result. Indeed, the present specifications of the CDIM are insufficient to detect the HeII signal, even in cross-correlation with H$\alpha$ at any redshift, but the cross-power is detectable in the CDIM+ configuration for a wide range of redshifts, $z\lesssim 16$ (assuming an S/N threshold of 5). We note that a marginal detection of the cross power may be possible with the present CDIM specifications if the signal is measurable at a few different redshifts. That is, though the signal in any one $z$ bin is weak, introducing the redshift evolution of the signal, which depends on the underlying physical model has the potential to offer powerful parameter constraints. With an eye towards future survey design, we further explore how the instrument and survey specifications can improve the detectability of this signal in Appendix~\ref{app:survey_design}. 

All of these conclusions, however, rest on the assumed fiducial Pop~III model, which, while motivated by the latest theoretical work, is ultimately unconstrained. Thus, a Pop~III scenario that heightens the level of star formation or efficiency of ionizing photon production could in principle make these forecasts more optimistic, as we explore further below.

\section{Parameter constraints}\label{sec:parameter_constraints} 

\subsection{Understanding the parameter dependence}\label{sec:param_dependence_classicalIII}
Before moving towards a truly statistical forecast on the underlying Pop~III parameters, it is instructive to first understand the sensitivity of the signal (and measurement uncertainty) to the parameters at play. In what follows, we focus on three key Pop~III parameters, the normalization of the star formation efficiency $f_\star^{\III}$, and the two characteristic IMF parameters, $\alpha_{\rm IMF}$ and $m_{\rm char}^{\rm IMF}$. As has been noted in past work (see e.g., \cite{parsons_probing_2022}), the Pop~III LIM signal is sensitive both to the \textit{level} of Pop~III star formation and the \textit{efficiency} of ionizing photon production. This is evident from Eq.~(\ref{eq:mean_intensity}) --- the Pop~III mean intensity, with which the power spectrum scales quadratically, is set by the product of the line coefficient and SFRD. Thus, for a fixed Pop~II SFRD/mean intensity, the total mean intensity will increase with both an increase in $\langle f_{\nu}^\III\rangle$ and $f_\star^\III$ (with the magnitude of the increase depending on the relative levels of Pop~II and III star formation at that redshift).

In Fig.~\ref{fig:model_var_classicalIII} we quantitatively showcase these effects. Namely, each of the three aforementioned components heightens the magnitude of the Pop~III contribution to the HeII and H$\alpha$ LIM signals. For example, the SFE directly normalizes the line luminosity in each halo (Eq.~(\ref{eq:line_lum_from_sfr})), so variations in the SFE are directly translated to variations in the auto- and cross-power. Varying $f_\star^{\III}$ by two orders of magnitude thus introduces 2-3 orders of magnitude in signal variation, though the relatively larger Pop~II contribution sets a floor on the signal for H$\alpha$. We additionally note a nuance in our results here. Because we self-consistently incorporate feedback from early star formation on subsequent generations of star formation (through the metagalactic H$_2$-dissociating LW background), an increase in star formation efficiency brings with it an increase in LW intensity and thus an associated decrease in later star formation. As a result, a two order of magnitude increase in $f_\star^\III$ corresponds to only a tenfold increase in the Pop~III SFRD, and in turn the effects on the line intensity and power spectrum scale accordingly (comparatively less than might be naively expected, due to this self regulation).

Similarly, the IMF parameters modulate the signal qualitatively as might be expected\footnote{Variation in the IMF can in principle modulate the star formation efficiency and thus feed into the overall level of Pop~III star formation, but this effect is subdominant relative to the other parameter variations, so we omit it for simplicity in this work \citep{hegde_efficient_2025}.} (as is represented in Fig.~\ref{fig:line_coeffs}), but the level of Pop~II star formation places an important lower limit on the variation of the signal with the IMF. For example, though the IMF slope varies considerably in Fig.~\ref{fig:model_var_classicalIII} (which should introduce a factor of 10-20 increase in line luminosity per unit star formation from Fig.~\ref{fig:line_coeffs}), the higher Pop~II SFRD (see Fig.~\ref{fig:model_SFRD_ex}) compensates for some of the difference and thus the signal only varies by one order of magnitude.

Given the nontrivial dependence of the signal on these parameters, in Fig.~\ref{fig:parameter_var_classicalIII} we show the dependence of the magnitude of the signal at $k=0.1\ {\rm Mpc^{-1}}$ at $z=5$ on the IMF slope and SFE. Based on this analysis, we first note that the signal grows by a factor of a few over a broad range of IMF slopes and only by 2-3 orders of magnitude over a factor of 100 in SFE, consistent with the preceding discussion and the results highlighted in Fig.~\ref{fig:model_var_classicalIII}, reinforcing the conclusion that variations in SFE are suppressed when translated to SFRD and mean intensity. Indeed, the high level of Pop~II star formation at this time (Fig.~\ref{fig:model_SFRD_ex}) sets a floor on the cross power such that it varies minimally with Pop~III parameters unless $f_\star^\III\gtrsim 6-7\times 10^{-2}$. Thus, to truly sensitively explore the relevant Pop~III parameter space, an instrument needs to be able to probe earlier in cosmic time, when the Pop~II contribution is less significant.

Next, because the measurement uncertainty depends both on the instrument specifications (principally the thermal noise) and the sample variance (which is set by the magnitude of the signal), the locus of detectable signals evolves nontrivially with the model parameters as well. In particular, we find that the deep SPHEREx survey will be sensitive to the most extreme Pop~III scenarios, those with a very high SFE or top-heavy IMF slope. CDIM lowers that limit by a factor of $\sim 2$, but it will require a revised next generation instrument such as our so-called CDIM+ to truly be sensitive to the entire parameter space (including the regime suggested by our current theoretical understanding of Pop~III star formation, which sets our fiducial model). In this sense, if the other contributions to the signal can be well characterized by other inputs, SPHEREx could place joint limits on the Pop~III SFE and IMF slope, even as late as $z\sim 5$. While this is promising, we note that such extremal values for the Pop~III SFE would likely have important implications for reionization and the cosmological 21-cm signal, for example, though a detailed quantification of such limits is beyond the scope of this work. In addition, as we discuss below, extensions to `classical' Pop~III models  and the contribution of Pop~II and other confounding sources can complicate the strength of such limits, so we emphasize that the forecasts outlined in this section are primarily presented to build intuition.

\subsection{Fisher forecasts}\label{sec:fisher}
While the preceding estimates characterize the overall detectability of a signal of interest, they do not quantify the significance with which one could extract detailed parameter constraints from an observation, as this requires accounting for the correlations between individual parameters. To this end, and to expand our analysis to beyond two parameters, we employ the Fisher matrix formalism to forecast the best-possible constraints on the parameters that would be possible given a particular survey configuration and the uncertainty estimate summarized in Section~\ref{sec:estimating_SNR} \citep{tegmark_karhunen_1997, tegmark_CMB_1997, heavens_statistical_2009}.

For a posterior distribution $P(\mathbf{\theta}|\mathcal{S})$ of our model parameters $\mathbf{\theta}$ given a collection of measured statistics $\mathcal{S}$ for an observable $\mathcal{O}$, the Fisher information formalism enables us to derive the covariance matrix by taking derivatives of the posterior distribution (assuming Gaussian uncertainties $\sigma_\mathcal{O}$ and uncorrelated statistics\footnote{Though the H$\alpha$ and H$\alpha\times$HeII power spectra share the H$\alpha$ information and are thus not formally uncorrelated, the HeII measurements are dominated by instrumental noise (see Figs.~\ref{fig:line_PS_noise_ex} and \ref{fig:SNR_forecast_fid}), so the covariance between these two observables is relatively small and can neglected in this analysis.}). The elements of the matrix are given by
\begin{equation}
    \label{eq:fisher}
    \mathcal{F}_{ij} = \sum_{\mathcal{O}}\frac{1}{\sigma_\mathcal{O}^2}\frac{\partial \mathcal{S}(\mathcal{O})}{\partial \mathbf{\theta}_i}\frac{\partial \mathcal{S}(\mathcal{O})}{\partial \mathbf{\theta}_j} + \frac{\delta_{ij}}{\sigma_{\rm prior}^2(\theta_i)},
\end{equation}
where $\sigma_{\rm prior}(\theta_i)$ represents the assumed prior range on each parameter. The resulting covariance matrix---the inverse of $\mathcal{F}$--- represents the \textit{minimum} achievable uncertainties for a given observation.

\subsection{Constraining `classical' Pop~III}\label{sec:classical_popIII}
The SNR forecasts in Fig.~\ref{fig:SNR_forecast_fid} suggest that a LIM measurement of the H$\alpha\times$HeII cross power at high significance is possible with the improved CDIM configuration. Indeed, by design, CDIM will be able to probe Pop~III star formation at its peak, at $z\sim 10-15$, the latest times for which Pop~III star formation is expected to still dominate the cosmic star formation budget (Fig.~\ref{fig:model_SFRD_ex}; see e.g., Refs.~\cite{hegde_self-consistent_2023, hegde_efficient_2025, feathers_global_2024, ventura_semi_2025}). As noted in Section~\ref{sec:param_dependence_classicalIII}, below $z\lesssim 10$, Pop~II SFRs will have grown to a level that they are likely to confound any probe of early star formation. Thus, for this discussion, we focus on the constraints achievable at the peak of Pop~III star formation, balancing the feasibility of the measurement with the evolution of global star formation rates.

\begin{figure}
    \centering
    \includegraphics[width=\linewidth]{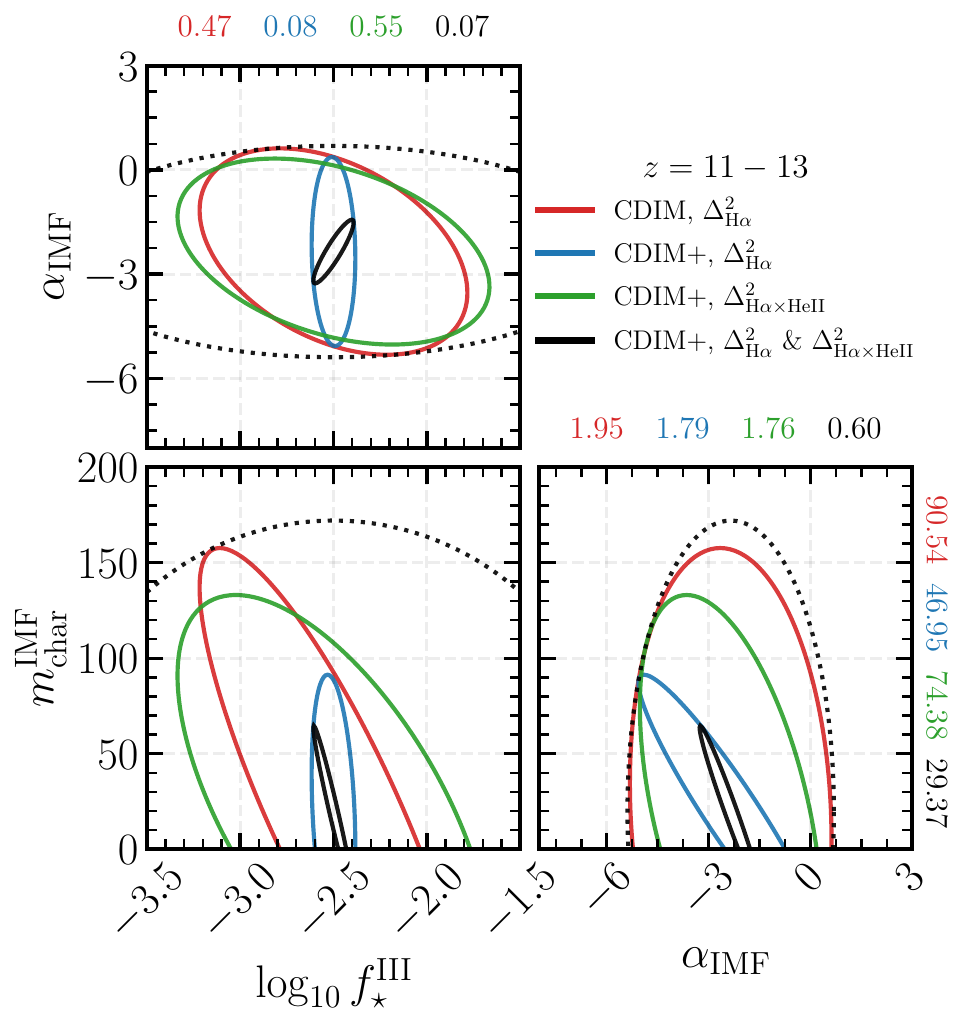}
    \caption{\textbf{With their increased sensitivity, next-generation intensity mapping experiments have the potential to tightly constrain the Pop~III IMF.} Fisher confidence ellipses for the covariance between the normalization of the Pop~III SFE, the IMF slope, and the IMF characteristic mass, based on observational forecasts of the $\Delta^2_{\rm H\alpha\times HeII}$ and $\Delta^2_{\rm H\alpha}$ power at $z=11-13$ with the two CDIM instrumental configurations presented in Ref.~\cite{parsons_probing_2022}. Colored curves correspond to 1$\sigma$ confidence ellipses for each parameter (with the marginalized $1\sigma$ uncertainties annotated atop each column or at the end of the final row) and the dotted black curve represents our chosen prior. Note that though the cross power is a more sensitive probe of the IMF than the H$\alpha$ auto power, it is less constraining on its own because the overall amplitude of that signal is weaker and the uncertainties are larger.}
    \label{fig:fisher_fidIII_surveyComp}
\end{figure}

In Fig.~\ref{fig:fisher_fidIII_surveyComp} we quantify these variations with a Fisher forecast for the same three parameters discussed in the preceding sections, simulating a mock survey targeting the H$\alpha$ and H$\alpha\times$HeII LIM signals at $z=11-13$. The current specification of the CDIM is able to moderately constrain the Pop~III SFE --- the parameter to which the signal is most sensitive --- but has little power when it comes to the IMF, though it does again highlight the degeneracy between IMF slope and SFE in the LIM power. This is consistent with the overall SNR forecast discussed previously (Fig.~\ref{fig:SNR_forecast_fid} and Section~\ref{sec:param_dependence_classicalIII}); namely, for a low SFE, the present CDIM configuration will only be sensitive to the H$\alpha$ signal, which itself is only mildly modulated by the IMF.

If instead we consider a revised CDIM, the heightened SNR in the cross power translates to exquisite constraints on the Pop~III parameters, allowing for a truly statistical parameter survey of Pop~III physics and showcasing the power of such a next-generation LIM study. In fact, the CDIM+ also demonstrates the complementary power of the auto and cross power measurements. That is, the H$\alpha$ auto power alone is sensitive to the Pop~III SFE but is comparatively less sensitive to the IMF parameters. This induces the shrinking of the red confidence ellipse along the horizontal axis to produce the blue in the upper left panel of Fig.~\ref{fig:fisher_fidIII_surveyComp}. Adding in the more IMF-sensitive cross power tightens the constraints considerably and yields a forecast that can sensitively discriminate between many classical Pop~III models.

\begin{figure}
    \centering
    \includegraphics[width=\linewidth]{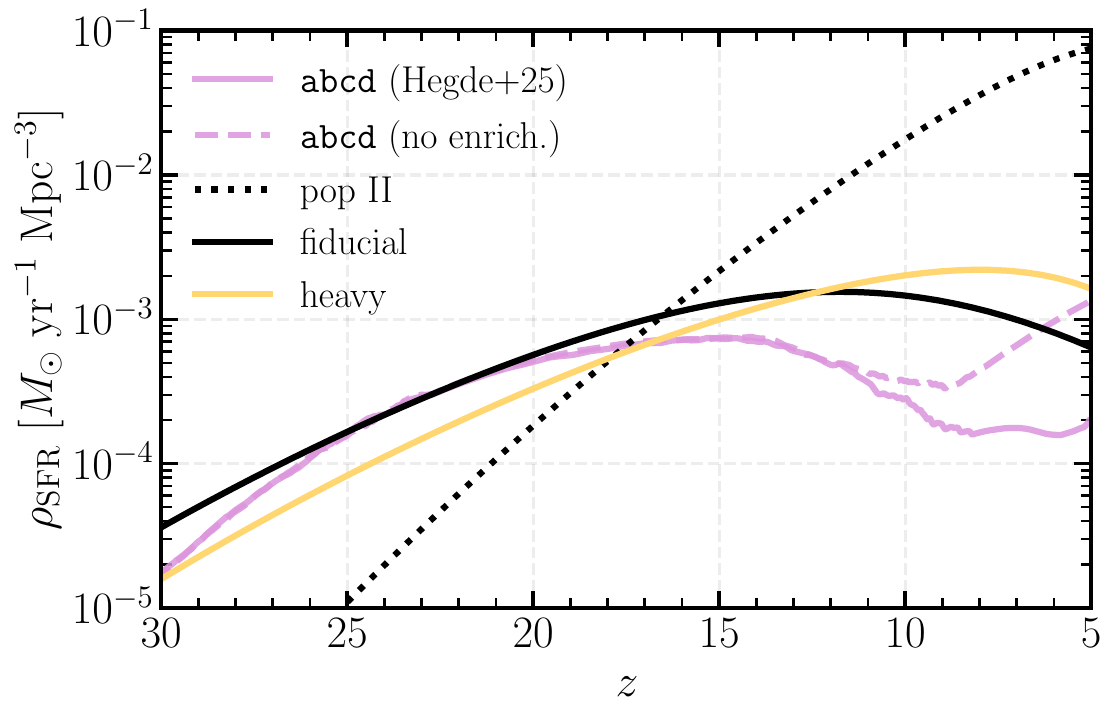}
    \caption{\textbf{Our simple SFRD predictions broadly agree with the latest theoretical models, though they overestimate the late-time SFRD by a factor of a few.} The Pop~III SFRD as a function of redshift for two different model iterations: the fiducial model (black) and one that allows Pop~III star formation to extend to more massive halos with a slightly lower star formation efficiency (`heavy', yellow). We compare these variations against the theoretical semi-analytic prediction of the Pop~III SFRD from the \texttt{abcd} model, which self-consistently tracks Pop~III star formation across this redshift range and includes various environmental feedback effects at the latest times (with and without the effects of enrichment --- purple solid and dashed; \cite{hegde_efficient_2025}). We integrate the SFRD over all halo masses in this comparison.}
    \label{fig:model_SFRD_ex}
\end{figure}

\section{Expanding the base Pop~III model}\label{sec:model_extensions}
The discussion thus far has focused on Pop~III star formation in the `classical' model. While it is unlikely that this purely minihalo-driven Pop~III signal will be detectable with current-generation LIM instruments like SPHEREx and may even be difficult to detect with proposed next-generation surveys, if the details of Pop~III star formation deviate from the assumptions made in our fiducial model, then these measurements could hold more constraining power. 

The latest theoretical work suggests that Pop~III star formation at $z\lesssim 10$ could be dominated by star-forming clumps hosted in more massive, atomic cooling halos, rather than minihalos, as has been conventionally assumed \citep{venditti_needle_2023, venditti_first_2024, hegde_efficient_2025, venditti_bursty_2025, zier_thesan_2025, katz_megatron_2025, storck_megatron_2025}. In turn, deep observational searches for Pop~III star formation in the Epoch of Reionization have inferred a tentative measurement of the Pop~III luminosity function that may require deviations from typical models to achieve the unprecedented observed abundance of systems \citep{fujimoto_glimpse_2025, venditti_bursty_2025, morishita_pristine_2025, fujimoto_glimpsed_2025}. Given the challenge of detecting Pop~III star formation with LIM alone and the current promise of possibly constraining the UVLF with JWST, here we forecast the power of utilizing these probes together.

\begin{figure*}
    \centering
    \includegraphics[width=0.9\linewidth]{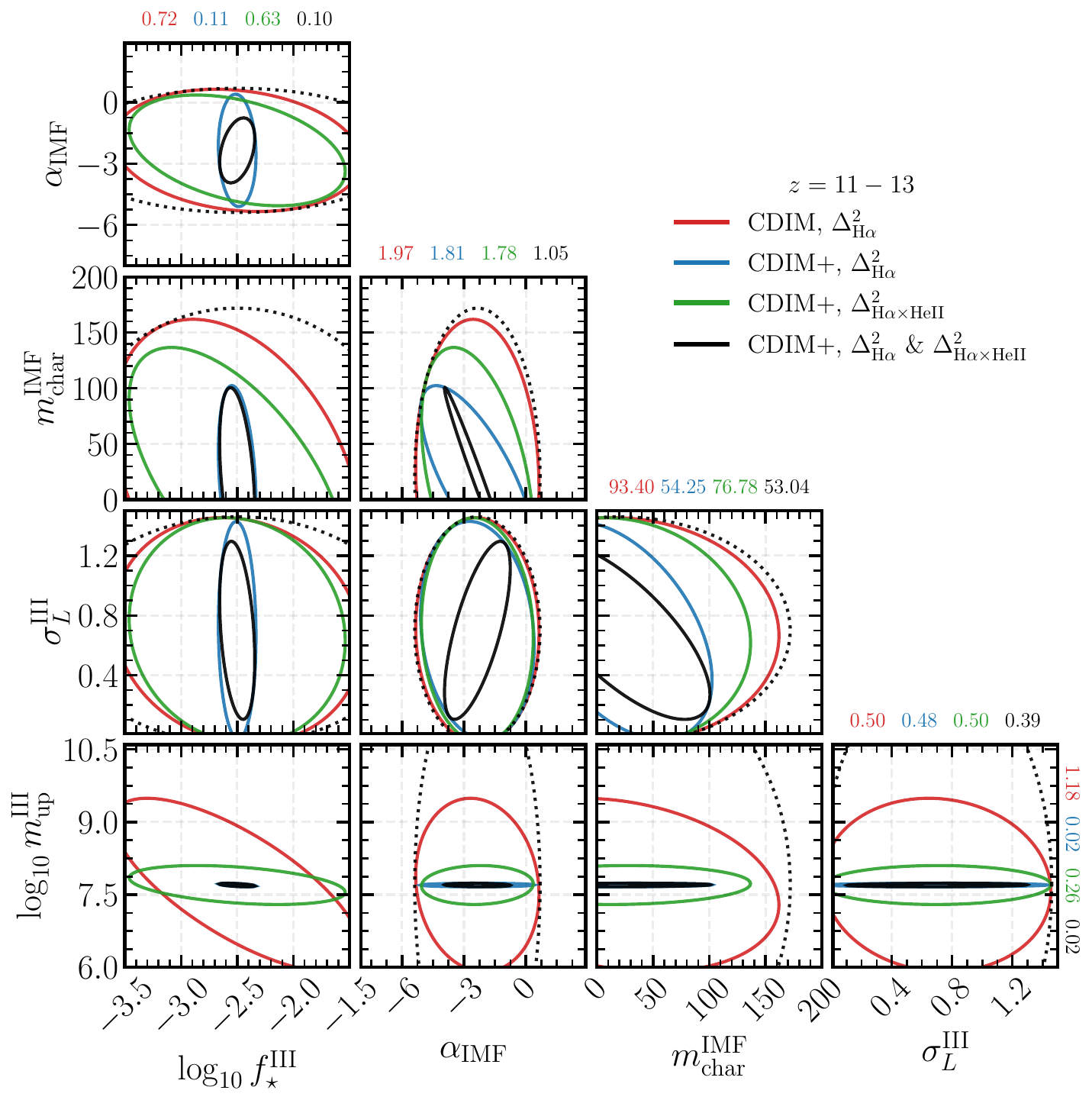}
    \caption{\textbf{Next-generation LIM surveys have the potential to constrain extensions to classical Pop~III models.} Fisher confidence ellipses for the covariance between the normalization of the Pop~III SFE, the IMF slope, the IMF characteristic mass, the stochasticity of the Pop~III $m_h-L_\nu^{\III}$ relation, and the upper limit on the Pop~III host halo mass, based on observational forecasts of the $\Delta^2_{\rm H\alpha\times HeII}$ and $\Delta^2_{\rm H\alpha}$ power at $z=11-13$ with the two CDIM instrumental configurations presented in Ref.~\cite{parsons_probing_2022}. Colored curves correspond to 1$\sigma$ confidence ellipses for each parameter (with the marginalized $1\sigma$ uncertainties annotated atop each column or at the end of the final row) and the dotted black curve represents our chosen prior.}
    \label{fig:fisher_exoticIII_highz}
\end{figure*}

In practice, deviations from the fiducial Pop~III model manifest as variations in the $m_h-L$ relation, which can be a result of modifications to the star formation efficiency (i.e., changes to the mass dependence of $f_\star^\III(m_h, z)$) or stochasticity in the halo-galaxy connection (i.e., invoking `bursty' star formation). These two modes of expanding the Pop~III model are explored in Ref.~\cite{venditti_bursty_2025} to understand the UVLF measurement first introduced in Ref.~\cite{fujimoto_glimpse_2025} (though see Ref.~\cite{fujimoto_glimpsed_2025} for a refinement to this work).

Ref.~\cite{venditti_bursty_2025} extend the star formation law used in Ref.~\cite{cruz_effective_2025} and earlier in this work by varying the fraction of halos hosting Pop~III star formation; i.e.,
\begin{equation}\label{eq:duty_cycle}
    f_{\rm duty}^{\III}(m_h, z) =  \exp\Bigg(-\frac{m_{\rm mol}(z)}{m_h}\Bigg)\exp\Bigg(-\frac{m_h}{m_{\rm up}^{\III}}\Bigg),
\end{equation}
where $m_{\rm mol}(z)$ is the critical halo mass for molecular cooling \citep{kulkarni_critical_2021, schauer_influence_2021, munoz_impact_2022, hegde_self-consistent_2023, nebrin_starbursts_2023} and $m_{\rm up}^\III$ sets the upper halo mass cutoff, which is generally assumed to be set by the atomic cooling mass. Rather than fixing it to this threshold, Ref.~\cite{venditti_bursty_2025} allow the latter mass limit to vary and fit for it with the UVLF observations, finding satisfactory agreement if Pop~III star formation can occur in halos up to $m_{\rm up}^\III \sim 10^{10.5}M_\odot$. Similarly, by heightening the degree of scatter in the $L_{\rm UV}-m_h$ relation with a reduced mass threshold of $m_{\rm up}^\III \sim 10^{8.5}M_\odot$, they demonstrate comparably good agreement with the UVLF. 

In Fig.~\ref{fig:model_SFRD_ex} we compare the SFRD predicted by our fiducial model, one that allows for star formation in more massive halos (the `heavy' model from Ref.~\cite{venditti_bursty_2025}) with the predictions of the \texttt{abcd} semi-analytic model for Pop~III star formation \citep{hegde_efficient_2025}. \texttt{abcd} comprehensively traces Pop~III star formation from minihalos in the very early universe to pristine star clusters in the reionization era, and self-consistently accounts for many of the feedback processes affecting star formation, such as local and global metal enrichment, radiation backgrounds, and reionization. Thus, it provides a useful benchmark for the more phenomenological double power law SFE used here. This comparison highlights that while the simpler models agree well with the predicted SFRD at early times (in the minihalo-dominated era), they overpredict the SFRD when feedback effects can become significant ($z\lesssim 15$). Indeed, at late times they are roughly comparable to the \texttt{abcd} estimates with no enrichment (i.e., which provide a rough upper limit on the level of pristine star formation we expect to see).

With this in mind, here we vary the fiducial framework to accommodate this expanded Pop~III model. In particular, in addition to allowing the upper host halo mass limit $m_{\rm up}^\III$ to vary, we additionally vary the stochasticity parameter $\sigma_L$, which quantifies the spread in the mapping from halo mass to line luminosity (Eq.~(\ref{eq:stochastic_L})).

\subsection{Constraining `exotic' Pop~III with LIM cross-correlations}
We revisit the LIM forecasts discussed in Section~\ref{sec:classical_popIII} in the context of these `exotic' Pop~III models. In this case, we quantify the power of high-redshift LIM surveys (the proposed CDIM and our enhanced CDIM+ configuration) to constrain extensions to classical Pop~III models. We expand the preceding analysis to consider a five-parameter Fisher forecast, varying $\theta = \{f_\star^\III, \alpha_{\rm IMF}, m_{\rm char}^{\rm IMF}, \sigma_L^\III, m_{\rm up}^\III\}$, holding the Pop~II parameters fixed to the fiducial values used in Ref.~\cite{cruz_effective_2025}. The results of this analysis are shown in Fig.~\ref{fig:fisher_exoticIII_highz}. In general, we note that extensions to classical Pop~III models tend to increase the Pop~III luminosity or expand Pop~III star formation to a wider range of host halos, which would in turn manifest as an increase in the signal generated by these stellar populations, corresponding to an SNR increase of a factor of $3-5$ between the fiducial model and the `heavy' model at $z\sim 11-13$, for example. However, because we introduce more parameters to the model in this instance (compared with the analysis in Fig.~\ref{fig:fisher_fidIII_surveyComp}), we expect that the constraints achievable on any one parameter are likely to be weaker than in the classical case.

By introducing new degrees of freedom to the Pop~III demographics, the most promising constraints on classical Pop~III parameters weaken (compare the marginalized 1$\sigma$ parameter uncertainties in Figs.~\ref{fig:fisher_fidIII_surveyComp} and \ref{fig:fisher_exoticIII_highz}), though a measurement of the H$\alpha$ and H$\alpha\times$HeII power spectra can still pin down the Pop~III SFE to within 0.25 dex in an optimistic observing configuration. The IMF parameters remain poorly constrained, though rough limits are achievable with the improved CDIM survey discussed here.

In kind, the LIM signal is sensitive to both the Pop~III burstiness and host halo mass threshold, though limits on the former are fairly weak. Indeed, the overall normalization of the signal is fairly sensitive to the host halo mass limit, and this limit could be constrained very precisely, assuming the signal is well-measured. We also find that the Pop~III burstiness and host halo mass threshold are both somewhat degenerate with the IMF parameters and SFE, as all have the effect of scaling the normalization of the LIM power. With this in mind, we conclude that a CDIM-like experiment could reasonably well discriminate between different theoretical models for Pop~III star formation that extend beyond the classical regime and thus provide an important ancillary diagnostic for star formation models at cosmic dawn.

\subsection{Inferring the UVLF}\label{sec:uvlf}
To this point we have focused our interest on line intensity observables. With our framework, however, we can simultaneously predict the $1500$ \AA\ UV luminosity function associated with both Pop~II and III stars. That is, the UVLF is given by 
\begin{equation}\label{eq:UVLF}
    \Phi_{\rm UV}(M_{\rm UV}) = \int dm_h\frac{dn}{dm_h}p^{\rm (i)}(M_{\rm UV}|m_h),
\end{equation}
where $p(M_{\rm UV}|m_h)$ is taken to be a normal distribution with the same variance $\sigma_L^2$ centered on a mean relation between halo mass and UV luminosity $L_{\rm UV}^{\rm (i)}(m_h) = \dot{m}_\star^{\rm (i)}(m_h)/\kappa_{\rm UV}^{\rm (i)}$ for an IMF-dependent UV luminosity-SFR conversion factor $\kappa_{\rm UV}^{\rm (i)}$. For our Pop~III IMFs, we can infer this factor following a similar procedure to the line intensity coefficients summarized in Section~\ref{sec:line_lum}, but now must use the Pop~III stellar SEDs provided in Ref.~\cite{costa_evolutionary_2025} directly.\footnote{https://github.com/lluism/seds} In particular, for a given IMF $\phi(m)$, the total stellar lifetime-weighted energy density output per unit stellar mass in a population is
\begin{equation}\label{eq:tot_energy_dens}
    \bigg\langle \frac{E_{\rm \nu}}{m_\star} \bigg\rangle = \frac{\int_{m_{\rm min}}^{m_{\rm max}} \overline{L}_{\nu}(m)t_{\rm life}(m)\phi(m)dm}{\int_{m_{\rm min}}^{m_{\rm max}}m\phi(m)dm},
\end{equation}
where $\overline{L}_\nu(m)$ is the mean (lifetime-averaged) luminosity density for a zero metallicity star of mass $m$ in units of ${\rm erg\ s^{-1}\ Hz^{-1}}$:
\begin{equation}
    \label{eq:Lbar_nu}
    \overline{L}_{\nu}(m) = \frac{1}{t_{\rm life}}\int_{t_{\rm ZAMS}}^{t_{\rm ZAMS}+t_{\rm life}}L_\nu(t|m) dt,
\end{equation}
with $L_{\nu}(t|m)$ denoting the stellar luminosity density for a star of mass $m$ at a time $t$ after the zero-age main sequence (ZAMS) given by the tracks computed in Ref.~\cite{costa_evolutionary_2025} (and compiled by Ref.~\cite{zackrisson_detection_2024}).

\begin{figure}
    \centering
    \includegraphics[width=\linewidth]{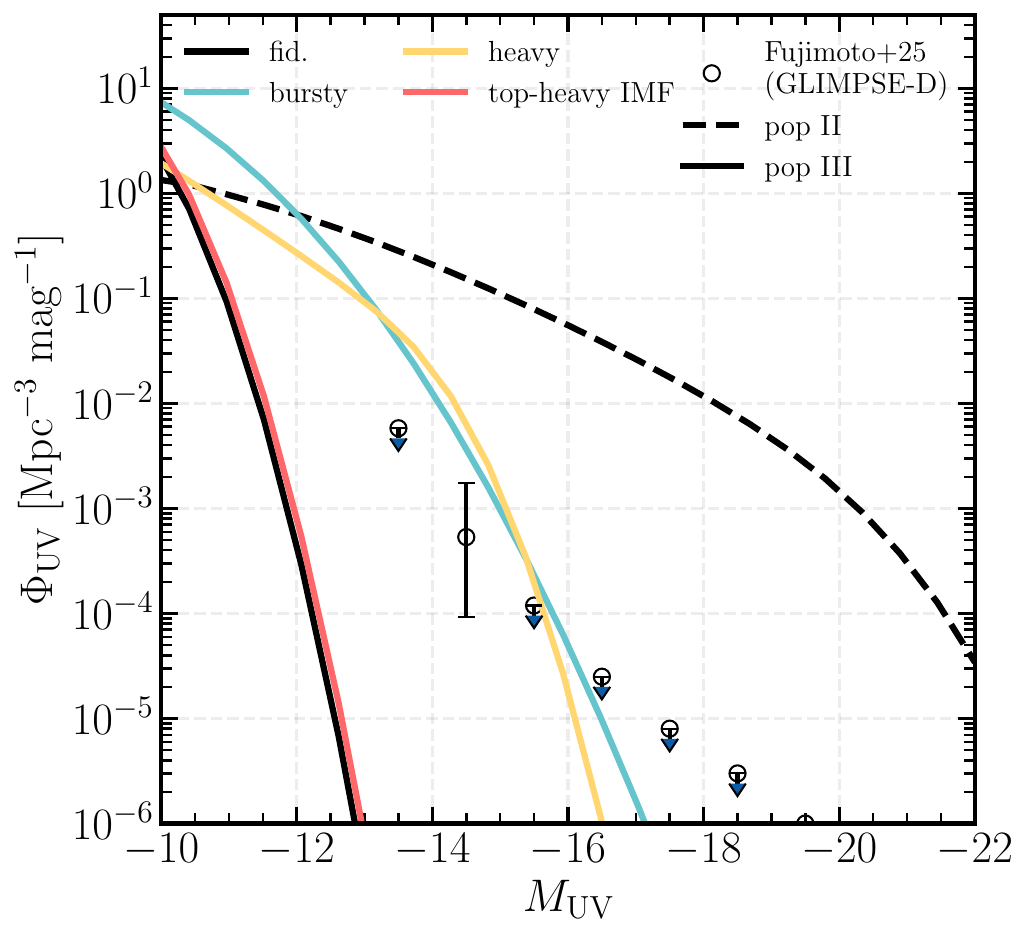}
    \caption{\textbf{Increased Pop~III `burstiness' and extensions of the SFE from minihalos to more massive halos are degenerate at the level of the UVLF.} The predicted Pop~II UVLF (dashed), and Pop~III UVLF (solid) for the fiducial model (black) and three representative model variations at $z=6$: increased stochasticity in the $L_{\rm UV}-m_h$ relation (blue), an extension allowing Pop~III star formation in massive halos (yellow), and a top-heavy IMF (red). We refer the interested reader to Ref.~\cite{venditti_bursty_2025} for a detailed discussion of the bursty and heavy curves. For comparison, we also show the Pop~III UVLF estimates discussed in Refs.~\cite{fujimoto_glimpse_2025, fujimoto_glimpsed_2025}.}
    \label{fig:uvlf_examples_z6}
\end{figure}

\begin{figure*}
    \centering
    \includegraphics[width=0.9\linewidth]{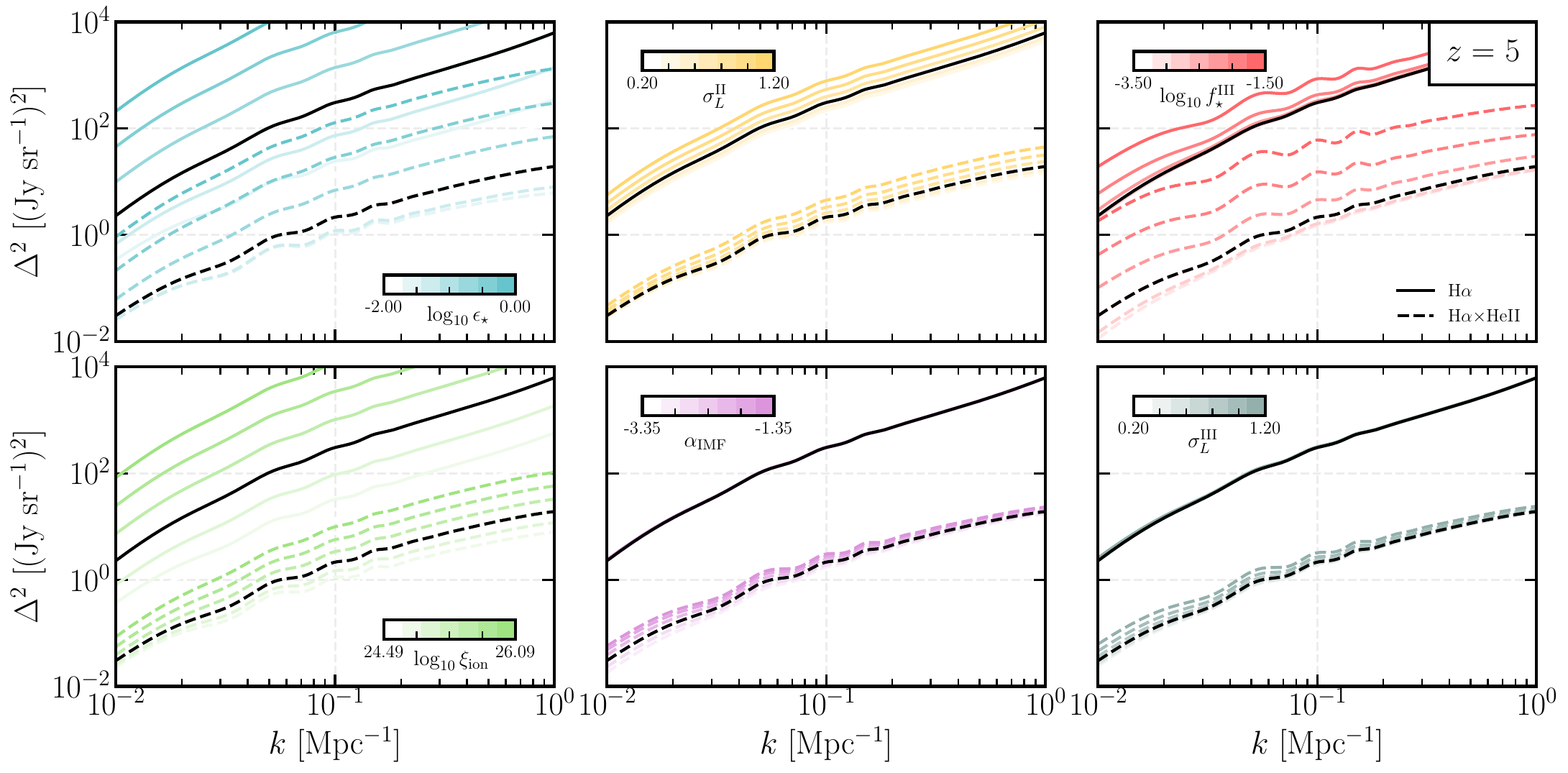}
    \caption{\textbf{The intensity mapping signal is most sensitive to variations in the overall magnitude of star formation.} The H$\alpha$ auto- and ${\rm H\alpha\times HeII}$ cross power at $z=5$ subject to variations in the parameters characterizing the underlying Pop~III or II model, with a darkening color indicating a larger value of that parameter (see colorbar). From left to right, the top row highlights the normalization of the Pop~II SFE, the stochasticity in the $m_h-L_\nu^{\rm II}$ relation, and the normalization of the Pop~III SFE. The lower row showcases the sensitivity to the ionizing photon production efficiency, the Pop~III IMF slope, and the stochasticity in the $m_h-L_\nu^{\rm III}$ relation.}
    \label{fig:model_variations_fisher}
\end{figure*}

Then the luminosity density associated with a given SFR is $L_{\nu} = \langle E_\nu/m_\star \rangle \dot{m}_\star$ and 
\begin{equation}
    \label{eq:kappa_UV}
    \kappa_{\rm UV}^{\III} = \big(L_{1500} \big)^{-1} = \frac{1}{\langle E_{1500} /m_\star\rangle (1\ M_\odot\ {\rm yr^{-1}})}.
\end{equation}
This procedure yields an SFR-to-UV conversion factor $\kappa_{\rm UV}^{\rm III, fid} = 6.56\times 10^{-29}\ M_\odot {\rm \ yr^{-1}\ (erg\ s^{-1}\ Hz^{-1})^{-1}}$ that is roughly 60\% smaller than the Pop~II case (corresponding to 1.75x higher luminosities at a fixed SFR for Pop~III compared to Pop~II and ranging up to a factor of 2 for the most extreme IMFs).\footnote{We note that this procedure for estimating the SFR-to-UV conversion is somewhat different from others employed in past work, which assume a form for the SFH (usually a burst or constant star formation), integrate over the IMF-averaged SEDs of the evolving stellar population given this SFH, and evaluate the conversion factor at some fixed time (usually on the order of a few hundred Myr or 1 Gyr). We do not adopt the same calculation for simplicity --- in our approach, we do not assume a SFH --- and validate that our procedure yields $\kappa_{\rm UV}^{\rm III}$ within a factor of 2 of these more detailed calculations (which is smaller than the uncertainty introduced in the choice of SFH; see e.g., Refs.~\cite{raiter_predicted_2010, zackrisson_spectral_2011, inayoshi_lower_2022})}

Given a survey configuration (limiting magnitude, target redshift, and observing area), we then estimate the uncertainty on a UVLF prediction as the quadrature sum of the Poisson noise in a particular bin (which is proportional to $\sqrt{N_{\rm gal}}$, where $N_{\rm gal}$ is the number of galaxies in each bin) and the contribution from cosmic variance, which we estimate following the method described in Ref.~\cite{trapp_flexible_2020}.

For concreteness, we show example luminosity function variations generated with the representative models described in Ref.~\cite{venditti_bursty_2025} in Fig.~\ref{fig:uvlf_examples_z6}. In particular, we reproduce the results of that work, demonstrating that expanding the fiducial (classical) Pop~III model to allow for star formation in halos above the atomic cooling threshold ($m_{\rm up}^\III \sim 10^{10.5} M_\odot$) and increasing the scatter in the mapping from halo mass to UV luminosity ($\sigma_{L}^\III\sim 1.5$ dex) are degenerate at the level of the detectable UVLF (comparing the yellow and blue curves at $M_{\rm UV}\lesssim -14$), and both can satisfactorily reproduce the large abundance of candidate Pop~III systems inferred in Ref.~\cite{fujimoto_glimpsed_2025}. Finally, we highlight the lack of sensitivity of the UVLF to IMF variations --- because massive stars have roughly constant, near-Eddington light-to-mass ratios, flattening the slope to $\alpha = 0$ only mildly modifies the derived value of $\kappa_{\rm UV}^\III$ and thus minimally boosts the predicted UVLF (consistent with what is found in Ref.~\cite{raiter_predicted_2010}, which also includes nebular emission). 

\subsection{The Pop~III contribution to line intensity and the UVLF}\label{sec:popIII_UVLF}
In what follows, we show Fisher forecasts for two observables of interest, the LIM power and UVLF. In this forecast we consider a seven-parameter model for star formation and line luminosity, varying $\mathbf{\theta} = \{\epsilon_\star, \sigma_{L}^{\II}, f_\star^{\rm III}, \xi_{\rm ion}, \alpha_{\rm IMF}, \sigma_L^{\III}, m_{\rm up}^\III\}$. All other parameters are held fixed at the fiducial values specified in Ref.~\cite{cruz_effective_2025}. The first and third these parameters $\epsilon_\star$ and $f_\star^{\III}$ represent the (redshift-independent) overall normalization of the star formation efficiency for Pop~II and III stars, respectively. The second parameter characterizes the burstiness of Pop~II star formation, which features in both the Pop~II UVLF and line luminosity calculation. The fourth parameter, $\xi_{\rm ion}$ represents the efficiency of ionizing photon production and quantifies the mapping from Pop~II SFR to H$\alpha$ luminosity, effectively allowing us to test variations in the Pop~II `IMF', which is otherwise fixed to the conventional Salpeter values (Section~\ref{sec:line_lum}). The fifth parameter, $\alpha_{\rm IMF}$ represents the mass power law slope of the Pop~III IMF defined in Eq.~(\ref{eq:chabrier_imf}), and the final two parameters characterize deviations from the `classical' Pop~III model as summarized in Section~\ref{sec:model_extensions}. 

\begin{figure*}
    \centering
    \includegraphics[width=0.9\linewidth]{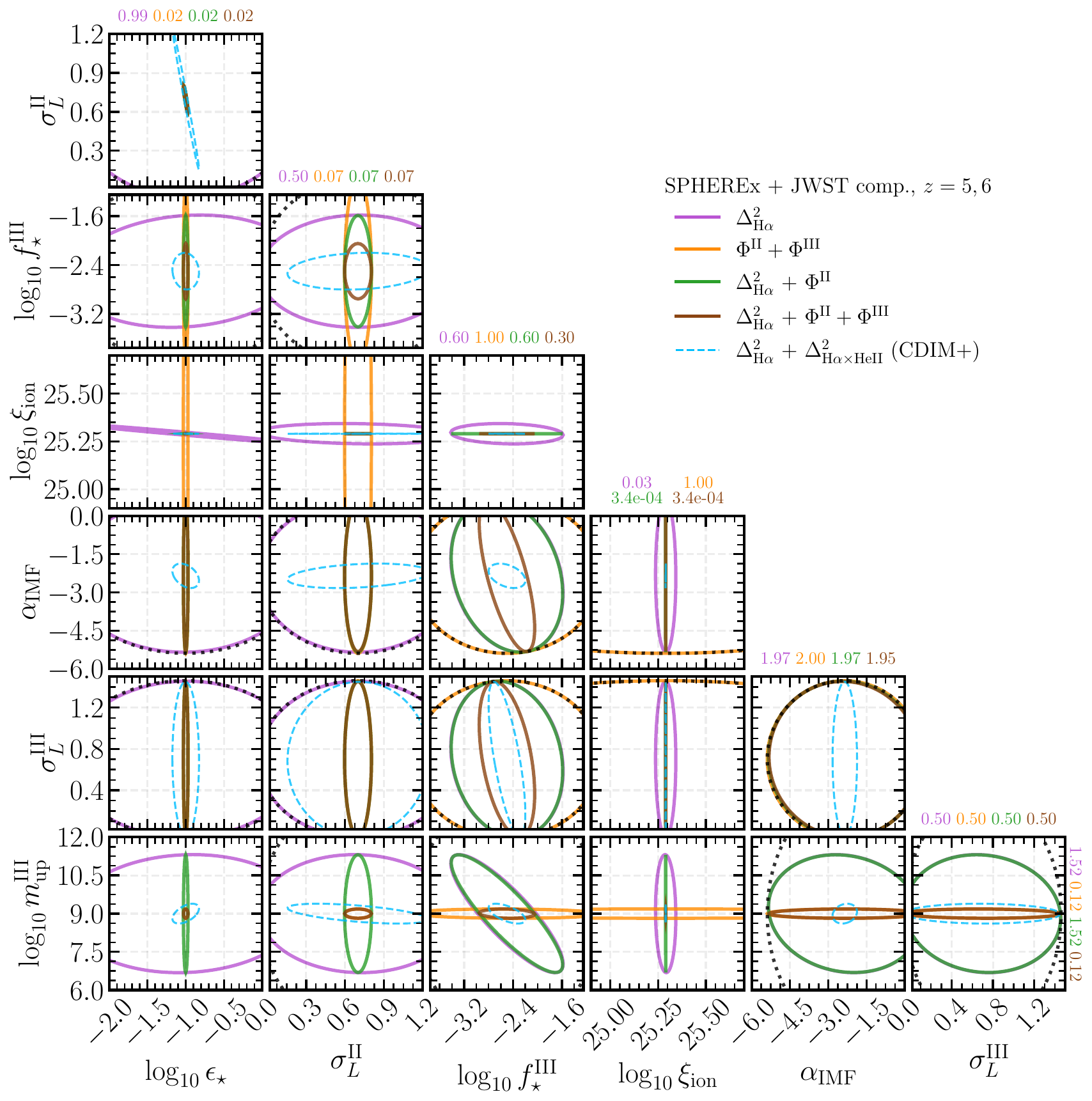}
    \caption{\textbf{Using the redshift evolution of multiple Pop~III probes can tighten constraints on parameters characterizing demographics of the population.} Fisher forecasts combining SPHEREx observations of the H$\alpha$ auto power with a Pop~II and III UVLF measurement at $z=5$ and 6 for our seven parameter model. We show confidence ellipses for LIM alone (purple), the Pop~II and III UVLFs alone (orange), the joint constraint with H$\alpha$ and the Pop~II UVLF (green), and the full joint constraint with both UVLFs and LIM (brown). We additionally show the forecasted constraints achievable by observing the auto and cross power with the improved CDIM+ configuration explored in this work (dashed blue), highlighting the potential sensitivity of such a survey to the details of Pop~III star formation. Colored curves correspond to 1$\sigma$ confidence ellipses for each parameter (with the marginalized $1\sigma$ uncertainties annotated atop each column or at the end of the final row) and the dotted black curve represents our chosen prior. The UVLF estimates used here rely on the compilation of JWST fields used in the Ref.~\cite{fujimoto_glimpse_2025} photometric Pop~III search --- at the deepest end, GLIMPSE, with an area of 9.4 ${\rm arcmin}^2$ and an apparent magnitude depth of 33 mag (in the most highly lensed regions; \cite{atek_jwst_2025}), and at the widest end, PRIMER, with an area of $\sim 300\ {\rm arcmin}^2$ and a depth of 28.4 mag \citep{donnan_jwst_2024}.}
    \label{fig:fisher_forecast_SPHEREx}
\end{figure*}

To build intuition for these parameter variations, in Fig.~\ref{fig:model_variations_fisher}, we show the H$\alpha$ and H$\alpha\times$HeII power spectra predicted by our framework for a range of representative parameter variations at $z=5$. For completeness, we show the related variations for the UVLF in Appendix~\ref{app:uvlf_variations}. As in Fig.~\ref{fig:model_var_classicalIII}, the LIM signals are most significantly modulated by the overall normalization of the SFEs. Even at the end of reionization, the Pop~III SFE still contributes to the H$\alpha$ auto power, suggesting that a limit in the SPHEREx era may be possible (Section~\ref{sec:param_dependence_classicalIII}). Indeed, in the most extreme cases, the Pop~III SFE introduces a distinct difference in the slope of the H$\alpha$ power at the largest scales above the Pop~II-set `floor' (owing to the lower typical host halo masses for Pop~III star formation in the classical picture). In contrast, modifications to the signal induced by variations in Pop~II parameters generally scale the amplitude of the signal at these lower redshifts, suggesting that the large-scale shape of the auto- and cross-power could hold crucial information to disentangle the nature of the stellar populations driving the observed signal in different epochs. We also find that the H$\alpha$ power is much less sensitive to other extensions to the fiducial Pop~III model --- the auto power is effectively unchanged in response to significant changes to the IMF and level of Pop~III burstiness, though there is a slight sensitivity to the upper mass limit of the Pop~III SFE. Instead, as noted before, this investigation suggests that tight Pop~III constraints in the reionization era from LIM alone are likely to require next-generation LIM instruments which can reach the cross power signal. Because Pop~III stars remain dominant in the cross power, we see increased sensitivity to the upper host halo mass threshold, even at $z\sim 5$. Despite this sensitivity, the magnitude of the signal only increases by a factor of 10 over 5 dex in halo mass, so even in these extremal cases, the cross power likely would not be detectable with SPHEREx. 

If we are able to robustly quantify the (dominant) Pop~II contribution to the LIM signal, then the additional contribution from Pop~III could possibly be identified. To this end, we also consider the UVLF, which could provide such a lever.

For each model variation summarized above, we then estimate the noise for a mock observation following the procedure outlined in Section~\ref{sec:estimating_SNR} for the LIM power spectra (along with the survey configurations summarized in Table~\ref{tab:survey_specs}) and in Section~\ref{sec:uvlf} for the UV luminosity function. For the latter, we use the compilation of surveys analyzed in Refs.~\cite{fujimoto_glimpse_2025, fujimoto_glimpsed_2025} with the depths and observing areas reported therein (see Table~2 in Ref.~\cite{fujimoto_glimpse_2025}) and assume that the Pop~II and III UVLFs are measured separately, as is done in those works. Constraining the Pop~III UVLF requires exquisite survey sensitivity and serendipity, as these halos likely have magnitudes fainter than $M_{\rm UV}\lesssim 15$ and reside beyond the faint end of most UVLF measurements. Thus, to demonstrate the constraining power of the UVLF in tandem with a LIM survey, in what follows, we modify the depth of the GLIMPSE survey to the deepest effective depths (due to lensing) quoted in their analysis (i.e., $m_{\rm AB, lim}^{\rm GLIMPSE}\approx 33$).

\subsubsection{Near-term forecasts}\label{sec:SPHEREx_fisher}
In Fig.~\ref{fig:fisher_forecast_SPHEREx}, we show the potential parameter constraints that can be achieved by combining a JWST measurement of the UVLF with a SPHEREx measurement of the H$\alpha$ auto power at $z=5-6$ (the highest redshift range for which the H$\alpha$ line is detectable with SPHEREx)---a forecast of what could be currently achievable given continued deep JWST observations of faint Pop~III candidates in the EoR. In this case, we have increased the Pop~III host halo upper mass threshold to $m_{\rm up}^\III\sim 10^9 M_\odot$ in the fiducial demonstration as an intermediate between the classical Pop~III case and `heavy' model introduced in Ref.~\cite{venditti_bursty_2025}.

We find that the predicted SPHEREx LIM observations are somewhat sensitive to the overall normalization of the Pop~II and III star formation efficiencies, but at the level of H$\alpha$ alone, cannot provide much constraining power for the IMF slope, burstiness, or host halo demographics of Pop~III stellar populations (as is visually apparent from Figs.~\ref{fig:parameter_var_classicalIII} and \ref{fig:model_variations_fisher}). In kind, because Pop~II star formation dominates the $z\sim 5$ SFRD, SPHEREx observations alone should be able to constrain the ionizing photon production efficiency well (as the Pop~II H$\alpha$ signal can be highly sensitive to variations in this quantity) and can help inform models of reionization, though degeneracies with the overall SFE persist. However, we note that the constraints shown here are likely overestimating the precision with which we will be able to measure this and other parameters, as our treatment neglects foregrounds (see e.g., the discussion in Section~\ref{sec:caveats}) and other systematics that will play an important role in, for example, mapping $\xi_{\rm ion}$ to H$\alpha$ luminosity. We thus encourage caution in interpreting these results and emphasize that this work is primarily intended to be illustrative and exploratory. The UVLF, on the other hand, is more sensitive to the brightest sources, and, together, these probes may be able to distinguish between the extensions to the classical Pop~III model discussed in Section~\ref{sec:model_extensions} and Ref.~\cite{venditti_bursty_2025}, especially if the Pop~III UVLF is detectable in the near term. 

Even if the Pop~III UVLF is not detectable, a robust measurement of the Pop~II UVLF to the faint end will still enable us to tightly constrain the Pop~II SFE and burstiness. Grounded by this Pop~II contribution, the H$\alpha$ LIM signal could constrain the Pop~III SFE to within 1 dex. If we are able to continue constraining the Pop~III UVLF with JWST surveys like GLIMPSE, the joint LIM + UV measurements at $z\sim 5-6$ could be able to place constraints on the upper mass threshold of Pop~III host halos to within 2 dex and simultaneously distinguish the burstiness of this population to within 5\%. From this investigation, we again reiterate the power of making such measurements at \textit{multiple redshifts} --- even though the individual measurements may have low SNR, the redshift evolution of the predictions limits the available parameter space and provides a crucial knob to tighten our forecasts.

\section{Discussion and Conclusion}
\subsection{Caveats and limitations}\label{sec:caveats}
Though the forecasts outlined in this work suggest that line intensity mapping cross correlations may be a promising avenue to place limits on Pop~III demographics, there are simplifications inherent to the modeling and important observational and data processing considerations that can complicate the picture. 

With respect to the former, here we have employed a highly simplified framework that relies on the assumption that galaxies grow their stellar populations in a quasi-equilibrium state, resulting in the double power law star formation efficiencies used in Refs.~\cite{munoz_effective_2023, cruz_effective_2025} (see e.g., Refs.~\cite{sun_constraints_2016, furlanetto_minimalist_2017, mirocha_effects_2020} for a demonstration that such models can satisfactorily reproduce galaxy luminosity functions). However, more recent work suggests that young galaxies in the early universe likely exist in dynamic, `bursty' states, introducing stochasticity around the mean SFR/line luminosity-halo mass relations explored here. SFR stochasticity is only one of several sources (e.g., variations in ISM properties) that can introduce line-to-line scatter, which can decorrelate the observed LIM signals of interest to this work. 

In addition, there are potential non-stellar sources that can contribute to HeII emission. For example, quasars and other hard ionizing sources can confound interpretation of the observed HeII mean intensity. Ref.~\cite{visbal_looking_2015} estimate that such sources would produce a signal orders of magnitude less luminous than that of Pop~III-hosting galaxies at early times (especially for the most top-heavy IMFs), but suggest that the contribution would rapidly grow at $z\lesssim 10$, as the quasar population builds up. While differences in the host DM halo demographics of quasars compared with Pop~III sources would imprint unique signatures in the HeII power spectra, we reserve an investigation of these effects to future work.

From an observational standpoint, the analysis presented here assumes that instrumental and observational systematics are well-described by the formalism described in Section~\ref{sec:estimating_SNR}. However, it can be challenging to separate continuum foregrounds and comprehensively account for lower redshift line interlopers, both of which will be relevant for any measurement of the autocorrelation LIM signal. In particular, we emphasize that many of the tight mock parameter constraints shown here crucially rest on the idealized signal variance treatment discussed in Section~\ref{sec:estimating_SNR}. With the aforementioned foreground challenges and other systematics, we expect that constraints achievable in a real survey will likely be weaker than those shown here and thus caution readers against interpreting the precise values too literally. 

Nevertheless, because many of our conclusions rely on signals that are instrument-noise dominated, we expect that the effect of some of these uncertainties will be modest.  For example, to estimate the robustness of our results to such uncertainties, we carry out a simple phenomenological test. Namely, we inflate the sample variance contribution by a factor of 5 (to mimic the additional variance that may result from residual foregrounds) and verify that the resulting S/N and marginalized parameter constraints change by $<5\%$.

While we do not attempt to account for either effect in any more detail this work, we briefly summarize some of the relevant literature for completeness. Treatments of continuum emission removal --- which often rely on the spectral smoothness-based subtraction or avoidance --- have been explored extensively in the context of the cosmological 21-cm signal \citep{furlanetto_cosmology_2006, mcquinn_cosmological_2006, harker_non-parametric_2009, liu_method_2011, chapman_effect_2016, kittiwisit_measurements_2022}, and such methods will likely be effective in the context of LIM studies as well. 

Lower-redshift line interloper signals are comparatively more challenging to combat (see Ref.~\cite{ambrose_cross_2026} for a discussion of this contamination for SPHEREx cross-correlations). To this end, Refs.~\cite{visbal_measuring_2010, visbal_looking_2015, breysse_high_2016} suggest cross-correlating the target line with other probes of the same volume, such as other lines (e.g., 21-cm) or galaxy surveys. `Blind' and `guided' voxel flux-masking techniques have also been proposed as potential strategies for removing line interlopers, especially if the interloper signal is dominated by luminous sources \citep{gong_intensity_2012, gong_foreground_2014, sun_foreground_2018}. With current-term survey telescopes such as Euclid, the Rubin Observatory, and the Roman Space Telescope providing detailed measurements of the lower-redshift (interloper) galaxy population, we expect such cleaning techniques to become increasingly more effective \citep{parsons_probing_2022}. While interloper contamination from background sources is also a possible systematic, line `de-confusion' strategies, which leverage anisotropies that appear in the mapping from the observed frequency and angle to physical distances, offer a promising mitigation technique \citep{cheng_spectral_2016, lidz_removing_2016, gong_cosmological_2020}. 

Given these challenges, measuring cross-correlations of the H$\alpha$ and HeII lines at high redshift may be the most straightforward approach to probing the Pop~III signal, but if observational systematics are well controlled, then a constraint could be possible in the autocorrelation signal as well.

\subsection{Conclusion}\label{sec:conclusion}
We extend previous work introducing an efficient, fully analytical framework for spatial correlations of star formation and line luminosities to study the prospects for detecting and constraining Population~III stars with current and next-generation LIM experiments. We combine an effective model for star formation in molecular and atomic cooling halos at cosmic dawn with a flexible calculation of the recombination line luminosities of metal free stellar populations, self-consistently incorporating the inhomogeneous feedback processes regulating star formation in the smallest halos at the earliest times. 

Our phenomenological analytic model for high-$z$ star formation agrees well with more comprehensive calculations for Pop~III and II star formation, both numerical and semi-analytic. However, when compared with previous LIM forecasts for Pop~III stars, our model predicts global SFRs with a markedly distinct redshift evolution and magnitude, in general yielding more pessimistic forecasts for the detection of early star formation with real LIM surveys. Building from previous explorations of key signposts of low metallicity, top-heavy star formation, we generate a range of forecasts for auto- and cross-correlation power spectra for the H$\alpha$ and HeII recombination lines.

We leverage the flexibility of our framework to explore the sensitivity of existing and next-generation LIM measurements of these power spectra to various star formation parameters driving the LIM signal. We find that the Pop~II and III star formation efficiency and Pop~III IMF shape are the strongest levers modulating the LIM power, but the strength of this variation is highly redshift-dependent and subtly sensitive to the astrophysical and cosmological effects regulating Pop~III star formation, underscoring the need for models that can self-consistently trace star formation and feedback to make realistic forecasts of any large scale signature of high-redshift star formation.

Incorporating a flexible model to estimate the measurement uncertainties associated with a particular LIM instrument and survey design, we identify the configurations necessary to broadly survey the uncertain parameter space of minihalo-hosted pristine star formation. We find that SPHEREx will likely only be able to place joint limits on the Pop~III SFE and IMF slope, and a CDIM-like instrument will only modestly improve those constraints, suggesting that a deeper survey with more sensitive instruments will be necessary to discriminate between different Pop~III models.

Current-generation forecasts are somewhat more optimistic for models that expand Pop~III star formation beyond the `classical’ regime and allow for pristine star formation alongside enriched star formation in more massive, atomic cooling halos. In this limit, the Pop~III UVLF may be detectable in highly lensed fields with JWST, and we demonstrate that a combination of such measurements with the forthcoming SPHEREx-measured H$\alpha$ power can heighten the constraining power of both instruments. These extensions to classical Pop~III models more efficiently modulate the high-redshift LIM signal in both auto- and cross-correlations, and a CDIM-like experiment could provide important diagnostic limits for theoretical models grounded in the UVLF in the EoR.

Finally, we present this new model as an update to the publicly available \texttt{oLIMpus} intensity mapping code, which itself builds on the \texttt{Zeus21} 21-cm framework. By incorporating Pop~III star formation into \texttt{oLIMpus}, we present the first flexible, efficient, and comprehensive LIM forecasting tool for Pop~III stars, unlocking a new regime of observational inference through which we can study the early universe. The proof-of-concept exploration and detailed forecasts carried out in this work present an optimistic --- yet physically rigorous --- outlook on the power of LIM to shed light on the era of the first stars. Indeed, with the advent of SPHEREx and the power of JWST, we are poised to make some of the first detections of Pop~III star formation and statistical constraints on their demographics may be closer than ever before.

\section*{Acknowledgments}
We thank Erik Zackrisson for providing the processed Ref.~\cite{costa_evolutionary_2025} tracks used in Eq.~(\ref{eq:kappa_UV}) and Hector Cruz for helpful conversations. S.H. is supported by the National Science Foundation Graduate Research Fellowship Program under Grant No. DGE-2034835. Any opinions, findings, and conclusions or recommendations expressed in this material are those of the authors and do not necessarily reflect the views of the National Science Foundation. S.H. acknowledges support from the Future Investigators in NASA Earth and Space Science and Technology (FINESST) Grant No. 80NSSC23K1432, the UCLA Michael A. Jura Fellowship, and the UCLA Consortium for Developing Leadership in Science (CDLS) Fellowship. S.H. and S.R.F. were supported by NASA through award 80NSSC22K0818 and by the National Science Foundation through award AST-2510939. GS acknowledges support from a CIERA Postdoctoral Fellowship, with additional support provided by NSF through grant AST-2307327; by NASA through grant 23-ATP23-0008; and by STScI through grant JWST-AR-03252.001-A. JBM acknowledges support from NSF Grants AST-2307354 and AST-2408637, and NASA through grant JWST-GO-03224. 

We acknowledge that the location where this work took place, the University of California, Los Angeles, lies on indigenous land. The Gabrielino/Tongva peoples are the traditional land caretakers of Tovaangar (the Los Angeles basin and So. Channel Islands). 

This work has made extensive use of NASA's Astrophysics Data System (\href{http://ui.adsabs.harvard.edu/}{http://ui.adsabs.harvard.edu/}) and the arXiv e-Print service (\href{http://arxiv.org}{http://arxiv.org}), as well as the following softwares: \textsc{matplotlib} \citep{Matplotlib}, \textsc{numpy} \citep{numpy}, \textsc{astropy} \citep{Astropy}, \textsc{Zeus21} \citep{munoz_effective_2023, cruz_effective_2025}, \textsc{olimpus} \citep{libanore_olimpus_2025}, and \textsc{scipy} \citep{Scipy}.

\appendix

\section{Comparison with past work}\label{app:parsons_comparison}
\begin{figure*}
    \centering
    \includegraphics[width=\linewidth]{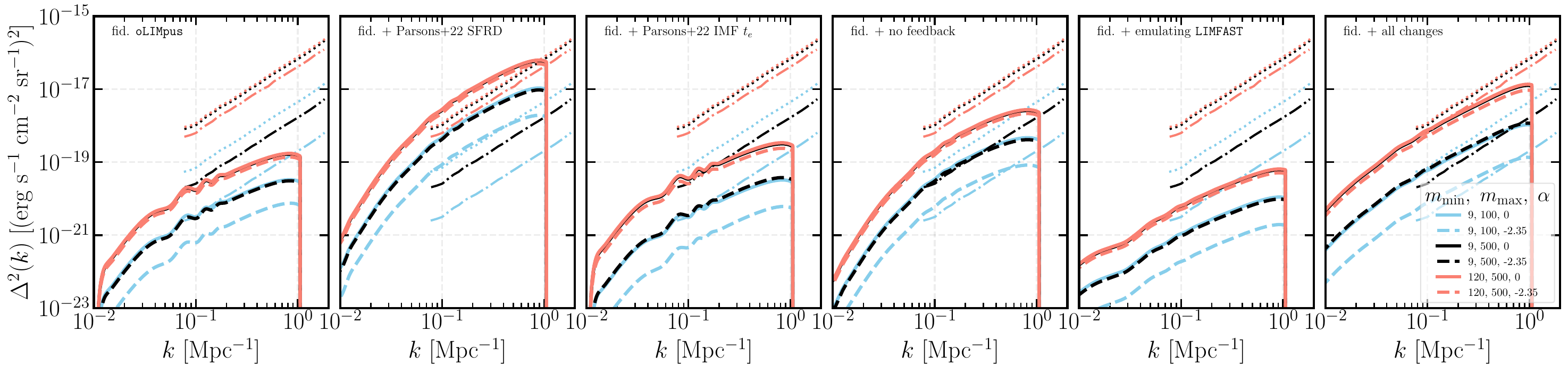}
    \caption{\textbf{Differences between the predictions made in Ref.~\cite{parsons_probing_2022} and those in this work can be attributed to well-motivated differences in assumptions.} $z=13$ H$\alpha\times$ HeII power spectra following Eq.~(\ref{eq:CF_popIIIandII}) (thick curves) for the power-law IMFs (different linestyles) used in Ref.~\cite{parsons_probing_2022}, compared with the results from the semi-numerical simulation \texttt{LIMFAST} (thin curves). The panels highlight the contribution of each of the modifications to the Ref.~\cite{parsons_probing_2022} that we incorporate here. From left to right: our fiducial model, changing the SFRD to the Ref.~\cite{parsons_probing_2022} phenomenological Pop~III SFRD, computing the line coefficients assuming a mass-independent stellar lifetime of $t_{\rm life}=3\ {\rm Myr}$ in Eq.~(\ref{eq:ionizing_photon_counts}), removing Pop~III feedback effects, applying the \texttt{Flag\_emulate\_21cmfast} flag to match the astrophysical model of \texttt{21cmFAST}/\texttt{LIMFAST}, and finally, including all of the preceding changes.}
    \label{fig:line_PS_parsons_comparison}
\end{figure*}

Here we highlight the contribution of different modeling choices to the resulting line intensity power spectrum by comparison with Ref.~\cite{parsons_probing_2022} in Fig.~\ref{fig:line_PS_parsons_comparison} and identify the key sources of the 2 orders of magnitude difference between our fiducial model and that of Ref.~\cite{parsons_probing_2022} (see leftmost panel). Ref.~\cite{parsons_probing_2022} makes a number of different methodological assumptions relative to this work, so here we isolate each difference and quantify the effect it has on the LIM power.

First, we apply a more sophisticated computation of the Pop~III SFRD motivated by simulations of the Pop~III SFE and including feedback effects such as ${\rm H}_2$-dissociating Lyman-Werner radiation and the dark matter-baryon relative `stream' velocity \citep{kulkarni_critical_2021, schauer_influence_2021, munoz_impact_2022, hegde_self-consistent_2023, nebrin_starbursts_2023}. The effects of these changes can be seen by comparing the second and fourth panels of the figure to the first --- the physically-motivated star formation efficiency reduces the overall level of Pop~III star formation considerably (by nearly an order of magnitude at $z=13$). Comparing the fourth panel to the first, we see that feedback flattens the power spectrum and the relative velocities introduce VAO `wiggles' \citep{munoz_robust_2019, munoz_impact_2022, cruz_effective_2025, cruz_rise_2026}. 

Next, Ref.~\cite{parsons_probing_2022} also model the Pop~III IMF in a slightly different manner from the procedure outlined in Section~\ref{sec:line_lum}. That is, though they use the same stellar tables, Ref.~\cite{parsons_probing_2022} assume that every Pop~III star contributes ionizing photons for a fixed time $t_e=3\ {\rm Myr}$. This amounts to taking $t_{\rm life}(m)=3\ {\rm Myr}$ in our Eq.~(\ref{eq:ionizing_photon_counts}) and thus \textit{overestimates} the contribution of the most massive, most efficiently-ionizing stars in a given population, thus heightening the sensitivity of the signal to the Pop~III IMF. 

Finally, we adjust the astrophysical model to match that of \texttt{21cmFAST}/\texttt{LIMFAST} (i.e., we change the Pop~II and III parameters as in Ref.~\cite{munoz_effective_2023, cruz_effective_2025}; see the fifth panel), which alters the magnitude of the signal and the feedback effects. 

Altogether, applying the changes to the SFE, SFRD, line luminosity, and astrophysical model, we see excellent agreement between our analytically computed power spectra and the semi-numerical predictions from Ref.~\cite{parsons_probing_2022}, confirming the robustness of our approach.

Revisiting the first panel, the cumulative effect of the different modeling assumptions is a reduction in power of nearly two orders of magnitude relative to the predictions made in Ref.~\cite{parsons_probing_2022} for a given IMF. Thus, we expect (and find) that our SNR forecasts and possible parameter constraints (and the associated conclusions drawn) will be necessarily more pessimistic than the results found in Ref.~\cite{parsons_probing_2022}. However, we emphasize that each of these choices is well-motivated given the latest theoretical understanding of Pop~III star formation and thus argue that a relatively more pessimistic outlook is nevertheless more realistic.

In addition to providing a critical validation of the analytic calculation, this comparison also demonstrates the flexibility and utility of our approach. We can efficiently vary any of the model components with minimal computational overhead and can thus tease out subtle sensitivities of the signal to the underlying modeling prescription.

\section{Derivation of the auto- and cross-power variance}\label{app:cross_power_noise}
Here we summarize the derivation for the auto- and cross-power variance introduced in Section~\ref{sec:estimating_SNR} and used throughout the observational forecasts (specifically Eqs.~(\ref{eq:auto_power_uncertainty}) and (\ref{eq:cross_power_uncertainty})). 

For a line with rest frequency $\nu$, let an individual mode of the Fourier transform of the fluctuations in the observed intensity field (i.e., including instrumental noise) be represented by $\tilde{\delta}_\nu(\mathbf{k})$. By definition, the ensemble average of a single mode estimator of the auto power spectrum $\hat{P}_{\nu\nu}$ (i.e., an observed realization of the intensity) satisfies
\begin{equation}\label{eq:single_mode_estimator}
	\big\langle \hat{P}_{\nu\nu} \big\rangle \equiv \big\langle {\rm Re}[\tilde{\delta}_\nu \tilde{\delta}_\nu^*] \big\rangle  = \big\langle \big|\tilde{\delta}_\nu(\mathbf{k})\big|^2 \big\rangle = P_{\nu\nu, \rm obs}(k,\mu).
\end{equation}

Because this is a complex Gaussian field with mean 0, we can also represent $\tilde{\delta}_\nu = x_\nu + iy_\nu$, where $x_\nu$ and $y_\nu$ are (real) independent Gaussian random fields. For an isotropic field, each of these components then has mean 0 and variance given by half of the total observed power $\langle x_\nu^2\rangle = \langle y_\nu^2\rangle = P_{\nu\nu, \rm obs}/2$. 

The variance in the observed power can be derived from the statistical variance in this single mode estimator. Using the definition in Eq.~(\ref{eq:single_mode_estimator}), the ensemble average of the squared estimator is
\begin{equation}\label{eq:PS_estimator_squared}
	\big\langle \hat{P}_{\nu\nu} ^2\big\rangle = \big\langle (x_\nu^2 + y_\nu^2)^2\big\rangle = 2 P_{\nu\nu, \rm obs}^2(k,\mu),
\end{equation}
where we have applied Wick's theorem to simplify the four-point terms as sums of the two-point terms. Therefore, the variance in observed realizations of the autocorrelation intensity field is equal to the square of the observed power:
\begin{equation}\label{eq:auto_variance}
	{\rm var}\big[\hat{P}_{\nu\nu}\big] = \langle \hat{P}_{\nu\nu} ^2\rangle - \langle \hat{P}_{\nu\nu}\rangle^2 = P_{\nu\nu, \rm obs}^2.
\end{equation}
This establishes the integrand of Eq.~(\ref{eq:auto_power_uncertainty_permode}) as the appropriate single mode uncertainty on the auto power.

\begin{figure*}
    \centering
    \includegraphics[width=\linewidth]{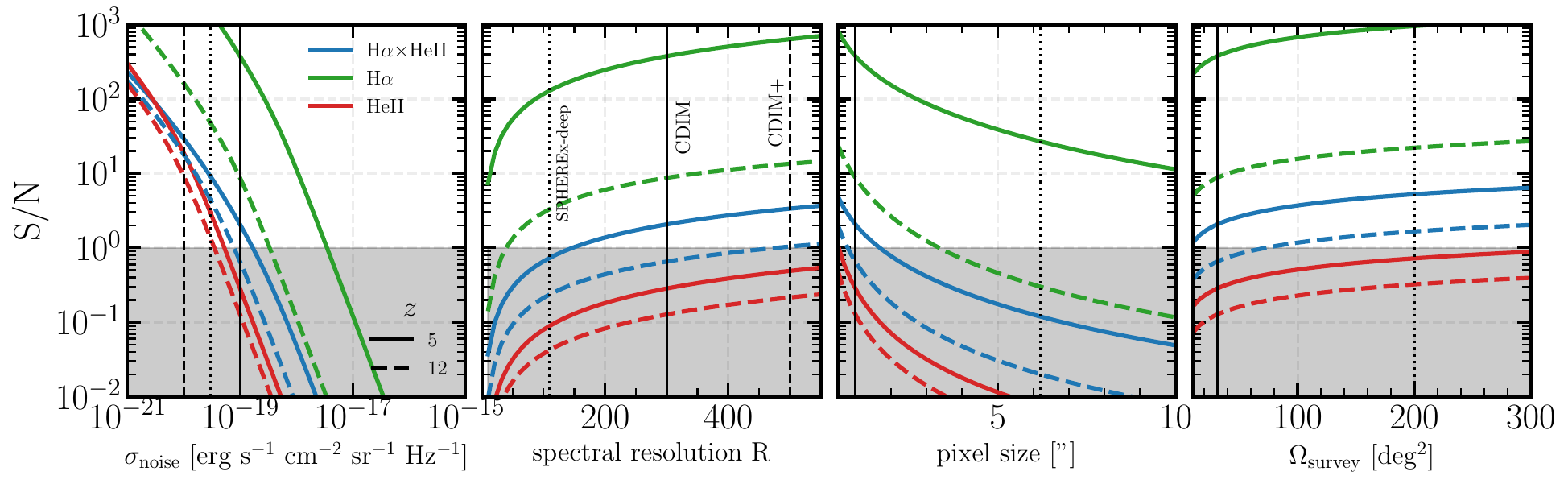}
    \caption{\textbf{The current CDIM configuration would enable a marginal detection of the H$\alpha\times$HeII cross-correlation signal, and a more sophisticated instrument or deeper integration time are necessary for a higher significance detection of top-heavy star formation.} The cumulative S/N ratio predicted for a range of survey configurations based on our fiducial model and IMF applied for each of the three LIM signals discussed in this work at $z=5$ and 12 (solid and dashed lines, respectively). From left to right, the four panels vary the instrument sensitivity, spectral resolution, pixel size, and survey area individually holding the other parameters fixed at the current CDIM specifications summarized in Table~\ref{tab:survey_specs}. The vertical lines in each panel correspond to the value of that parameter for each of the survey configurations explored in this work (SPHEREx is dotted, CDIM is solid, and CDIM+ is dashed).}
    \label{fig:SNR_survey_var}
\end{figure*}

\begin{figure*}
    \centering
    \includegraphics[width=\linewidth]{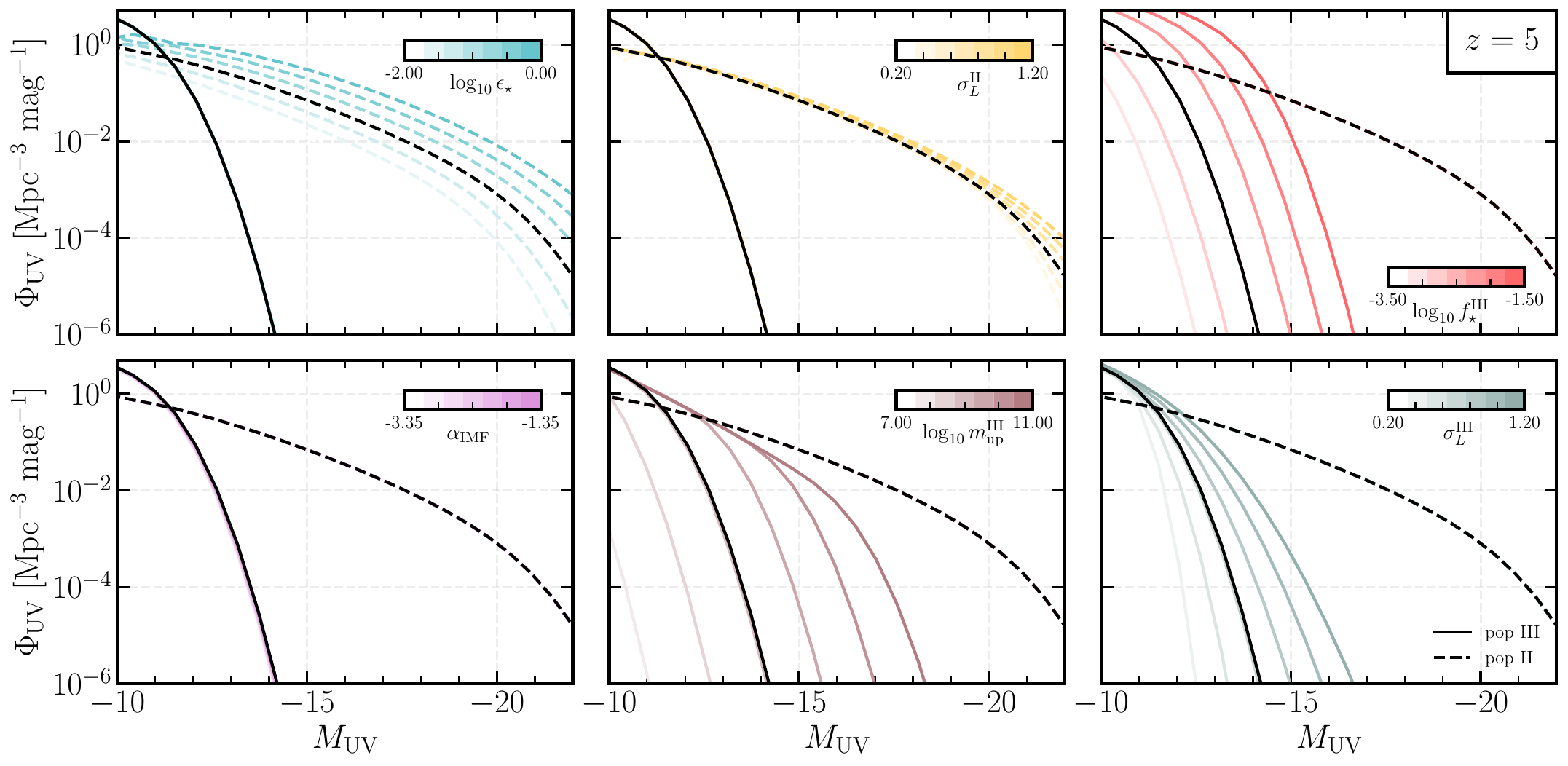}
    \caption{The Pop~III (solid) and II (dashed) UVLF predictions made with \texttt{Zeus21} varying several of the underlying astrophysical parameters: the Pop~II star formation efficiency (upper left), stochasticity in the $m_h-L_{\rm UV}^{\II}$ relation (upper center), Pop~III star formation efficiency (upper right), Pop~III IMF slope (lower left), Pop~III host halo upper mass limit (lower center), and stochasticity in the $m_h-L_{\rm UV}^{\III}$ relation (lower right). We note the lack of sensitivity  of the Pop~III UVLF to the IMF slope (lower left) panel, owing to the fact that the light-to-mass ratio is approximately constant for the most massive stars (see e.g., Ref.~\cite{bromm_first_2013}; consistent with the results of Ref.~\cite{raiter_predicted_2010}). The fiducial model used for the Fisher forecasts presented in Fig.~\ref{fig:fisher_forecast_SPHEREx} is shown in black in each panel.}
    \label{fig:uvlf_variations}
\end{figure*}

The variance of the cross power estimator follows similarly. In this case, we consider an additional line with rest frequency $\nu'$, and represent fluctuations in the intensity field as $\tilde{\delta}_{\nu'} = x_{\nu'} + iy_{\nu'}$. Then, the cross power estimator (i.e., an observed realization of the cross power field) is defined as 
\begin{equation}\label{eq:cross_power_estimator}
	\hat{P}_{\nu\nu'}(\mathbf{k}) \equiv {\rm Re}\big[\tilde{\delta}_\nu\tilde{\delta}^*_{\nu'}\big] = x_\nu x_{\nu'} + y_\nu y_{\nu'}\ .
\end{equation} 
In this case, the ensemble average is
\begin{equation}\label{eq:cross_power_estimator_average}
	\big\langle \hat{P}_{\nu\nu'}\big\rangle = P_{\nu\nu'}\ ,
\end{equation}	
so the mean of the estimator is unbiased with respect to the underlying cross signal and does \textit{not} include contributions from the instrumental noise, because thermal noise is uncorrelated between different frequency channels. However, in order to estimate the variance, we must also evaluate the average of the squared estimator, which will include cross terms between the signal and noise from the two lines, and can be expressed as:
\begin{equation}\label{eq:cross_power_squared_average}
	\big\langle \hat{P}_{\nu\nu'}^2\big\rangle = \big\langle ( x_\nu x_{\nu'} + y_\nu y_{\nu'})^2\big\rangle = \frac{1}{2}\big(\sigma_{\nu\nu}\sigma_{\nu'\nu'} + 3 P_{\nu\nu'}^2\big)\ ,
\end{equation}
where again we have applied Wick's theorem to simplify the four-point terms, and $\sigma_{\nu\nu}\equiv P_{\nu\nu,\rm obs}$ is the single mode standard deviation introduced in Eq.~(\ref{eq:auto_power_uncertainty_permode}). Therefore, we find the variance in the cross power is
\begin{equation}
    \label{eq:cross_power_variance_deriv}
    {\rm var}\big[\hat{P}_{\nu\nu'}\big] = \big\langle \hat{P}_{\nu\nu'} ^2\big\rangle - \big\langle \hat{P}_{\nu\nu'}\big\rangle^2 = \frac{1}{2}\big(\sigma_{\nu\nu}\sigma_{\nu'\nu'} + P_{\nu\nu'}^2\big)\ ,
\end{equation}
which similarly reproduces the integrand of Eq.~(\ref{eq:cross_power_uncertainty}).

In reality, we average over the many independent modes probed in the observed volume, so the variance of the total observed auto- or cross-power is smaller by a factor of $N_{\rm mode}(k)$, as defined in Eq.~(\ref{eq:N_modes}) and applied in Eqs.~(\ref{eq:auto_power_uncertainty}-\ref{eq:cross_power_uncertainty}).

\section{Optimizing survey design}\label{app:survey_design}
Given the need for next generation LIM studies to probe low-metallicity, top-heavy star formation (Section~\ref{sec:classical_popIII}), here we summarize the dependence of the cumulative SNR for each of the lines of interest in this work on the survey design choices made. In Fig.~\ref{fig:SNR_survey_var}, we isolate and vary each of the instrumental or survey specifications used in Table~\ref{tab:survey_specs} over a representative range and for each of the lines of interest in this work. For the analysis summarized in the Fig., we have assumed our fiducial model and a fixed baseline corresponding to the proposed CDIM specifications (Table~\ref{tab:survey_specs}) and varied each instrument or survey design choice individually.

In general, because it is dominated by Pop~II stars, the H$\alpha$ auto power is detectable across a wide range of survey specifications at $z\lesssim 10$. Indeed, the SPHEREx-deep survey specifications (dotted lines in the Fig.) are consistent with a measurement of the H$\alpha$ auto power at high significance at $z\lesssim 6$. 

In order to detect a signal featuring the far weaker HeII emission line, however, we will need next generation instruments. In terms of instrument sensitivity, we find that a thermal noise threshold below $\sigma_{\rm noise}\lesssim 3\times 10^{-20}\ (1\times 10^{-20})\ {\rm erg\ s^{-1}\ cm^{-2}\ sr^{-1}\ Hz^{-1}}$ is necessary to detect the H$\alpha\times$HeII cross (HeII auto) power over a range $z\sim 5-12$. These are both marginally below the currently quoted sensitivity for CDIM, even in the current deep configuration \citep{cooray_CDIM_2019}. Given a sensitivity of $\sigma_{\rm noise}\sim 10^{-19}\ {\rm erg\ s^{-1}\ cm^{-2}\ sr^{-1}\ Hz^{-1}}$, we see the same trends reflected in the spectral resolution, pixel size, and survey area. That is, the current configuration would, at best, enable a marginal detection of the cross power at $z\sim 5$ and would be unlikely to detect the signal at higher redshifts, for our fiducial IMF. This, however, indicates that the current specifications of the CDIM could allow us to place joint constraints on the Pop~III IMF and SFE, but a more sophisticated instrument, or deeper integration time, would be necessary to place any tighter constraints.

\section{Appendix D: UVLF parameter variations}\label{app:uvlf_variations}
We present the sensitivity of the UVLF to a subset of the parameters varied in Section~\ref{sec:popIII_UVLF} in Fig.~\ref{fig:uvlf_variations}. We note that the burstiness/stochasticity in the $m_h-L_{\rm UV}$ relations modify the shape of the UVLF in a distinct way compared with models that increase the upper host halo mass limit for Pop~III stars, which preserves the faint-end slope more strictly. We also find that the Pop~III UVLF is fairly insensitive to the IMF slope because the most massive stars are close to the Eddington limit and have roughly constant mass-to-light ratios \citep{bromm_first_2013}.

\clearpage

\bibliography{main}

@ARTICLE{cassata_HeII_2013,
       author = {{Cassata}, P. and {Le F{\`e}vre}, O. and {Charlot}, S. and {Contini}, T. and {Cucciati}, O. and {Garilli}, B. and {Zamorani}, G. and {Adami}, C. and {Bardelli}, S. and {Le Brun}, V. and {Lemaux}, B. and {Maccagni}, D. and {Pollo}, A. and {Pozzetti}, L. and {Tresse}, L. and {Vergani}, D. and {Zanichelli}, A. and {Zucca}, E.},
        title = "{He II emitters in the VIMOS VLT Deep Survey: Population III star formation or peculiar stellar populations in galaxies at 2 < z < 4.6?}",
      journal = {\aap},
         year = 2013,
        month = aug,
       volume = {556},
          eid = {A68},
        pages = {A68},
          doi = {10.1051/0004-6361/201220969},
archivePrefix = {arXiv},
       eprint = {1212.5270},
 primaryClass = {astro-ph.CO},
       adsurl = {https://ui.adsabs.harvard.edu/abs/2013A&A...556A..68C}
}

@ARTICLE{schaan_multi_2021,
       author = {{Schaan}, Emmanuel and {White}, Martin},
        title = "{Multi-tracer intensity mapping: cross-correlations, line noise \& decorrelation}",
      journal = {\jcap},
         year = 2021,
        month = may,
       volume = {2021},
       number = {5},
          eid = {068},
        pages = {068},
          doi = {10.1088/1475-7516/2021/05/068},
archivePrefix = {arXiv},
       eprint = {2103.01964},
 primaryClass = {astro-ph.CO},
       adsurl = {https://ui.adsabs.harvard.edu/abs/2021JCAP...05..068S}
}

@ARTICLE{yang_multitracer_2021,
       author = {{Yang}, Shengqi and {Somerville}, Rachel S. and {Pullen}, Anthony R. and {Popping}, Gerg{\"o} and {Breysse}, Patrick C. and {Maniyar}, Abhishek S.},
        title = "{Multitracer Cosmological Line Intensity Mapping Mock Light-cone Simulation}",
      journal = {\apj},
         year = 2021,
        month = apr,
       volume = {911},
       number = {2},
          eid = {132},
        pages = {132},
          doi = {10.3847/1538-4357/abec75},
archivePrefix = {arXiv},
       eprint = {2009.11933},
 primaryClass = {astro-ph.GA},
       adsurl = {https://ui.adsabs.harvard.edu/abs/2021ApJ...911..132Y}
}

@ARTICLE{zaldarriaga_21_2004,
       author = {{Zaldarriaga}, Matias and {Furlanetto}, Steven R. and {Hernquist}, Lars},
        title = "{21 Centimeter Fluctuations from Cosmic Gas at High Redshifts}",
      journal = {\apj},
         year = 2004,
        month = jun,
       volume = {608},
       number = {2},
        pages = {622-635},
          doi = {10.1086/386327},
archivePrefix = {arXiv},
       eprint = {astro-ph/0311514},
 primaryClass = {astro-ph},
       adsurl = {https://ui.adsabs.harvard.edu/abs/2004ApJ...608..622Z}
}

@ARTICLE{furlanetto_cross_2007,
       author = {{Furlanetto}, Steven R. and {Lidz}, Adam},
        title = "{The Cross-Correlation of High-Redshift 21 cm and Galaxy Surveys}",
      journal = {\apj},
         year = 2007,
        month = may,
       volume = {660},
       number = {2},
        pages = {1030-1038},
          doi = {10.1086/513009},
archivePrefix = {arXiv},
       eprint = {astro-ph/0611274},
 primaryClass = {astro-ph},
       adsurl = {https://ui.adsabs.harvard.edu/abs/2007ApJ...660.1030F}
}

@ARTICLE{lidz_probing_2009,
       author = {{Lidz}, Adam and {Zahn}, Oliver and {Furlanetto}, Steven R. and {McQuinn}, Matthew and {Hernquist}, Lars and {Zaldarriaga}, Matias},
        title = "{Probing Reionization with the 21 cm Galaxy Cross-Power Spectrum}",
      journal = {\apj},
         year = 2009,
        month = jan,
       volume = {690},
       number = {1},
        pages = {252-266},
          doi = {10.1088/0004-637X/690/1/252},
archivePrefix = {arXiv},
       eprint = {0806.1055},
 primaryClass = {astro-ph},
       adsurl = {https://ui.adsabs.harvard.edu/abs/2009ApJ...690..252L}
}

@ARTICLE{visbal_cross_2023,
       author = {{Visbal}, Eli and {McQuinn}, Matthew},
        title = "{Cross Correlation of Pencil-beam Galaxy Surveys and Line-intensity Maps: An Application of the James Webb Space Telescope}",
      journal = {\apj},
         year = 2023,
        month = oct,
       volume = {956},
       number = {2},
          eid = {84},
        pages = {84},
          doi = {10.3847/1538-4357/ace435},
archivePrefix = {arXiv},
       eprint = {2212.05096},
 primaryClass = {astro-ph.CO},
       adsurl = {https://ui.adsabs.harvard.edu/abs/2023ApJ...956...84V}
}

@ARTICLE{zackrisson_detection_2024,
       author = {{Zackrisson}, Erik and {Hultquist}, Adam and {Kordt}, Aron and {Diego}, Jose M. and {Nabizadeh}, Armin and {Vikaeus}, Anton and {Meena}, Ashish Kumar and {Zitrin}, Adi and {Volpato}, Guglielmo and {Lundqvist}, Emma and {Welch}, Brian and {Costa}, Guglielmo and {Windhorst}, Rogier A.},
        title = "{The detection and characterization of highly magnified stars with JWST: prospects of finding Population III}",
      journal = {\mnras},
         year = 2024,
        month = sep,
       volume = {533},
       number = {3},
        pages = {2727-2746},
          doi = {10.1093/mnras/stae1881},
archivePrefix = {arXiv},
       eprint = {2312.09289},
 primaryClass = {astro-ph.GA},
       adsurl = {https://ui.adsabs.harvard.edu/abs/2024MNRAS.533.2727Z}
}

@ARTICLE{costa_evolutionary_2025,
       author = {{Costa}, G. and {Shepherd}, K.~G. and {Bressan}, A. and {Addari}, F. and {Chen}, Y. and {Fu}, X. and {Volpato}, G. and {Nguyen}, C.~T. and {Girardi}, L. and {Marigo}, P. and {Mazzi}, A. and {Pastorelli}, G. and {Trabucchi}, M. and {Bossini}, D. and {Zaggia}, S.},
        title = "{Evolutionary tracks, ejecta, and ionizing photons from intermediate-mass to very massive stars with PARSEC}",
      journal = {\aap},
         year = 2025,
        month = feb,
       volume = {694},
          eid = {A193},
        pages = {A193},
          doi = {10.1051/0004-6361/202452573},
archivePrefix = {arXiv},
       eprint = {2501.12917},
 primaryClass = {astro-ph.SR},
       adsurl = {https://ui.adsabs.harvard.edu/abs/2025A&A...694A.193C}
}

@ARTICLE{sun_limfastIII_2025,
       author = {{Sun}, Guochao and {Lidz}, Adam and {Chang}, Tzu-Ching and {Mirocha}, Jordan and {Furlanetto}, Steven R.},
        title = "{LIMFAST. III. Timing Cosmic Reionization with the 21 cm and Near-infrared Backgrounds}",
      journal = {\apj},
         year = 2025,
        month = mar,
       volume = {981},
       number = {1},
          eid = {92},
        pages = {92},
          doi = {10.3847/1538-4357/adae12},
archivePrefix = {arXiv},
       eprint = {2410.21410},
 primaryClass = {astro-ph.CO},
       adsurl = {https://ui.adsabs.harvard.edu/abs/2025ApJ...981...92S}
}

@ARTICLE{libanore_effective_2025,
       author = {{Libanore}, Sarah and {Mu{\~n}oz}, Julian B. and {Kovetz}, Ely D.},
        title = "{Effective model for line intensity mapping: Auto- and cross-power spectra in the cosmic dawn and reionization}",
      journal = {\prd},
         year = 2025,
        month = oct,
       volume = {112},
       number = {8},
          eid = {083552},
        pages = {083552},
          doi = {10.1103/xq1r-zh51},
archivePrefix = {arXiv},
       eprint = {2507.15922},
 primaryClass = {astro-ph.CO},
       adsurl = {https://ui.adsabs.harvard.edu/abs/2025PhRvD.112h3552L}
}

@ARTICLE{libanore_new_2025,
       author = {{Libanore}, Sarah and {Kovetz}, Ely D. and {Munoz}, Julian B. and {Sklansky}, Yonatan and {Th{\'e}lie}, Emilie},
        title = "{A New Boundary Condition on Reionization}",
      journal = {arXiv e-prints},
         year = 2025,
        month = sep,
          eid = {arXiv:2509.08886},
        pages = {arXiv:2509.08886},
          doi = {10.48550/arXiv.2509.08886},
archivePrefix = {arXiv},
       eprint = {2509.08886},
 primaryClass = {astro-ph.CO},
       adsurl = {https://ui.adsabs.harvard.edu/abs/2025arXiv250908886L}
}

@ARTICLE{sato-polito_multitracer_2023,
       author = {{Sato-Polito}, Gabriela and {Kokron}, Nickolas and {Bernal}, Jos{\'e} Luis},
        title = "{A multitracer empirically driven approach to line-intensity mapping light cones}",
      journal = {\mnras},
         year = 2023,
        month = dec,
       volume = {526},
       number = {4},
        pages = {5883-5899},
          doi = {10.1093/mnras/stad2498},
archivePrefix = {arXiv},
       eprint = {2212.08056},
 primaryClass = {astro-ph.CO},
       adsurl = {https://ui.adsabs.harvard.edu/abs/2023MNRAS.526.5883S}
}

@ARTICLE{schaan_astrophysics_2021,
       author = {{Schaan}, Emmanuel and {White}, Martin},
        title = "{Astrophysics \& cosmology from line intensity mapping vs galaxy surveys}",
      journal = {\jcap},
         year = 2021,
        month = may,
       volume = {2021},
       number = {5},
          eid = {067},
        pages = {067},
          doi = {10.1088/1475-7516/2021/05/067},
archivePrefix = {arXiv},
       eprint = {2103.01971},
 primaryClass = {astro-ph.CO},
       adsurl = {https://ui.adsabs.harvard.edu/abs/2021JCAP...05..067S}
}

@ARTICLE{fujimoto_glimpse_2025,
       author = {{Fujimoto}, Seiji and {Naidu}, Rohan P. and {Chisholm}, John and {Atek}, Hakim and {Endsley}, Ryan and {Kokorev}, Vasily and {Furtak}, Lukas J. and {Pan}, Richard and {Liu}, Boyuan and {Bromm}, Volker and {Venditti}, Alessandra and {Visbal}, Eli and {Sarmento}, Richard and {Weibel}, Andrea and {Oesch}, Pascal A. and {Brammer}, Gabriel and {Schaerer}, Daniel and {Adamo}, Angela and {Berg}, Danielle A. and {Bezanson}, Rachel and {Bouwens}, Rychard and {Chemerynska}, Iryna and {Claeyssens}, Ad{\'e}la{\"\i}de and {Dessauges-Zavadsky}, Miroslava and {Frebel}, Anna and {Korber}, Damien and {Labbe}, Ivo and {Marques-Chaves}, Rui and {Matthee}, Jorryt and {McQuinn}, Kristen B.~W. and {Mu{\~n}oz}, Julian B. and {Natarajan}, Priyamvada and {Saldana-Lopez}, Alberto and {Suess}, Katherine A. and {Volonteri}, Marta and {Zitrin}, Adi},
        title = "{GLIMPSE: An Ultrafaint $\simeq${}10$^{5}$ M$_{{\ensuremath{\odot}}}$ Pop III Galaxy Candidate and First Constraints on the Pop III UV Luminosity Function at z $\simeq$ 6-7}",
      journal = {\apj},
         year = 2025,
        month = aug,
       volume = {989},
       number = {1},
          eid = {46},
        pages = {46},
          doi = {10.3847/1538-4357/ade9a1},
archivePrefix = {arXiv},
       eprint = {2501.11678},
 primaryClass = {astro-ph.GA},
       adsurl = {https://ui.adsabs.harvard.edu/abs/2025ApJ...989...46F}
}

@ARTICLE{sheth_large_1999,
       author = {{Sheth}, Ravi K. and {Tormen}, Giuseppe},
        title = "{Large-scale bias and the peak background split}",
      journal = {\mnras},
         year = 1999,
        month = sep,
       volume = {308},
       number = {1},
        pages = {119-126},
          doi = {10.1046/j.1365-8711.1999.02692.x},
archivePrefix = {arXiv},
       eprint = {astro-ph/9901122},
 primaryClass = {astro-ph},
       adsurl = {https://ui.adsabs.harvard.edu/abs/1999MNRAS.308..119S}
}

@ARTICLE{atek_jwst_2025,
       author = {{Atek}, Hakim and {Chisholm}, John and {Kokorev}, Vasily and {Endsley}, Ryan and {Pan}, Richard and {Furtak}, Lukas and {Chemerynska}, Iryna and {Richard}, Johan and {Claeyssens}, Ad{\'e}la{\"\i}de and {Oesch}, Pascal and {Fujimoto}, Seiji and {Naidu}, Rohan and {Korber}, Damien and {Schaerer}, Daniel and {Blaizot}, Jeremy and {Rosdahl}, Joki and {Adamo}, Angela and {Asada}, Yoshihisa and {Basu}, Arghyadeep and {Beauchesne}, Benjamin and {Berg}, Danielle and {Bezanson}, Rachel and {Bouwens}, Rychard and {Brammer}, Gabriel and {Dessauges-Zavadsky}, Miroslava and {Ellien}, Ama{\"e}l and {Ezziati}, Meriam and {Fei}, Qinyue and {Goovaerts}, Ilias and {Heurtier}, Sylvain and {Hsiao}, Tiger Yu-Yang and {Jecmen}, Michelle and {Khullar}, Gourav and {Kneib}, Jean-Paul and {Labb{\'e}}, Ivo and {Leclercq}, Floriane and {Marques-Chaves}, Rui and {Mason}, Charlotte and {McQuinn}, Kristen B.~W. and {Mu{\~n}oz}, Julian B. and {Natarajan}, Priyamvada and {Saldana-Lopez}, Alberto and {Stephenson}, Mabel G. and {Trebitsch}, Maxime and {Volonteri}, Marta and {Weibel}, Andrea and {Zitrin}, Adi},
        title = "{JWST's GLIMPSE: an overview of the deepest probe of early galaxy formation and cosmic reionization}",
      journal = {arXiv e-prints},
         year = 2025,
        month = nov,
          eid = {arXiv:2511.07542},
        pages = {arXiv:2511.07542},
          doi = {10.48550/arXiv.2511.07542},
archivePrefix = {arXiv},
       eprint = {2511.07542},
 primaryClass = {astro-ph.GA},
       adsurl = {https://ui.adsabs.harvard.edu/abs/2025arXiv251107542A}
}

@misc{libanore_olimpus_2025,
  title = {{{oLIMpus}}: {{An Effective Model}} for {{Line Intensity Mapping Auto-}} and {{Cross- Power Spectra}} in {{Cosmic Dawn}} and {{Reionization}}},
  shorttitle = {{{oLIMpus}}},
  author = {Libanore, Sarah and Munoz, Julian B. and Kovetz, Ely D.},
  year = {2025},
  month = jul,
  number = {arXiv:2507.15922},
  eprint = {2507.15922},
  primaryclass = {astro-ph},
  publisher = {arXiv},
  doi = {10.48550/arXiv.2507.15922},
  urldate = {2025-07-23},
  archiveprefix = {arXiv}
}

@article{cruz_effective_2025,
  title = {Effective Model for the 21-Cm Signal with Population {{III}} Stars},
  author = {Cruz, Hector Afonso G. and Mu{\~n}oz, Julian B. and Sabti, Nashwan and Kamionkowski, Marc},
  year = {2025},
  month = apr,
  journal = {Physical Review D},
  volume = {111},
  number = {8},
  pages = {083503},
  publisher = {American Physical Society},
  doi = {10.1103/PhysRevD.111.083503},
  urldate = {2025-12-08}
}

@ARTICLE{coles_lognormal_1991,
       author = {{Coles}, Peter and {Jones}, Bernard},
        title = "{A lognormal model for the cosmological mass distribution.}",
      journal = {\mnras},
         year = 1991,
        month = jan,
       volume = {248},
        pages = {1-13},
          doi = {10.1093/mnras/248.1.1},
       adsurl = {https://ui.adsabs.harvard.edu/abs/1991MNRAS.248....1C}
}

@ARTICLE{press_formation_1974,
       author = {{Press}, William H. and {Schechter}, Paul},
        title = "{Formation of Galaxies and Clusters of Galaxies by Self-Similar Gravitational Condensation}",
      journal = {\apj},
         year = 1974,
        month = feb,
       volume = {187},
        pages = {425-438},
          doi = {10.1086/152650},
       adsurl = {https://ui.adsabs.harvard.edu/abs/1974ApJ...187..425P}
}

@ARTICLE{bond_excursion_1991,
       author = {{Bond}, J.~R. and {Cole}, S. and {Efstathiou}, G. and {Kaiser}, N.},
        title = "{Excursion Set Mass Functions for Hierarchical Gaussian Fluctuations}",
      journal = {\apj},
         year = 1991,
        month = oct,
       volume = {379},
        pages = {440},
          doi = {10.1086/170520},
       adsurl = {https://ui.adsabs.harvard.edu/abs/1991ApJ...379..440B}
}

@ARTICLE{williams_lighting_2024,
       author = {{Williams}, Claire E. and {Lake}, William and {Naoz}, Smadar and {Burkhart}, Blakesley and {Treu}, Tommaso and {Marinacci}, Federico and {Nakazato}, Yurina and {Vogelsberger}, Mark and {Yoshida}, Naoki and {Chiaki}, Gen and {Chiou}, Yeou S. and {Chen}, Avi},
        title = "{The Supersonic Project: Lighting Up the Faint End of the JWST UV Luminosity Function}",
      journal = {\apjl},
         year = 2024,
        month = jan,
       volume = {960},
       number = {2},
          eid = {L16},
        pages = {L16},
          doi = {10.3847/2041-8213/ad1491},
archivePrefix = {arXiv},
       eprint = {2310.03799},
 primaryClass = {astro-ph.GA},
       adsurl = {https://ui.adsabs.harvard.edu/abs/2024ApJ...960L..16W}
}

@ARTICLE{lake_early_2024,
       author = {{Lake}, William and {Williams}, Claire E. and {Naoz}, Smadar and {Marinacci}, Federico and {Burkhart}, Blakesley and {Vogelsberger}, Mark and {Yoshida}, Naoki and {Chiaki}, Gen and {Chen}, Avi and {Chiou}, Yeou S.},
        title = "{The Supersonic Project: Early Star Formation with the Streaming Velocity}",
      journal = {\apj},
         year = 2024,
        month = oct,
       volume = {973},
       number = {2},
          eid = {115},
        pages = {115},
          doi = {10.3847/1538-4357/ad6762},
archivePrefix = {arXiv},
       eprint = {2405.14938},
 primaryClass = {astro-ph.GA},
       adsurl = {https://ui.adsabs.harvard.edu/abs/2024ApJ...973..115L}
}

@ARTICLE{kubat_spherically_2012,
       author = {{Kub{\'a}t}, Ji{\v{r}}{\'\i}},
        title = "{Spherically Symmetric NLTE Model Atmospheres of Hot Hydrogen-Helium First Stars}",
      journal = {\apjs},
         year = 2012,
        month = dec,
       volume = {203},
       number = {2},
          eid = {20},
        pages = {20},
          doi = {10.1088/0067-0049/203/2/20},
archivePrefix = {arXiv},
       eprint = {1209.6581},
 primaryClass = {astro-ph.SR},
       adsurl = {https://ui.adsabs.harvard.edu/abs/2012ApJS..203...20K}
}

@ARTICLE{marigo_zero_2001,
       author = {{Marigo}, P. and {Girardi}, L. and {Chiosi}, C. and {Wood}, P.~R.},
        title = "{Zero-metallicity stars. I. Evolution at constant mass}",
      journal = {\aap},
         year = 2001,
        month = may,
       volume = {371},
        pages = {152-173},
          doi = {10.1051/0004-6361:20010309},
archivePrefix = {arXiv},
       eprint = {astro-ph/0102253},
 primaryClass = {astro-ph},
       adsurl = {https://ui.adsabs.harvard.edu/abs/2001A&A...371..152M}
}

@ARTICLE{mas-ribas_boosting_2016,
       author = {{Mas-Ribas}, Llu{\'\i}s and {Dijkstra}, Mark and {Forero-Romero}, Jaime E.},
        title = "{Boosting Ly{\ensuremath{\alpha}} and He II {\ensuremath{\lambda}}1640 Line Fluxes from Population III Galaxies: Stochastic IMF Sampling and Departures from Case-B}",
      journal = {\apj},
         year = 2016,
        month = dec,
       volume = {833},
       number = {1},
          eid = {65},
        pages = {65},
          doi = {10.3847/1538-4357/833/1/65},
archivePrefix = {arXiv},
       eprint = {1609.02150},
 primaryClass = {astro-ph.GA},
       adsurl = {https://ui.adsabs.harvard.edu/abs/2016ApJ...833...65M}
}

@ARTICLE{hegde_efficient_2025,
       author = {{Hegde}, Sahil and {Furlanetto}, Steven R.},
        title = "{Efficient semi-analytic modelling of Pop III star formation from Cosmic Dawn to Reionization}",
      journal = {The Open Journal of Astrophysics},
         year = 2025,
        month = oct,
       volume = {8},
          eid = {147},
        pages = {147},
          doi = {10.33232/001c.145070},
archivePrefix = {arXiv},
       eprint = {2507.19581},
 primaryClass = {astro-ph.GA},
       adsurl = {https://ui.adsabs.harvard.edu/abs/2025OJAp....8E.147H}
}

@ARTICLE{venditti_bursty_2025,
       author = {{Venditti}, Alessandra and {Mu{\~n}oz}, Julian B. and {Bromm}, Volker and {Fujimoto}, Seiji and {Finkelstein}, Steven L. and {Chisholm}, John},
        title = "{Bursty or Heavy? The Surprise of Bright Population III Systems in the Reionization Era}",
      journal = {\apj},
         year = 2025,
        month = nov,
       volume = {994},
       number = {1},
          eid = {32},
        pages = {32},
          doi = {10.3847/1538-4357/ae0610},
archivePrefix = {arXiv},
       eprint = {2505.20263},
 primaryClass = {astro-ph.GA},
       adsurl = {https://ui.adsabs.harvard.edu/abs/2025ApJ...994...32V}
}

@ARTICLE{storck_megatron_2025,
       author = {{Storck}, Anatole and {Katz}, Harley and {Devriendt}, Julien and {Slyz}, Adrianne and {Cadiou}, Corentin and {Choustikov}, Nicholas and {Rey}, Martin P. and {Saxena}, Aayush and {Agertz}, Oscar and {Kimm}, Taysun},
        title = "{MEGATRON: The environments of Population III stars at Cosmic Dawn and their connection to present day galaxies}",
      journal = {arXiv e-prints},
         year = 2025,
        month = oct,
          eid = {arXiv:2510.06853},
        pages = {arXiv:2510.06853},
          doi = {10.48550/arXiv.2510.06853},
archivePrefix = {arXiv},
       eprint = {2510.06853},
 primaryClass = {astro-ph.GA},
       adsurl = {https://ui.adsabs.harvard.edu/abs/2025arXiv251006853S}
}

@ARTICLE{katz_megatron_2025,
       author = {{Katz}, Harley and {Rey}, Martin P. and {Cadiou}, Corentin and {Agertz}, Oscar and {Blaizot}, Jeremy and {Cameron}, Alex J. and {Choustikov}, Nicholas and {Devriendt}, Julien and {Hauk}, Uliana and {Jones}, Gareth C. and {Kimm}, Taysun and {Laseter}, Isaac and {Martin-Alvarez}, Sergio and {Matsumoto}, Kosei and {Pearce}, Autumn and {Rodr{\'\i}guez Montero}, Francisco and {Rosdahl}, Joki and {Sanati}, Mahsa and {Saxena}, Aayush and {Slyz}, Adrianne and {Stiskalek}, Richard and {Storck}, Anatole and {Veenema}, Oscar and {Yee}, Wonjae},
        title = "{MEGATRON: Reproducing the Diversity of High-Redshift Galaxy Spectra with Cosmological Radiation Hydrodynamics Simulations}",
      journal = {arXiv e-prints},
         year = 2025,
        month = oct,
          eid = {arXiv:2510.05201},
        pages = {arXiv:2510.05201},
          doi = {10.48550/arXiv.2510.05201},
archivePrefix = {arXiv},
       eprint = {2510.05201},
 primaryClass = {astro-ph.GA},
       adsurl = {https://ui.adsabs.harvard.edu/abs/2025arXiv251005201K}
}

@ARTICLE{zier_thesan_2025,
       author = {{Zier}, Oliver and {Kannan}, Rahul and {Smith}, Aaron and {Puchwein}, Ewald and {Vogelsberger}, Mark and {Borrow}, Josh and {Garaldi}, Enrico and {Keating}, Laura and {McClymont}, William and {Shen}, Xuejian and {Hernquist}, Lars},
        title = "{The THESAN-ZOOM project: Population III star formation continues until the end of reionization}",
      journal = {\mnras},
         year = 2025,
        month = nov,
       volume = {544},
       number = {1},
        pages = {410-429},
          doi = {10.1093/mnras/staf1053},
archivePrefix = {arXiv},
       eprint = {2503.03806},
 primaryClass = {astro-ph.GA},
       adsurl = {https://ui.adsabs.harvard.edu/abs/2025MNRAS.544..410Z}
}

@ARTICLE{morishita_pristine_2025,
       author = {{Morishita}, Takahiro and {Liu}, Zhaoran and {Stiavelli}, Massimo and {Treu}, Tommaso and {Bergamini}, Pietro and {Zhang}, Yechi},
        title = "{Pristine Massive Star Formation Caught at the Break of Cosmic Dawn}",
      journal = {arXiv e-prints},
         year = 2025,
        month = jul,
          eid = {arXiv:2507.10521},
        pages = {arXiv:2507.10521},
          doi = {10.48550/arXiv.2507.10521},
archivePrefix = {arXiv},
       eprint = {2507.10521},
 primaryClass = {astro-ph.CO},
       adsurl = {https://ui.adsabs.harvard.edu/abs/2025arXiv250710521M}
}

@ARTICLE{fujimoto_glimpsed_2025,
       author = {{Fujimoto}, Seiji and {Asada}, Yoshihisa and {Naidu}, Rohan P. and {Chisholm}, John and {Atek}, Hakim and {Brammer}, Gabriel and {Berg}, Danielle A. and {Schaerer}, Daniel and {Kokorev}, Vasily and {Furtak}, Lukas J. and {Richard}, Johan and {Venditti}, Alessandra and {Bromm}, Volker and {Adamo}, Angela and {Claeyssens}, Adelaide and {Dessauges-Zavadsky}, Miroslava and {Fei}, Qinyue and {Hsiao}, Tiger Yu-Yang and {Korber}, Damien and {Munoz}, Julian B. and {Pan}, Richard and {Saldana-Lopez}, Alberto},
        title = "{GLIMPSE-D: An Exotic Balmer-Jump Object at z=6.20? Revisiting Photometric Selection and the Cosmic Abundance of Pop III Galaxies}",
      journal = {arXiv e-prints},
         year = 2025,
        month = dec,
          eid = {arXiv:2512.11790},
        pages = {arXiv:2512.11790},
          doi = {10.48550/arXiv.2512.11790},
archivePrefix = {arXiv},
       eprint = {2512.11790},
 primaryClass = {astro-ph.GA},
       adsurl = {https://ui.adsabs.harvard.edu/abs/2025arXiv251211790F}
}

@ARTICLE{sun_LIMFAST_2026,
       author = {{Sun}, Guochao and {Nguyen}, Tri and {Faucher-Gigu{\`e}re}, Claude-Andr{\'e} and {Lidz}, Adam and {Starkenburg}, Tjitske and {Scott}, Bryan R. and {Chang}, Tzu-Ching and {Furlanetto}, Steven R.},
        title = "{LIMFAST. IV. Learning high-redshift galaxy formation from multiline intensity mapping with implicit likelihood inference}",
      journal = {\jcap},
         year = 2026,
        month = feb,
       volume = {2026},
       number = {2},
          eid = {008},
        pages = {008},
          doi = {10.1088/1475-7516/2026/02/008},
archivePrefix = {arXiv},
       eprint = {2509.07060},
 primaryClass = {astro-ph.GA},
       adsurl = {https://ui.adsabs.harvard.edu/abs/2026JCAP...02..008S}
}

@BOOK{loeb_first_2013,
       author = {{Loeb}, Abraham and {Furlanetto}, Steven R.},
        title = "{The First Galaxies in the Universe}",
         year = 2013,
         publisher = {Princeton University Press},
       adsurl = {https://ui.adsabs.harvard.edu/abs/2013fgu..book.....L}
}

@ARTICLE{bromm_formation_2013,
       author = {{Bromm}, Volker},
        title = "{Formation of the first stars}",
      journal = {Reports on Progress in Physics},
         year = 2013,
        month = nov,
       volume = {76},
       number = {11},
          eid = {112901},
        pages = {112901},
          doi = {10.1088/0034-4885/76/11/112901},
archivePrefix = {arXiv},
       eprint = {1305.5178},
 primaryClass = {astro-ph.CO},
       adsurl = {https://ui.adsabs.harvard.edu/abs/2013RPPh...76k2901B}
}

@article{chon_impact_2024,
	title = {Impact of radiative feedback on the initial mass function of metal-poor stars},
	volume = {530},
	issn = {0035-8711},
	url = {https://ui.adsabs.harvard.edu/abs/2024MNRAS.530.2453C},
	doi = {10.1093/mnras/stae1027},
	urldate = {2024-08-02},
	journal = {Monthly Notices of the Royal Astronomical Society},
	author = {Chon, Sunmyon and Hosokawa, Takashi and Omukai, Kazuyuki and Schneider, Raffaella},
	month = may,
	year = {2024},
	note = {Publisher: OUP
ADS Bibcode: 2024MNRAS.530.2453C},
	pages = {2453--2474},
}

@article{feathers_global_2024,
	title = {A {Global} {Semianalytic} {Model} of the {First} {Stars} and {Galaxies} {Including} {Dark} {Matter} {Halo} {Merger} {Histories}},
	volume = {962},
	issn = {0004-637X},
	url = {https://ui.adsabs.harvard.edu/abs/2024ApJ...962...62F},
	doi = {10.3847/1538-4357/ad1688},
	urldate = {2024-07-10},
	journal = {The Astrophysical Journal},
	author = {Feathers, Colton R. and Kulkarni, Mihir and Visbal, Eli and Hazlett, Ryan},
	month = feb,
	year = {2024},
	note = {Publisher: IOP
ADS Bibcode: 2024ApJ...962...62F},
	pages = {62},
}

@article{zackrisson_spectral_2011,
	title = {The {Spectral} {Evolution} of the {First} {Galaxies}. {I}. {James} {Webb} {Space} {Telescope} {Detection} {Limits} and {Color} {Criteria} for {Population} {III} {Galaxies}},
	volume = {740},
	issn = {0004-637X},
	url = {https://ui.adsabs.harvard.edu/abs/2011ApJ...740...13Z},
	doi = {10.1088/0004-637X/740/1/13},
	urldate = {2024-07-04},
	journal = {The Astrophysical Journal},
	author = {Zackrisson, Erik and Rydberg, Claes-Erik and Schaerer, Daniel and Östlin, Göran and Tuli, Manan},
	month = oct,
	year = {2011},
	note = {Publisher: IOP
ADS Bibcode: 2011ApJ...740...13Z},
	pages = {13},
}

@misc{gelli_impact_2024,
	title = {The impact of mass-dependent stochasticity at cosmic dawn},
	url = {https://ui.adsabs.harvard.edu/abs/2024arXiv240513108G},
	doi = {10.48550/arXiv.2405.13108},
	urldate = {2024-07-02},
	author = {Gelli, Viola and Mason, Charlotte and Hayward, Christopher C.},
	month = may,
	year = {2024},
	note = {Publication Title: arXiv e-prints
ADS Bibcode: 2024arXiv240513108G},
}

@article{oh_he_2001,
	title = {{HE} {II} {Recombination} {Lines} from the {First} {Luminous} {Objects}},
	volume = {553},
	issn = {0004-637X},
	url = {https://ui.adsabs.harvard.edu/abs/2001ApJ...553...73O},
	doi = {10.1086/320650},
	urldate = {2024-06-19},
	journal = {The Astrophysical Journal},
	author = {Oh, S. Peng and Haiman, Zoltán and Rees, Martin J.},
	month = may,
	year = {2001},
	note = {Publisher: IOP
ADS Bibcode: 2001ApJ...553...73O},
	pages = {73--77},
}

@article{heger_how_2003,
	title = {How {Massive} {Single} {Stars} {End} {Their} {Life}},
	volume = {591},
	issn = {0004-637X},
	url = {https://ui.adsabs.harvard.edu/abs/2003ApJ...591..288H},
	doi = {10.1086/375341},
	urldate = {2024-05-14},
	journal = {The Astrophysical Journal},
	author = {Heger, A. and Fryer, C. L. and Woosley, S. E. and Langer, N. and Hartmann, D. H.},
	month = jul,
	year = {2003},
	note = {Publisher: IOP
ADS Bibcode: 2003ApJ...591..288H},
	pages = {288--300},
}

@misc{furlanetto_cosmology_2006,
	title = {Cosmology at low frequencies: {The} 21 cm transition and the high-redshift {Universe}},
	shorttitle = {doi},
	url = {https://reader.elsevier.com/reader/sd/pii/S0370157306002730?token=1FAC730BFCE76F970BAE2BE224100F78ECD0633F8E8F14A39C5A3965FFEA3A75FE6948A6840D9F66CC4BFAEF47EC9845&originRegion=us-east-1&originCreation=20221023230446},
	
	urldate = {2022-10-23},
	author = {Furlanetto, Steven R and Oh, S. Peng and Briggs, Frank},
	month = aug,
	year = {2006},
	doi = {10.1016/j.physrep.2006.08.002},
}

@article{parsons_probing_2022,
	title = {Probing {Population} {III} {Initial} {Mass} {Functions} with {He} {II}/{H$\alpha$} {Intensity} {Mapping}},
	volume = {933},
	issn = {0004-637X},
	url = {https://ui.adsabs.harvard.edu/abs/2022ApJ...933..141P},
	doi = {10.3847/1538-4357/ac746b},
	urldate = {2024-04-25},
	journal = {The Astrophysical Journal},
	author = {Parsons, Jasmine and Mas-Ribas, Lluís and Sun, Guochao and Chang, Tzu-Ching and Gonzalez, Michael O. and Mebane, Richard H.},
	month = jul,
	year = {2022},
	note = {Publisher: IOP
ADS Bibcode: 2022ApJ...933..141P},
	pages = {141},
}

@article{sun_revealing_2021,
	title = {Revealing the formation histories of the first stars with the cosmic near-infrared background},
	volume = {508},
	issn = {0035-8711},
	url = {https://ui.adsabs.harvard.edu/abs/2021MNRAS.508.1954S},
	doi = {10.1093/mnras/stab2697},
	urldate = {2024-04-11},
	journal = {Monthly Notices of the Royal Astronomical Society},
	author = {Sun, Guochao and Mirocha, Jordan and Mebane, Richard H. and Furlanetto, Steven R.},
	month = dec,
	year = {2021},
	note = {ADS Bibcode: 2021MNRAS.508.1954S},
	pages = {1954--1972},
}

@article{lidz_intensity_2011,
	title = {Intensity {Mapping} with {Carbon} {Monoxide} {Emission} {Lines} and the {Redshifted} 21 cm {Line}},
	volume = {741},
	issn = {0004-637X},
	url = {https://ui.adsabs.harvard.edu/abs/2011ApJ...741...70L},
	doi = {10.1088/0004-637X/741/2/70},
	urldate = {2024-04-09},
	journal = {The Astrophysical Journal},
	author = {Lidz, Adam and Furlanetto, Steven R. and Oh, S. Peng and Aguirre, James and Chang, Tzu-Ching and Doré, Olivier and Pritchard, Jonathan R.},
	month = nov,
	year = {2011},
	note = {ADS Bibcode: 2011ApJ...741...70L},
	pages = {70},
}

@article{sheth_ellipsoidal_2001,
	title = {Ellipsoidal collapse and an improved model for the number and spatial distribution of dark matter haloes},
	volume = {323},
	issn = {0035-8711},
	url = {https://doi.org/10.1046/j.1365-8711.2001.04006.x},
	doi = {10.1046/j.1365-8711.2001.04006.x},
	number = {1},
	urldate = {2024-04-08},
	journal = {Monthly Notices of the Royal Astronomical Society},
	author = {Sheth, Ravi K. and Mo, H. J. and Tormen, Giuseppe},
	month = may,
	year = {2001},
	pages = {1--12},
}

@article{mirocha_effects_2020,
	title = {Effects of self-consistent rest-ultraviolet colours in semi-empirical galaxy formation models},
	volume = {498},
	issn = {0035-8711},
	url = {https://ui.adsabs.harvard.edu/abs/2020MNRAS.498.2645M},
	doi = {10.1093/mnras/staa2586},
	urldate = {2024-03-20},
	journal = {Monthly Notices of the Royal Astronomical Society},
	author = {Mirocha, Jordan and Mason, Charlotte and Stark, Daniel P.},
	month = oct,
	year = {2020},
	note = {ADS Bibcode: 2020MNRAS.498.2645M},
	pages = {2645--2661},
}

@ARTICLE{cruz_rise_2026,
       author = {{Cruz}, Hector Afonso G. and {Montefalcone}, Gabriele and {Mu{\~n}oz}, Julian B. and {Kovetz}, Ely D. and {Venditti}, Alessandra},
        title = "{The Rise and Fall of Acoustic Oscillations at Cosmic Dawn}",
      journal = {arXiv e-prints},
         year = 2026,
        month = jul,
          eid = {arXiv:2607.09846},
        pages = {arXiv:2607.09846},
          doi = {10.48550/arXiv.2607.09846},
archivePrefix = {arXiv},
       eprint = {2607.09846},
 primaryClass = {astro-ph.CO},
       adsurl = {https://ui.adsabs.harvard.edu/abs/2026arXiv260709846C}
}

@ARTICLE{bromm_first_2013,
       author = {{Bromm}, V.},
        title = "{The first stars and galaxies - Basic principles}",
      journal = {Asociacion Argentina de Astronomia La Plata Argentina Book Series},
         year = 2013,
        month = jan,
       volume = {4},
        pages = {3},
          doi = {10.48550/arXiv.1203.3824},
archivePrefix = {arXiv},
       eprint = {1203.3824},
 primaryClass = {astro-ph.CO},
       adsurl = {https://ui.adsabs.harvard.edu/abs/2013AAABS...4....3B}
}

@ARTICLE{bracks_forecasting_2026,
       author = {{Bracks}, Justin S. and {Keenan}, Ryan P. and {Agrawal}, Shubh and {Keating}, Garrett K. and {Aguirre}, James E. and {Lidz}, Adam and {Bradford}, Charles M. and {Brendal}, Brockton and {Filippini}, Jeffrey P. and {Fu}, Jianyang and {Garcia}, Karolina and {Groppi}, Christopher and {Hailey-Dunsheath}, Steve and {Janssen}, Reinier M.~J. and {Kang}, Wooseok and {Liu}, Lun-jun and {Lowe}, Ian and {Manduca}, Alex and {Marrone}, Daniel P. and {Mauskopf}, Philip and {Mayer}, Evan C. and {O'Donnell}, Sydnee and {Saeid}, Talia and {Tartakovsky}, Simon and {Van Cuyck}, Mathilde and {Vieira}, Joaquin and {Zebrowski}, Jessica A.},
        title = "{Forecasting the Cross Correlation of Terahertz Intensity Mapper [C II] Line Intensity Maps with Euclid Galaxies}",
      journal = {\apj},
         year = 2026,
        month = jul,
       volume = {1005},
       number = {1},
          eid = {17},
        pages = {17},
          doi = {10.3847/1538-4357/ae75b5},
       adsurl = {https://ui.adsabs.harvard.edu/abs/2026ApJ..1005...17B}
}

@ARTICLE{donnan_jwst_2024,
       author = {{Donnan}, C.~T. and {McLure}, R.~J. and {Dunlop}, J.~S. and {McLeod}, D.~J. and {Magee}, D. and {Arellano-C{\'o}rdova}, K.~Z. and {Barrufet}, L. and {Begley}, R. and {Bowler}, R.~A.~A. and {Carnall}, A.~C. and {Cullen}, F. and {Ellis}, R.~S. and {Fontana}, A. and {Illingworth}, G.~D. and {Grogin}, N.~A. and {Hamadouche}, M.~L. and {Koekemoer}, A.~M. and {Liu}, F. -Y. and {Mason}, C. and {Santini}, P. and {Stanton}, T.~M.},
        title = "{JWST PRIMER: a new multifield determination of the evolving galaxy UV luminosity function at redshifts z $\simeq$ 9 - 15}",
      journal = {\mnras},
         year = 2024,
        month = sep,
       volume = {533},
       number = {3},
        pages = {3222-3237},
          doi = {10.1093/mnras/stae2037},
archivePrefix = {arXiv},
       eprint = {2403.03171},
 primaryClass = {astro-ph.GA},
       adsurl = {https://ui.adsabs.harvard.edu/abs/2024MNRAS.533.3222D}
}

@article{venditti_first_2024,
	title = {The first fireworks: {A} roadmap to {Population} {III} stars during the epoch of reionization through pair-instability supernovae},
	volume = {527},
	issn = {0035-8711},
	shorttitle = {The first fireworks},
	url = {https://ui.adsabs.harvard.edu/abs/2024MNRAS.527.5102V},
	doi = {10.1093/mnras/stad3513},
	urldate = {2024-02-22},
	journal = {Monthly Notices of the Royal Astronomical Society},
	author = {Venditti, Alessandra and Bromm, Volker and Finkelstein, Steven L. and Graziani, Luca and Schneider, Raffaella},
	month = jan,
	year = {2024},
	note = {ADS Bibcode: 2024MNRAS.527.5102V},
	pages = {5102--5116},
}

@misc{kovetz_line-intensity_2017,
	title = {Line-{Intensity} {Mapping}: 2017 {Status} {Report}},
	shorttitle = {Line-{Intensity} {Mapping}},
	url = {https://ui.adsabs.harvard.edu/abs/2017arXiv170909066K},
	doi = {10.48550/arXiv.1709.09066},
	urldate = {2024-02-01},
	author = {Kovetz, Ely D. and Viero, Marco P. and Lidz, Adam and Newburgh, Laura and Rahman, Mubdi and Switzer, Eric and Kamionkowski, Marc and Aguirre, James and Alvarez, Marcelo and Bock, James and Bond, J. Richard and Bower, Goeffry and Bradford, C. Matt and Breysse, Patrick C. and Bull, Philip and Chang, Tzu-Ching and Cheng, Yun-Ting and Chung, Dongwoo and Cleary, Kieran and Corray, Asantha and Crites, Abigail and Croft, Rupert and Doré, Olivier and Eastwood, Michael and Ferrara, Andrea and Fonseca, José and Jacobs, Daniel and Keating, Garrett K. and Lagache, Guilaine and Lakhlani, Gunjan and Liu, Adrian and Moodley, Kavilan and Murray, Norm and Pénin, Aurélie and Popping, Gergö and Pullen, Anthony and Reichers, Dominik and Saito, Shun and Saliwanchik, Ben and Santos, Mario and Somerville, Rachel and Stacey, Gordon and Stein, George and Villaescusa-Navarro, Francesco and Visbal, Eli and Weltman, Amanda and Wolz, Laura and Zemcov, Micheal},
	month = sep,
	year = {2017},
	note = {Publication Title: arXiv e-prints
ADS Bibcode: 2017arXiv170909066K},
}

@ARTICLE{liu_effects_2024,
       author = {{Liu}, Lun-Jun and {Sun}, Guochao and {Chang}, Tzu-Ching and {Furlanetto}, Steven R. and {Bradford}, Charles M.},
        title = "{Effects of Bursty Star Formation on [C II] Line Intensity Mapping of High-redshift Galaxies}",
      journal = {\apj},
         year = 2024,
        month = oct,
       volume = {974},
       number = {2},
          eid = {175},
        pages = {175},
          doi = {10.3847/1538-4357/ad73d5},
archivePrefix = {arXiv},
       eprint = {2401.04204},
 primaryClass = {astro-ph.GA},
       adsurl = {https://ui.adsabs.harvard.edu/abs/2024ApJ...974..175L}
}

@ARTICLE{bernal_line_2022,
       author = {{Bernal}, Jos{\'e} Luis and {Kovetz}, Ely D.},
        title = "{Line-intensity mapping: theory review with a focus on star-formation lines}",
      journal = {\aapr},
         year = 2022,
        month = dec,
       volume = {30},
       number = {1},
          eid = {5},
        pages = {5},
          doi = {10.1007/s00159-022-00143-0},
archivePrefix = {arXiv},
       eprint = {2206.15377},
 primaryClass = {astro-ph.CO},
       adsurl = {https://ui.adsabs.harvard.edu/abs/2022A&ARv..30....5B}
}

@ARTICLE{kannan_THESAN_2022,
       author = {{Kannan}, Rahul and {Smith}, Aaron and {Garaldi}, Enrico and {Shen}, Xuejian and {Vogelsberger}, Mark and {Pakmor}, R{\"u}diger and {Springel}, Volker and {Hernquist}, Lars},
        title = "{The THESAN project: predictions for multitracer line intensity mapping in the epoch of reionization}",
      journal = {\mnras},
         year = 2022,
        month = aug,
       volume = {514},
       number = {3},
        pages = {3857-3878},
          doi = {10.1093/mnras/stac1557},
archivePrefix = {arXiv},
       eprint = {2111.02411},
 primaryClass = {astro-ph.CO},
       adsurl = {https://ui.adsabs.harvard.edu/abs/2022MNRAS.514.3857K}
}

@ARTICLE{wang_strong_2024,
       author = {{Wang}, Xin and {Cheng}, Cheng and {Ge}, Junqiang and {Meng}, Xiao-Lei and {Daddi}, Emanuele and {Yan}, Haojing and {Ji}, Zhiyuan and {Jin}, Yifei and {Jones}, Tucker and {Malkan}, Matthew A. and {Arrabal Haro}, Pablo and {Brammer}, Gabriel and {Oguri}, Masamune and {Hou}, Meicun and {Zhang}, Shiwu},
        title = "{A Strong He II {\ensuremath{\lambda}}1640 Emitter with an Extremely Blue UV Spectral Slope at z = 8.16: Presence of Population III Stars?}",
      journal = {\apjl},
         year = 2024,
        month = jun,
       volume = {967},
       number = {2},
          eid = {L42},
        pages = {L42},
          doi = {10.3847/2041-8213/ad4ced},
archivePrefix = {arXiv},
       eprint = {2212.04476},
 primaryClass = {astro-ph.GA},
       adsurl = {https://ui.adsabs.harvard.edu/abs/2024ApJ...967L..42W}
}

@ARTICLE{reumert_hot_2026,
       author = {{Reumert}, Henriette and {Heintz}, Kasper E. and {Pollock}, Clara L. and {Cameron}, Alex J. and {Brammer}, Gabriel B. and {Katz}, Harley and {Sneppen}, Albert and {Witstok}, Joris and {Terp}, Chamilla and {Watson}, Darach},
        title = "{A hot, nebular-dominated galaxy interacting with a pristine PopIII system uncovered by JWST}",
      journal = {arXiv e-prints},
         year = 2026,
        month = mar,
          eid = {arXiv:2603.13471},
        pages = {arXiv:2603.13471},
          doi = {10.48550/arXiv.2603.13471},
archivePrefix = {arXiv},
       eprint = {2603.13471},
 primaryClass = {astro-ph.GA},
       adsurl = {https://ui.adsabs.harvard.edu/abs/2026arXiv260313471R}
}

@ARTICLE{chang_line_2026,
       author = {{Chang}, Tzu-Ching and {Lidz}, Adam},
        title = "{Line-Intensity Mapping}",
      journal = {arXiv e-prints},
         year = 2026,
        month = feb,
          eid = {arXiv:2602.03011},
        pages = {arXiv:2602.03011},
          doi = {10.48550/arXiv.2602.03011},
archivePrefix = {arXiv},
       eprint = {2602.03011},
 primaryClass = {astro-ph.CO},
       adsurl = {https://ui.adsabs.harvard.edu/abs/2026arXiv260203011C}
}

@ARTICLE{katz_challenges_2023,
       author = {{Katz}, Harley and {Kimm}, Taysun and {Ellis}, Richard S. and {Devriendt}, Julien and {Slyz}, Adrianne},
        title = "{The challenges of identifying Population III stars in the early Universe}",
      journal = {\mnras},
         year = 2023,
        month = sep,
       volume = {524},
       number = {1},
        pages = {351-360},
          doi = {10.1093/mnras/stad1903},
archivePrefix = {arXiv},
       eprint = {2207.04751},
 primaryClass = {astro-ph.GA},
       adsurl = {https://ui.adsabs.harvard.edu/abs/2023MNRAS.524..351K}
}

@ARTICLE{trussler_on_2023,
       author = {{Trussler}, James A.~A. and {Conselice}, Christopher J. and {Adams}, Nathan J. and {Maiolino}, Roberto and {Nakajima}, Kimihiko and {Zackrisson}, Erik and {Austin}, Duncan and {Ferreira}, Leonardo and {Harvey}, Tom},
        title = "{On the observability and identification of Population III galaxies with JWST}",
      journal = {\mnras},
         year = 2023,
        month = nov,
       volume = {525},
       number = {4},
        pages = {5328-5352},
          doi = {10.1093/mnras/stad2553},
archivePrefix = {arXiv},
       eprint = {2211.02038},
 primaryClass = {astro-ph.GA},
       adsurl = {https://ui.adsabs.harvard.edu/abs/2023MNRAS.525.5328T}
}

@INPROCEEDINGS{cooray_CDIM_2019,
       author = {{Cooray}, Asantha and {Chang}, Tzu-Ching and {Unwin}, Stephen and {Zemcov}, Michael and {Coffey}, Andrew and {Morrissey}, Patrick and {Raouf}, Nasrat and {Lipscy}, Sarah and {Shannon}, Mark and {Wu}, Gordon and {Cen}, Renyue and {Chary}, Ranga Ram and {Dor{\'e}}, Olivier and {Fan}, Xiaohui and {Fazio}, Giovanni G. and {Finkelstein}, Steven L. and {Heneka}, Caroline and {Lee}, Bomee and {Linden}, Philip and {Nayyeri}, Hooshang and {Rhodes}, Jason and {Sadoun}, Raphael and {Silva}, Marta B. and {Trac}, Hy and {Wu}, Hao-Yi and {Zheng}, Zheng},
        title = "{Cosmic Dawn Intensity Mapper}",
    booktitle = {Bulletin of the American Astronomical Society},
         year = 2019,
       volume = {51},
        month = sep,
          eid = {23},
        pages = {23},
          doi = {10.48550/arXiv.1903.03144},
archivePrefix = {arXiv},
       eprint = {1903.03144},
 primaryClass = {astro-ph.GA},
       adsurl = {https://ui.adsabs.harvard.edu/abs/2019BAAS...51g..23C}
}

@ARTICLE{chiti_enrichment_2026,
       author = {{Chiti}, Anirudh and {Placco}, Vinicius M. and {Pace}, Andrew B. and {Ji}, Alexander P. and {Prabhu}, Deepthi S. and {Cerny}, William and {Limberg}, Guilherme and {Stringfellow}, Guy S. and {Drlica-Wagner}, Alex and {Atzberger}, Kaia R. and {Choi}, Yumi and {Crnojevi{\'c}}, Denija and {Ferguson}, Peter S. and {Kallivayalil}, Nitya and {No{\"e}l}, Noelia E.~D. and {Riley}, Alexander H. and {Sand}, David J. and {Simon}, Joshua D. and {Walker}, Alistair R. and {Bom}, Clecio R. and {Carballo-Bello}, Julio A. and {James}, David J. and {Mart{\'\i}nez-V{\'a}zquez}, Clara E. and {Medina}, Gustavo E. and {Vivas}, A. Katherina},
        title = "{Enrichment by the first stars in a relic dwarf galaxy}",
      journal = {Nature Astronomy},
         year = 2026,
        month = mar,
          doi = {10.1038/s41550-026-02802-z},
archivePrefix = {arXiv},
       eprint = {2508.04053},
 primaryClass = {astro-ph.GA},
       adsurl = {https://ui.adsabs.harvard.edu/abs/2026NatAs.tmp...57C}
}

@ARTICLE{ji_nearly_2026,
       author = {{Ji}, Alexander P. and {Chandra}, Vedant and {Mejias-Torres}, Selenna and {Zhang}, Zhongyuan and {Eitner}, Philipp and {Schlaufman}, Kevin C. and {Andales}, Hillary Diane and {Do}, Ha and {Orrantia}, Natalie M. and {Tudmilla}, Rithika and {Thibodeaux}, Pierre N. and {Stassun}, Keivan G. and {Howell}, Madeline and {Tayar}, Jamie and {Bergemann}, Maria and {Casey}, Andrew R. and {Johnson}, Jennifer A. and {Carlberg}, Joleen K. and {Cerny}, William and {Fern{\'a}ndez-Trincado}, Jos{\'e} G. and {Hawkins}, Keith and {Kollmeier}, Juna A. and {Laporte}, Chervin F.~P. and {Limberg}, Guilherme and {Matsuno}, Tadafumi and {M{\'e}sz{\'a}ros}, Szabolcs and {Morrison}, Sean and {Nidever}, David L. and {Stringfellow}, Guy S. and {Schneider}, Donald P. and {Thai}, Riley},
        title = "{A nearly pristine star from the Large Magellanic Cloud}",
      journal = {Nature Astronomy},
         year = 2026,
        month = apr,
          doi = {10.1038/s41550-026-02816-7},
archivePrefix = {arXiv},
       eprint = {2509.21643},
 primaryClass = {astro-ph.SR},
       adsurl = {https://ui.adsabs.harvard.edu/abs/2026NatAs.tmp...67J}
}

@INPROCEEDINGS{frebel_metal_2026,
       author = {{Frebel}, Anna},
        title = "{Metal-poor stars in the Milky Way system}",
    booktitle = {Encyclopedia of Astrophysics, Volume 2},
         year = 2026,
       volume = {2},
        month = jan,
        pages = {533-557},
          doi = {10.1016/B978-0-443-21439-4.00052-3},
archivePrefix = {arXiv},
       eprint = {2411.15415},
 primaryClass = {astro-ph.GA},
       adsurl = {https://ui.adsabs.harvard.edu/abs/2026enap....2..533F}
}

@ARTICLE{bonifacio_most_2025,
       author = {{Bonifacio}, Piercarlo and {Caffau}, Elisabetta and {Fran{\c{c}}ois}, Patrick and {Spite}, Monique},
        title = "{The most metal-poor stars}",
      journal = {\aapr},
         year = 2025,
        month = jul,
       volume = {33},
       number = {1},
          eid = {2},
        pages = {2},
          doi = {10.1007/s00159-025-00159-2},
archivePrefix = {arXiv},
       eprint = {2504.06335},
 primaryClass = {astro-ph.GA},
       adsurl = {https://ui.adsabs.harvard.edu/abs/2025A&ARv..33....2B}
}

@BOOK{peebles_principles_1993,
       author = {{Peebles}, P.~J.~E.},
        title = "{Principles of Physical Cosmology}",
         year = 1993,
         publisher = {Princeton University Press},
          doi = {10.1515/9780691206721},
       adsurl = {https://ui.adsabs.harvard.edu/abs/1993ppc..book.....P}
}

@BOOK{dodelson_modern_2003,
       author = {{Dodelson}, Scott},
       publisher = {Academic Press},
        title = "{Modern Cosmology}",
         year = 2003,
       adsurl = {https://ui.adsabs.harvard.edu/abs/2003moco.book.....D}
}

@ARTICLE{kaiser_clustering_1987,
       author = {{Kaiser}, Nick},
        title = "{Clustering in real space and in redshift space}",
      journal = {\mnras},
         year = 1987,
        month = jul,
       volume = {227},
        pages = {1-21},
          doi = {10.1093/mnras/227.1.1},
       adsurl = {https://ui.adsabs.harvard.edu/abs/1987MNRAS.227....1K}
}

@ARTICLE{murray_powerbox_2018,
       author = {{Murray}, Steven G.},
        title = "{powerbox: A Python package for creating structured fields with isotropic power spectra}",
      journal = {The Journal of Open Source Software},
         year = 2018,
        month = aug,
       volume = {3},
       number = {28},
          eid = {850},
        pages = {850},
          doi = {10.21105/joss.00850},
archivePrefix = {arXiv},
       eprint = {1809.05030},
 primaryClass = {astro-ph.IM},
       adsurl = {https://ui.adsabs.harvard.edu/abs/2018JOSS....3..850M}
}

@ARTICLE{bock_spherex_2026,
       author = {{Bock}, James J. and {Aboobaker}, Asad M. and {Adamo}, Joseph and {Akeson}, Rachel and {Alred}, John M. and {Alibay}, Farah and {Ashby}, Matthew L.~N. and {Bach}, Yoonsoo P. and {Bleem}, Lindsey E. and {Bolton}, Douglas and {Braun}, David F. and {Bruton}, Sean and {Bryan}, Sean A. and {Chang}, Tzu-Ching and {Chen}, Shuang-Shuang and {Cheng}, Yun-Ting and {Cheshire}, IV, James R. and {Chiang}, Yi-Kuan and {de Janvry}, Jean Choppin and {Condon}, Samuel and {Cook}, Walter R. and {Cooray}, Asantha and {Crill}, Brendan P. and {Cukierman}, Ari J. and {Dor{\'e}}, Olivier and {Dowell}, C. Darren and {Dubois-Felsmann}, Gregory P. and {Eifler}, Tim and {Everett}, Spencer and {Fabinsky}, Beth E. and {Faisst}, Andreas L. and {Fanson}, James L. and {Farrington}, Allen H. and {Fatahi}, Tamim and {Fazar}, Candice M. and {Feder}, Richard M. and {Frater}, Eric H. and {Grasshorn Gebhardt}, Henry S. and {Giri}, Utkarsh and {Goldina}, Tatiana and {Gorjian}, Varoujan and {Habib}, Salman and {Hart}, William G. and {Heinrich}, Chen and {Hora}, Joseph L. and {Huai}, Zhaoyu and {Hui}, Howard and {Jo}, Young-Soo and {Jeong}, Woong-Seob and {Kang}, Jae Hwan and {Kang}, Miju and {Kecman}, Branislav and {Kim}, Chul-Hwan and {Kim}, Jaeyeong and {Kim}, Minjin and {Kim}, Young-Jun and {Kim}, Yongjung and {Kirkpatrick}, J. Davy and {Kobayashi}, Yosuke and {Korngut}, Phil M. and {Krause}, Elisabeth and {Lee}, Bomee and {Lee}, Ho-Gyu and {Lee}, Jae-Joon and {Lee}, Jeong-Eun and {Lisse}, Carey M. and {Mariani}, Giacomo and {Masters}, Daniel C. and {Mauskopf}, Philip D. and {Melnick}, Gary J. and {Minasyan}, Mary H. and {Mirocha}, Jordan and {Miyasaka}, Hiromasa and {Moore}, Anne and {Moore}, Bradley D. and {Murgia}, Giulia and {Naylor}, Bret J. and {Nelson}, Christina and {Nguyen}, Chi H. and {Nguyen}, Hien T. and {Noh}, Jinyoung K. and {Padin}, Stephen and {Paladini}, Roberta and {Park}, Sung-Joon and {Penanen}, Konstantin I. and {Putnam}, Dustin S. and {Pyo}, Jeonghyun and {Ramachandra}, Nesar and {Ramanathan}, Keshav and {Rustamkulov}, Zafar and {Reiley}, Daniel J. and {Rice}, Eric B. and {Rocca}, Jennifer M. and {Seok}, Ji Yeon and {Smith}, Roger and {Stober}, Jeremy and {Susca}, Sara and {Teplitz}, Harry I. and {Thelen}, Michael P. and {Tolls}, Volker and {Torrini}, Gabriela and {Trangsrud}, Amy R. and {Unwin}, Stephen and {Velicheti}, Phani and {Wang}, Pao-Yu and {Wen}, Robin Y. and {Werner}, Michael W. and {Williams}, Abby E. and {Williamson}, Ross and {Wincentsen}, James and {Windhorst}, Rogier A. and {Yang}, Soung-Chul and {Yang}, Yujin and {Zemcov}, Michael},
        title = "{The SPHEREx Satellite Mission}",
      journal = {\apj},
         year = 2026,
        month = mar,
       volume = {999},
       number = {1},
          eid = {139},
        pages = {139},
          doi = {10.3847/1538-4357/ae2be2},
archivePrefix = {arXiv},
       eprint = {2511.02985},
 primaryClass = {astro-ph.IM},
       adsurl = {https://ui.adsabs.harvard.edu/abs/2026ApJ...999..139B}
}

@ARTICLE{heneka_optimal_2021,
       author = {{Heneka}, Caroline and {Cooray}, Asantha},
        title = "{Optimal survey parameters: Ly {\ensuremath{\alpha}} and H {\ensuremath{\alpha}} intensity mapping for synergy with the 21-cm signal during reionization}",
      journal = {\mnras},
         year = 2021,
        month = sep,
       volume = {506},
       number = {2},
        pages = {1573-1584},
          doi = {10.1093/mnras/stab1842},
archivePrefix = {arXiv},
       eprint = {2104.12739},
 primaryClass = {astro-ph.CO},
       adsurl = {https://ui.adsabs.harvard.edu/abs/2021MNRAS.506.1573H}
}

@article{inayoshi_lower_2022,
	title = {A {Lower} {Bound} of {Star} {Formation} {Activity} in {Ultra}-high-redshift {Galaxies} {Detected} with {JWST}: {Implications} for {Stellar} {Populations} and {Radiation} {Sources}},
	volume = {938},
	issn = {0004-637X},
	shorttitle = {A {Lower} {Bound} of {Star} {Formation} {Activity} in {Ultra}-high-redshift {Galaxies} {Detected} with {JWST}},
	url = {https://ui.adsabs.harvard.edu/abs/2022ApJ...938L..10I},
	doi = {10.3847/2041-8213/ac9310},
	urldate = {2023-11-17},
	journal = {The Astrophysical Journal},
	author = {Inayoshi, Kohei and Harikane, Yuichi and Inoue, Akio K. and Li, Wenxiu and Ho, Luis C.},
	month = oct,
	year = {2022},
	note = {ADS Bibcode: 2022ApJ...938L..10I},
	pages = {L10},
}

@article{lake_supersonic_2021,
	title = {The {Supersonic} {Project}: {SIGOs}, {A} {Proposed} {Progenitor} to {Globular} {Clusters}, and {Their} {Connections} to {Gravitational}-wave {Anisotropies}},
	volume = {922},
	issn = {0004-637X},
	shorttitle = {The {Supersonic} {Project}},
	url = {https://ui.adsabs.harvard.edu/abs/2021ApJ...922...86L},
	doi = {10.3847/1538-4357/ac20d0},
	urldate = {2023-10-26},
	journal = {The Astrophysical Journal},
	author = {Lake, William and Naoz, Smadar and Chiou, Yeou S. and Burkhart, Blakesley and Marinacci, Federico and Vogelsberger, Mark and Kremer, Kyle},
	month = nov,
	year = {2021},
	note = {ADS Bibcode: 2021ApJ...922...86L},
	pages = {86},
}

@article{schaerer_properties_2002,
	title = {On the properties of massive {Population} {III} stars and metal-free stellar populations},
	volume = {382},
	issn = {0004-6361},
	url = {https://ui.adsabs.harvard.edu/abs/2002A&A...382...28S},
	doi = {10.1051/0004-6361:20011619},
	urldate = {2023-10-22},
	journal = {Astronomy and Astrophysics},
	author = {Schaerer, D.},
	month = jan,
	year = {2002},
	note = {ADS Bibcode: 2002A\&A...382...28S},
	pages = {28--42},
}

@article{schaerer_transition_2003,
	title = {The transition from {Population} {III} to normal galaxies: {Lyalpha} and {He} {II} emission and the ionising properties of high redshift starburst galaxies},
	volume = {397},
	issn = {0004-6361},
	shorttitle = {The transition from {Population} {III} to normal galaxies},
	url = {https://ui.adsabs.harvard.edu/abs/2003A&A...397..527S},
	doi = {10.1051/0004-6361:20021525},
	urldate = {2023-10-17},
	journal = {Astronomy and Astrophysics},
	author = {Schaerer, D.},
	month = jan,
	year = {2003},
	note = {ADS Bibcode: 2003A\&A...397..527S},
	pages = {527--538},
}

@article{hegde_self-consistent_2023,
	title = {A self-consistent semi-analytic model for {Population} {III} star formation in minihaloes},
	volume = {525},
	issn = {0035-8711},
	url = {https://doi.org/10.1093/mnras/stad2308},
	doi = {10.1093/mnras/stad2308},
	number = {1},
	urldate = {2023-10-11},
	journal = {Monthly Notices of the Royal Astronomical Society},
	author = {Hegde, Sahil and Furlanetto, Steven R},
	month = oct,
	year = {2023},
	pages = {428--447},
}

@article{klessen_first_2023,
	title = {The {First} {Stars}: {Formation}, {Properties}, and {Impact}},
	volume = {61},
	issn = {0066-4146},
	shorttitle = {The {First} {Stars}},
	url = {https://ui.adsabs.harvard.edu/abs/2023ARA&A..61...65K},
	doi = {10.1146/annurev-astro-071221-053453},
	urldate = {2023-10-07},
	journal = {Annual Review of Astronomy and Astrophysics},
	author = {Klessen, Ralf S. and Glover, Simon C. O.},
	month = aug,
	year = {2023},
	note = {ADS Bibcode: 2023ARA\&A..61...65K},
	pages = {65--130},
}

@article{furlanetto_minimalist_2017,
	title = {A minimalist feedback-regulated model for galaxy formation during the epoch of reionization},
	volume = {472},
	issn = {0035-8711},
	url = {https://doi.org/10.1093/mnras/stx2132},
	doi = {10.1093/mnras/stx2132},
	number = {2},
	urldate = {2023-09-29},
	journal = {Monthly Notices of the Royal Astronomical Society},
	author = {Furlanetto, Steven R. and Mirocha, Jordan and Mebane, Richard H. and Sun, Guochao},
	month = dec,
	year = {2017},
	pages = {1576--1592},
}

@article{sun_limfast_2023,
	title = {{LIMFAST}. {II}. {Line} {Intensity} {Mapping} as a {Probe} of {High}-redshift {Galaxy} {Formation}},
	volume = {950},
	issn = {0004-637X},
	url = {https://ui.adsabs.harvard.edu/abs/2023ApJ...950...40S},
	doi = {10.3847/1538-4357/acc9b3},
	urldate = {2023-09-26},
	journal = {The Astrophysical Journal},
	author = {Sun, Guochao and Mas-Ribas, Lluís and Chang, Tzu-Ching and Furlanetto, Steven R. and Mebane, Richard H. and Gonzalez, Michael O. and Parsons, Jasmine and Trapp, A. C.},
	month = jun,
	year = {2023},
	note = {ADS Bibcode: 2023ApJ...950...40S},
	pages = {40},
}

@article{visbal_measuring_2010,
	title = {Measuring the {3D} clustering of undetected galaxies through cross correlation of their cumulative flux fluctuations from multiple spectral lines},
	volume = {2010},
	issn = {1475-7516},
	url = {https://dx.doi.org/10.1088/1475-7516/2010/11/016},
	doi = {10.1088/1475-7516/2010/11/016},
	
	number = {11},
	urldate = {2023-09-24},
	journal = {Journal of Cosmology and Astroparticle Physics},
	author = {Visbal, Eli and Loeb, Abraham},
	month = nov,
	year = {2010},
	pages = {016},
}

@article{visbal_looking_2015,
	title = {Looking for {Population} {III} stars with {He} {II} line intensity mapping},
	volume = {450},
	issn = {0035-8711},
	url = {https://ui.adsabs.harvard.edu/abs/2015MNRAS.450.2506V},
	doi = {10.1093/mnras/stv785},
	urldate = {2023-09-24},
	journal = {Monthly Notices of the Royal Astronomical Society},
	author = {Visbal, Eli and Haiman, Zoltán and Bryan, Greg L.},
	month = jul,
	year = {2015},
	note = {ADS Bibcode: 2015MNRAS.450.2506V},
	pages = {2506--2513},
}

@article{sun_constraints_2016,
	title = {Constraints on the star formation efficiency of galaxies during the epoch of reionization},
	volume = {460},
	issn = {0035-8711},
	url = {https://ui.adsabs.harvard.edu/abs/2016MNRAS.460..417S},
	doi = {10.1093/mnras/stw980},
	urldate = {2023-08-07},
	journal = {Monthly Notices of the Royal Astronomical Society},
	author = {Sun, G. and Furlanetto, S. R.},
	month = jul,
	year = {2016},
	note = {ADS Bibcode: 2016MNRAS.460..417S},
	pages = {417--433},
}

@article{bromm_formation_2002,
	title = {The {Formation} of the {First} {Stars}. {I}. {The} {Primordial} {Star}-forming {Cloud}},
	volume = {564},
	issn = {0004-637X},
	url = {https://ui.adsabs.harvard.edu/abs/2002ApJ...564...23B},
	doi = {10.1086/323947},
	urldate = {2023-07-07},
	journal = {The Astrophysical Journal},
	author = {Bromm, Volker and Coppi, Paolo S. and Larson, Richard B.},
	month = jan,
	year = {2002},
	note = {ADS Bibcode: 2002ApJ...564...23B},
	pages = {23--51},
}

@article{heger_nucleosynthesis_2010,
	title = {Nucleosynthesis and {Evolution} of {Massive} {Metal}-free {Stars}},
	volume = {724},
	issn = {0004-637X},
	url = {https://ui.adsabs.harvard.edu/abs/2010ApJ...724..341H},
	doi = {10.1088/0004-637X/724/1/341},
	urldate = {2023-07-06},
	journal = {The Astrophysical Journal},
	author = {Heger, Alexander and Woosley, S. E.},
	month = nov,
	year = {2010},
	note = {ADS Bibcode: 2010ApJ...724..341H},
	pages = {341--373},
}

@ARTICLE{munoz_relatively_2026,
       author = {{Mu{\~n}oz}, Julian B. and {Chisholm}, John and {Sun}, Guochao and {Samuel}, Jenna and {Mirocha}, Jordan and {Bregou}, Emily and {Venditti}, Alessandra and {Qezlou}, Mahdi and {Simmonds}, Charlotte and {Endsley}, Ryan},
        title = "{Relatively fast and reasonably furious: evidence for increased burstiness in smaller haloes at cosmic dawn}",
      journal = {\mnras},
         year = 2026,
        month = apr,
       volume = {547},
       number = {4},
          eid = {stag415},
        pages = {stag415},
          doi = {10.1093/mnras/stag415},
archivePrefix = {arXiv},
       eprint = {2601.07912},
 primaryClass = {astro-ph.GA},
       adsurl = {https://ui.adsabs.harvard.edu/abs/2026MNRAS.547ag415M}
}

@article{Munoz:2021psm,
    author = "Mu{\~n}oz, Julian B. and Qin, Yuxiang and Mesinger, Andrei and Murray, Steven G. and Greig, Bradley and Mason, Charlotte",
    title = "{The impact of the first galaxies on cosmic dawn and reionization}",
    eprint = "2110.13919",
    archivePrefix = "arXiv",
    primaryClass = "astro-ph.CO",
    doi = "10.1093/mnras/stac185",
    journal = "Mon. Not. Roy. Astron. Soc.",
    volume = "511",
    number = "3",
    pages = "3657--3681",
    year = "2022"
}

@article{Munoz:2019rhi,
    author = "Mu{\~n}oz, Julian B.",
    title = "{Robust Velocity-induced Acoustic Oscillations at Cosmic Dawn}",
    eprint = "1904.07881",
    archivePrefix = "arXiv",
    primaryClass = "astro-ph.CO",
    doi = "10.1103/PhysRevD.100.063538",
    journal = "Phys. Rev. D",
    volume = "100",
    number = "6",
    pages = "063538",
    year = "2019"
}

@ARTICLE{kovetz_when_2026,
       author = {{Kovetz}, Ely D. and {Lazare}, Hovav and {Libanore}, Sarah and {Mu{\~n}oz}, Julian B. and {Vanzan}, Eleonora},
        title = "{When galaxies burst: enhanced shot-noise for line-intensity mapping in the JWST era}",
      journal = {arXiv e-prints},
         year = 2026,
        month = may,
          eid = {arXiv:2605.13967},
        pages = {arXiv:2605.13967},
          doi = {10.48550/arXiv.2605.13967},
archivePrefix = {arXiv},
       eprint = {2605.13967},
 primaryClass = {astro-ph.GA},
       adsurl = {https://ui.adsabs.harvard.edu/abs/2026arXiv260513967K}
}

@ARTICLE{sharda_importance_2020,
       author = {{Sharda}, Piyush and {Federrath}, Christoph and {Krumholz}, Mark R.},
        title = "{The importance of magnetic fields for the initial mass function of the first stars}",
      journal = {\mnras},
         year = 2020,
        month = sep,
       volume = {497},
       number = {1},
        pages = {336-351},
          doi = {10.1093/mnras/staa1926},
archivePrefix = {arXiv},
       eprint = {2002.11502},
 primaryClass = {astro-ph.GA},
       adsurl = {https://ui.adsabs.harvard.edu/abs/2020MNRAS.497..336S}
}

@ARTICLE{sharda_role_2019,
       author = {{Sharda}, Piyush and {Krumholz}, Mark R. and {Federrath}, Christoph},
        title = "{The role of the H$_{2}$ adiabatic index in the formation of the first stars}",
      journal = {\mnras},
         year = 2019,
        month = nov,
       volume = {490},
       number = {1},
        pages = {513-526},
          doi = {10.1093/mnras/stz2618},
archivePrefix = {arXiv},
       eprint = {1909.06269},
 primaryClass = {astro-ph.GA},
       adsurl = {https://ui.adsabs.harvard.edu/abs/2019MNRAS.490..513S}
}

@ARTICLE{he_simulating_2019,
       author = {{He}, Chong-Chong and {Ricotti}, Massimo and {Geen}, Sam},
        title = "{Simulating star clusters across cosmic time - I. Initial mass function, star formation rates, and efficiencies}",
      journal = {\mnras},
         year = 2019,
        month = oct,
       volume = {489},
       number = {2},
        pages = {1880-1898},
          doi = {10.1093/mnras/stz2239},
archivePrefix = {arXiv},
       eprint = {1904.07889},
 primaryClass = {astro-ph.GA},
       adsurl = {https://ui.adsabs.harvard.edu/abs/2019MNRAS.489.1880H}
}

@ARTICLE{sharda_magnetic_2025,
       author = {{Sharda}, Piyush and {Menon}, Shyam H. and {Gerasimov}, Roman and {Bromm}, Volker and {Burkhart}, Blakesley and {Haemmerl{\'e}}, Lionel and {van Veenen}, Lisanne and {Wibking}, Benjamin D.},
        title = "{Magnetic fields limit the mass of Population III stars even before the onset of protostellar radiation feedback}",
      journal = {\mnras},
         year = 2025,
        month = jul,
       volume = {541},
       number = {1},
        pages = {L1-L7},
          doi = {10.1093/mnrasl/slaf043},
archivePrefix = {arXiv},
       eprint = {2501.12734},
 primaryClass = {astro-ph.GA},
       adsurl = {https://ui.adsabs.harvard.edu/abs/2025MNRAS.541L...1S}
}

@ARTICLE{yang_empirical_2022,
       author = {{Yang}, Shengqi and {Popping}, Gerg{\"o} and {Somerville}, Rachel S. and {Pullen}, Anthony R. and {Breysse}, Patrick C. and {Maniyar}, Abhishek S.},
        title = "{An Empirical Representation of a Physical Model for the ISM [C II], CO, and [C I] Emission at Redshift 1 {\ensuremath{\leq}} z {\ensuremath{\leq}} 9}",
      journal = {\apj},
         year = 2022,
        month = apr,
       volume = {929},
       number = {2},
          eid = {140},
        pages = {140},
          doi = {10.3847/1538-4357/ac5d57},
archivePrefix = {arXiv},
       eprint = {2108.07716},
 primaryClass = {astro-ph.GA},
       adsurl = {https://ui.adsabs.harvard.edu/abs/2022ApJ...929..140Y}
}

@ARTICLE{tacchella_stochastic_2020,
       author = {{Tacchella}, Sandro and {Forbes}, John C. and {Caplar}, Neven},
        title = "{Stochastic modelling of star-formation histories II: star-formation variability from molecular clouds and gas inflow}",
      journal = {\mnras},
         year = 2020,
        month = sep,
       volume = {497},
       number = {1},
        pages = {698-725},
          doi = {10.1093/mnras/staa1838},
archivePrefix = {arXiv},
       eprint = {2006.09382},
 primaryClass = {astro-ph.GA},
       adsurl = {https://ui.adsabs.harvard.edu/abs/2020MNRAS.497..698T}
}

@ARTICLE{caplar_stochastic_2019,
       author = {{Caplar}, Neven and {Tacchella}, Sandro},
        title = "{Stochastic modelling of star-formation histories I: the scatter of the star-forming main sequence}",
      journal = {\mnras},
         year = 2019,
        month = aug,
       volume = {487},
       number = {3},
        pages = {3845-3869},
          doi = {10.1093/mnras/stz1449},
archivePrefix = {arXiv},
       eprint = {1901.07556},
 primaryClass = {astro-ph.GA},
       adsurl = {https://ui.adsabs.harvard.edu/abs/2019MNRAS.487.3845C}
}

@ARTICLE{iyer_diversity_2020,
       author = {{Iyer}, Kartheik G. and {Tacchella}, Sandro and {Genel}, Shy and {Hayward}, Christopher C. and {Hernquist}, Lars and {Brooks}, Alyson M. and {Caplar}, Neven and {Dav{\'e}}, Romeel and {Diemer}, Benedikt and {Forbes}, John C. and {Gawiser}, Eric and {Somerville}, Rachel S. and {Starkenburg}, Tjitske K.},
        title = "{The diversity and variability of star formation histories in models of galaxy evolution}",
      journal = {\mnras},
         year = 2020,
        month = oct,
       volume = {498},
       number = {1},
        pages = {430-463},
          doi = {10.1093/mnras/staa2150},
archivePrefix = {arXiv},
       eprint = {2007.07916},
 primaryClass = {astro-ph.GA},
       adsurl = {https://ui.adsabs.harvard.edu/abs/2020MNRAS.498..430I}
}

@ARTICLE{ambrose_cross_2026,
       author = {{Ambrose}, Abigail E. and {Visbal}, Eli and {McQuinn}, Matthew},
        title = "{On Cross-Correlating Line Intensity Maps from SPHEREx during Reionization}",
      journal = {arXiv e-prints},
         year = 2026,
        month = jun,
          eid = {arXiv:2606.10005},
        pages = {arXiv:2606.10005},
          doi = {10.48550/arXiv.2606.10005},
archivePrefix = {arXiv},
       eprint = {2606.10005},
 primaryClass = {astro-ph.GA},
       adsurl = {https://ui.adsabs.harvard.edu/abs/2026arXiv260610005A}
}

@ARTICLE{sun_foreground_2018,
       author = {{Sun}, G. and {Moncelsi}, L. and {Viero}, M.~P. and {Silva}, M.~B. and {Bock}, J. and {Bradford}, C.~M. and {Chang}, T.-C. and {Cheng}, Y.-T. and {Cooray}, A.~R. and {Crites}, A. and {Hailey-Dunsheath}, S. and {Uzgil}, B. and {Hunacek}, J.~R. and {Zemcov}, M.},
        title = "{A Foreground Masking Strategy for [C II] Intensity Mapping Experiments Using Galaxies Selected by Stellar Mass and Redshift}",
      journal = {\apj},
         year = 2018,
        month = apr,
       volume = {856},
       number = {2},
          eid = {107},
        pages = {107},
          doi = {10.3847/1538-4357/aab3e3},
archivePrefix = {arXiv},
       eprint = {1610.10095},
 primaryClass = {astro-ph.GA},
       adsurl = {https://ui.adsabs.harvard.edu/abs/2018ApJ...856..107S}
}

@ARTICLE{mcquinn_cosmological_2006,
       author = {{McQuinn}, Matthew and {Zahn}, Oliver and {Zaldarriaga}, Matias and {Hernquist}, Lars and {Furlanetto}, Steven R.},
        title = "{Cosmological Parameter Estimation Using 21 cm Radiation from the Epoch of Reionization}",
      journal = {\apj},
         year = 2006,
        month = dec,
       volume = {653},
       number = {2},
        pages = {815-834},
          doi = {10.1086/505167},
archivePrefix = {arXiv},
       eprint = {astro-ph/0512263},
 primaryClass = {astro-ph},
       adsurl = {https://ui.adsabs.harvard.edu/abs/2006ApJ...653..815M}
}

@ARTICLE{chapman_effect_2016,
       author = {{Chapman}, Emma and {Zaroubi}, Saleem and {Abdalla}, Filipe B. and {Dulwich}, Fred and {Jeli{\'c}}, Vibor and {Mort}, Benjamin},
        title = "{The effect of foreground mitigation strategy on EoR window recovery}",
      journal = {\mnras},
         year = 2016,
        month = may,
       volume = {458},
       number = {3},
        pages = {2928-2939},
          doi = {10.1093/mnras/stw161},
       adsurl = {https://ui.adsabs.harvard.edu/abs/2016MNRAS.458.2928C}
}

@ARTICLE{harker_non-parametric_2009,
       author = {{Harker}, Geraint and {Zaroubi}, Saleem and {Bernardi}, Gianni and {Brentjens}, Michiel A. and {de Bruyn}, A.~G. and {Ciardi}, Benedetta and {Jeli{\'c}}, Vibor and {Koopmans}, Leon V.~E. and {Labropoulos}, Panagiotis and {Mellema}, Garrelt and {Offringa}, Andr{\'e} and {Pandey}, V.~N. and {Schaye}, Joop and {Thomas}, Rajat M. and {Yatawatta}, Sarod},
        title = "{Non-parametric foreground subtraction for 21-cm epoch of reionization experiments}",
      journal = {\mnras},
         year = 2009,
        month = aug,
       volume = {397},
       number = {2},
        pages = {1138-1152},
          doi = {10.1111/j.1365-2966.2009.15081.x},
archivePrefix = {arXiv},
       eprint = {0903.2760},
 primaryClass = {astro-ph.CO},
       adsurl = {https://ui.adsabs.harvard.edu/abs/2009MNRAS.397.1138H}
}

@article{liu_method_2011,
  title = {A method for 21 cm power spectrum estimation in the presence of foregrounds},
  author = {Liu, Adrian and Tegmark, Max},
  journal = {Phys. Rev. D},
  volume = {83},
  issue = {10},
  pages = {103006},
  numpages = {23},
  year = {2011},
  month = {May},
  publisher = {American Physical Society},
  doi = {10.1103/PhysRevD.83.103006},
  url = {https://link.aps.org/doi/10.1103/PhysRevD.83.103006}
}

@ARTICLE{kittiwisit_measurements_2022,
       author = {{Kittiwisit}, Piyanat and {Bowman}, Judd D. and {Murray}, Steven G. and {Gehlot}, Bharat K. and {Jacobs}, Daniel C. and {Beardsley}, Adam P.},
        title = "{Measurements of one-point statistics in 21-cm intensity maps via foreground avoidance strategy}",
      journal = {\mnras},
         year = 2022,
        month = dec,
       volume = {517},
       number = {2},
        pages = {2138-2150},
          doi = {10.1093/mnras/stac2826},
archivePrefix = {arXiv},
       eprint = {2204.01124},
 primaryClass = {astro-ph.CO},
       adsurl = {https://ui.adsabs.harvard.edu/abs/2022MNRAS.517.2138K}
}

@ARTICLE{gong_foreground_2014,
       author = {{Gong}, Yan and {Silva}, Marta and {Cooray}, Asantha and {Santos}, Mario G.},
        title = "{Foreground Contamination in Ly{\ensuremath{\alpha}} Intensity Mapping during the Epoch of Reionization}",
      journal = {\apj},
         year = 2014,
        month = apr,
       volume = {785},
       number = {1},
          eid = {72},
        pages = {72},
          doi = {10.1088/0004-637X/785/1/72},
archivePrefix = {arXiv},
       eprint = {1312.2035},
 primaryClass = {astro-ph.CO},
       adsurl = {https://ui.adsabs.harvard.edu/abs/2014ApJ...785...72G}
}

@ARTICLE{gong_intensity_2012,
       author = {{Gong}, Yan and {Cooray}, Asantha and {Silva}, Marta and {Santos}, Mario G. and {Bock}, James and {Bradford}, C. Matt and {Zemcov}, Michael},
        title = "{Intensity Mapping of the [C II] Fine Structure Line during the Epoch of Reionization}",
      journal = {\apj},
         year = 2012,
        month = jan,
       volume = {745},
       number = {1},
          eid = {49},
        pages = {49},
          doi = {10.1088/0004-637X/745/1/49},
archivePrefix = {arXiv},
       eprint = {1107.3553},
 primaryClass = {astro-ph.CO},
       adsurl = {https://ui.adsabs.harvard.edu/abs/2012ApJ...745...49G}
}

@ARTICLE{gong_cosmological_2020,
       author = {{Gong}, Yan and {Chen}, Xuelei and {Cooray}, Asantha},
        title = "{Cosmological Constraints from Line Intensity Mapping with Interlopers}",
      journal = {\apj},
         year = 2020,
        month = may,
       volume = {894},
       number = {2},
          eid = {152},
        pages = {152},
          doi = {10.3847/1538-4357/ab87a0},
archivePrefix = {arXiv},
       eprint = {2001.10792},
 primaryClass = {astro-ph.CO},
       adsurl = {https://ui.adsabs.harvard.edu/abs/2020ApJ...894..152G}
}

@ARTICLE{cheng_spectral_2016,
       author = {{Cheng}, Yun-Ting and {Chang}, Tzu-Ching and {Bock}, James and {Bradford}, C. Matt and {Cooray}, Asantha},
        title = "{Spectral Line De-confusion in an Intensity Mapping Survey}",
      journal = {\apj},
         year = 2016,
        month = dec,
       volume = {832},
       number = {2},
          eid = {165},
        pages = {165},
          doi = {10.3847/0004-637X/832/2/165},
archivePrefix = {arXiv},
       eprint = {1604.07833},
 primaryClass = {astro-ph.CO},
       adsurl = {https://ui.adsabs.harvard.edu/abs/2016ApJ...832..165C}
}

@ARTICLE{lidz_removing_2016,
       author = {{Lidz}, Adam and {Taylor}, Jessie},
        title = "{On Removing Interloper Contamination from Intensity Mapping Power Spectrum Measurements}",
      journal = {\apj},
         year = 2016,
        month = jul,
       volume = {825},
       number = {2},
          eid = {143},
        pages = {143},
          doi = {10.3847/0004-637X/825/2/143},
archivePrefix = {arXiv},
       eprint = {1604.05737},
 primaryClass = {astro-ph.CO},
       adsurl = {https://ui.adsabs.harvard.edu/abs/2016ApJ...825..143L}
}

@ARTICLE{tegmark_karhunen_1997,
       author = {{Tegmark}, Max and {Taylor}, Andy N. and {Heavens}, Alan F.},
        title = "{Karhunen-Lo{\`e}ve Eigenvalue Problems in Cosmology: How Should We Tackle Large Data Sets?}",
      journal = {\apj},
         year = 1997,
        month = may,
       volume = {480},
       number = {1},
        pages = {22-35},
          doi = {10.1086/303939},
archivePrefix = {arXiv},
       eprint = {astro-ph/9603021},
 primaryClass = {astro-ph},
       adsurl = {https://ui.adsabs.harvard.edu/abs/1997ApJ...480...22T}
}

@ARTICLE{tegmark_CMB_1997,
       author = {{Tegmark}, Max},
        title = "{How to measure CMB power spectra without losing information}",
      journal = {\prd},
         year = 1997,
        month = may,
       volume = {55},
       number = {10},
        pages = {5895-5907},
          doi = {10.1103/PhysRevD.55.5895},
archivePrefix = {arXiv},
       eprint = {astro-ph/9611174},
 primaryClass = {astro-ph},
       adsurl = {https://ui.adsabs.harvard.edu/abs/1997PhRvD..55.5895T}
}

@ARTICLE{heavens_statistical_2009,
       author = {{Heavens}, Alan},
        title = "{Statistical techniques in cosmology}",
      journal = {arXiv e-prints},
         year = 2009,
        month = jun,
          eid = {arXiv:0906.0664},
        pages = {arXiv:0906.0664},
          doi = {10.48550/arXiv.0906.0664},
archivePrefix = {arXiv},
       eprint = {0906.0664},
 primaryClass = {astro-ph.CO},
       adsurl = {https://ui.adsabs.harvard.edu/abs/2009arXiv0906.0664H}
}

@ARTICLE{maiolino_jwst-jades_2024,
       author = {{Maiolino}, Roberto and {{\"U}bler}, Hannah and {Perna}, Michele and {Scholtz}, Jan and {D'Eugenio}, Francesco and {Witten}, Callum and {Laporte}, Nicolas and {Witstok}, Joris and {Carniani}, Stefano and {Tacchella}, Sandro and {Baker}, William M. and {Arribas}, Santiago and {Nakajima}, Kimihiko and {Eisenstein}, Daniel J. and {Bunker}, Andrew J. and {Charlot}, St{\'e}phane and {Cresci}, Giovanni and {Curti}, Mirko and {Curtis-Lake}, Emma and {de Graaff}, Anna and {Egami}, Eiichi and {Ji}, Zhiyuan and {Johnson}, Benjamin D. and {Kumari}, Nimisha and {Looser}, Tobias J. and {Maseda}, Michael and {Nelson}, Erica and {Robertson}, Brant and {Rodr{\'\i}guez Del Pino}, Bruno and {Sandles}, Lester and {Simmonds}, Charlotte and {Smit}, Renske and {Sun}, Fengwu and {Venturi}, Giacomo and {Williams}, Christina C. and {Willmer}, Christopher N.~A.},
        title = "{JADES. Possible Population III signatures at z = 10.6 in the halo of GN-z11}",
      journal = {\aap},
         year = 2024,
        month = jul,
       volume = {687},
          eid = {A67},
        pages = {A67},
          doi = {10.1051/0004-6361/202347087},
archivePrefix = {arXiv},
       eprint = {2306.00953},
 primaryClass = {astro-ph.GA},
       adsurl = {https://ui.adsabs.harvard.edu/abs/2024A&A...687A..67M}
}

@article{munoz_ethos_2021,
	title = {{ETHOS} ‐ an effective theory of structure formation: {Impact} of dark acoustic oscillations on cosmic dawn},
	volume = {103},
	issn = {1550-79980556-2821},
	shorttitle = {{ETHOS} ‐ an effective theory of structure formation},
	url = {https://ui.adsabs.harvard.edu/abs/2021PhRvD.103d3512M},
	doi = {10.1103/PhysRevD.103.043512},
	urldate = {2023-06-01},
	journal = {Physical Review D},
	author = {Mu\~{n}oz, Julian B. and Bohr, Sebastian and Cyr-Racine, Francis-Yan and Zavala, Jesús and Vogelsberger, Mark},
	month = feb,
	year = {2021},
	note = {ADS Bibcode: 2021PhRvD.103d3512M},
	pages = {043512},
}

@misc{sun_seen_2023,
	title = {Seen and unseen: bursty star formation and its implications for observations of high-redshift galaxies with {JWST}},
	shorttitle = {Seen and unseen},
	url = {http://arxiv.org/abs/2305.02713},
	doi = {10.48550/arXiv.2305.02713},
	urldate = {2023-05-05},
	publisher = {arXiv},
	author = {Sun, Guochao and Faucher-Giguère, Claude-André and Hayward, Christopher C. and Shen, Xuejian},
	month = may,
	year = {2023},
	note = {arXiv:2305.02713 [astro-ph]},
}

@misc{munoz_effective_2023,
	title = {An {Effective} {Model} for the {Cosmic}-{Dawn} 21-cm {Signal}},
	url = {http://arxiv.org/abs/2302.08506},
	doi = {10.48550/arXiv.2302.08506},
	urldate = {2023-04-14},
	publisher = {arXiv},
	author = {Mu\~{n}oz, Julian B.},
	month = feb,
	year = {2023},
	note = {arXiv:2302.08506 [astro-ph, physics:hep-ph]},
}

@misc{sharda_interplay_2023,
	title = {The interplay between feedback, accretion, transport and winds in setting gas-phase metal distribution in galaxies},
	url = {http://arxiv.org/abs/2303.15853},
	doi = {10.48550/arXiv.2303.15853},
	urldate = {2023-03-29},
	publisher = {arXiv},
	author = {Sharda, Piyush and Ginzburg, Omri and Krumholz, Mark R. and Forbes, John C. and Wisnioski, Emily and Mingozzi, Matilde and Zovaro, Henry R. M. and Dekel, Avishai},
	month = mar,
	year = {2023},
	note = {arXiv:2303.15853 [astro-ph]},
}

@ARTICLE{breysse_high_2016,
       author = {{Breysse}, Patrick C. and {Kovetz}, Ely D. and {Kamionkowski}, Marc},
        title = "{The high-redshift star formation history from carbon-monoxide intensity maps}",
      journal = {\mnras},
         year = 2016,
        month = mar,
       volume = {457},
       number = {1},
        pages = {L127-L131},
          doi = {10.1093/mnrasl/slw005},
archivePrefix = {arXiv},
       eprint = {1507.06304},
 primaryClass = {astro-ph.CO},
       adsurl = {https://ui.adsabs.harvard.edu/abs/2016MNRAS.457L.127B}
}

@article{hummel_source_2012,
	title = {{THE} {SOURCE} {DENSITY} {AND} {OBSERVABILITY} {OF} {PAIR}-{INSTABILITY} {SUPERNOVAE} {FROM} {THE} {FIRST} {STARS}},
	volume = {755},
	issn = {0004-637X},
	url = {https://dx.doi.org/10.1088/0004-637X/755/1/72},
	doi = {10.1088/0004-637X/755/1/72},
	
	number = {1},
	urldate = {2023-03-21},
	journal = {The Astrophysical Journal},
	author = {Hummel, Jacob A. and Pawlik, Andreas H. and Milosavljević, Miloš and Bromm, Volker},
	month = jul,
	year = {2012},
	note = {Publisher: The American Astronomical Society},
	pages = {72},
}

@ARTICLE{nebrin_starbursts_2023,
       author = {{Nebrin}, Olof and {Giri}, Sambit K. and {Mellema}, Garrelt},
        title = "{Starbursts in low-mass haloes at Cosmic Dawn. I. The critical halo mass for star formation}",
      journal = {\mnras},
         year = 2023,
        month = sep,
       volume = {524},
       number = {2},
        pages = {2290-2311},
          doi = {10.1093/mnras/stad1852},
archivePrefix = {arXiv},
       eprint = {2303.08024},
 primaryClass = {astro-ph.CO},
       adsurl = {https://ui.adsabs.harvard.edu/abs/2023MNRAS.524.2290N}
}

@article{mcquinn_impact_2012,
	title = {{THE} {IMPACT} {OF} {THE} {SUPERSONIC} {BARYON}-{DARK} {MATTER} {VELOCITY} {DIFFERENCE} {ON} {THE} \textit{z} $\sim$ 20 21 cm {BACKGROUND}},
	volume = {760},
	issn = {0004-637X, 1538-4357},
	url = {https://iopscience.iop.org/article/10.1088/0004-637X/760/1/3},
	doi = {10.1088/0004-637X/760/1/3},
	
	number = {1},
	urldate = {2023-02-17},
	journal = {The Astrophysical Journal},
	author = {McQuinn, Matthew and O'Leary, Ryan M.},
	month = nov,
	year = {2012},
	pages = {3},
}

@ARTICLE{venditti_needle_2023,
       author = {{Venditti}, Alessandra and {Graziani}, Luca and {Schneider}, Raffaella and {Pentericci}, Laura and {Di Cesare}, Claudia and {Maio}, Umberto and {Omukai}, Kazuyuki},
        title = "{A needle in a haystack? Catching Population III stars in the epoch of reionization: I. Population III star-forming environments}",
      journal = {\mnras},
         year = 2023,
        month = jul,
       volume = {522},
       number = {3},
        pages = {3809-3830},
          doi = {10.1093/mnras/stad1201},
archivePrefix = {arXiv},
       eprint = {2301.10259},
 primaryClass = {astro-ph.GA},
       adsurl = {https://ui.adsabs.harvard.edu/abs/2023MNRAS.522.3809V}
}

@article{mesinger_21cmfast_2011,
	title = {21cmfast: a fast, seminumerical simulation of the high-redshift 21-cm signal: 21cmfast},
	volume = {411},
	issn = {00358711},
	shorttitle = {21cmfast},
	url = {https://academic.oup.com/mnras/article-lookup/doi/10.1111/j.1365-2966.2010.17731.x},
	doi = {10.1111/j.1365-2966.2010.17731.x},
	
	number = {2},
	urldate = {2022-11-04},
	journal = {Monthly Notices of the Royal Astronomical Society},
	author = {Mesinger, Andrei and Furlanetto, Steven and Cen, Renyue},
	month = feb,
	year = {2011},
	pages = {955--972},
}

@article{naoz_simulations_2012,
	title = {{SIMULATIONS} {OF} {EARLY} {BARYONIC} {STRUCTURE} {FORMATION} {WITH} {STREAM} {VELOCITY}. {I}. {HALO} {ABUNDANCE}},
	volume = {747},
	issn = {0004-637X, 1538-4357},
	url = {https://iopscience.iop.org/article/10.1088/0004-637X/747/2/128},
	doi = {10.1088/0004-637X/747/2/128},
	
	number = {2},
	urldate = {2022-11-02},
	journal = {The Astrophysical Journal},
	author = {Naoz, Smadar and Yoshida, Naoki and Gnedin, Nickolay Y.},
	month = mar,
	year = {2012},
	pages = {128},
}

@article{furlanetto_global_2006,
	title = {The global 21-centimeter background from high redshifts},
	volume = {371},
	issn = {0035-8711, 1365-2966},
	url = {https://academic.oup.com/mnras/article-lookup/doi/10.1111/j.1365-2966.2006.10725.x},
	doi = {10.1111/j.1365-2966.2006.10725.x},
	
	number = {2},
	urldate = {2022-09-16},
	journal = {Monthly Notices of the Royal Astronomical Society},
	author = {Furlanetto, S. R.},
	month = sep,
	year = {2006},
	pages = {867--878},
}

@article{furlanetto_bursty_2022,
	title = {Bursty star formation during the {Cosmic} {Dawn} driven by delayed stellar feedback},
	volume = {511},
	issn = {0035-8711, 1365-2966},
	url = {https://academic.oup.com/mnras/article/511/3/3895/6522193},
	doi = {10.1093/mnras/stac310},
	
	number = {3},
	urldate = {2022-09-10},
	journal = {Monthly Notices of the Royal Astronomical Society},
	author = {Furlanetto, Steven R and Mirocha, Jordan},
	month = feb,
	year = {2022},
	pages = {3895--3909},
}

@article{tolstov_multicolor_2016,
	title = {{MULTICOLOR} {LIGHT} {CURVE} {SIMULATIONS} {OF} {POPULATION} {III} {CORE}-{COLLAPSE} {SUPERNOVAE}: {FROM} {SHOCK} {BREAKOUT} {TO} $^{\textrm{56}}$ {CO} {DECAY}},
	volume = {821},
	issn = {1538-4357},
	shorttitle = {{MULTICOLOR} {LIGHT} {CURVE} {SIMULATIONS} {OF} {POPULATION} {III} {CORE}-{COLLAPSE} {SUPERNOVAE}},
	url = {https://iopscience.iop.org/article/10.3847/0004-637X/821/2/124},
	doi = {10.3847/0004-637X/821/2/124},
	
	number = {2},
	urldate = {2022-08-22},
	journal = {The Astrophysical Journal},
	author = {Tolstov, Alexey and Nomoto, Ken’ichi and Tominaga, Nozomu and Ishigaki, Miho N. and Blinnikov, Sergey and Suzuki, Tomoharu},
	month = apr,
	year = {2016},
	pages = {124},
}

@article{kasen_pair_2011,
	title = {{PAIR} {INSTABILITY} {SUPERNOVAE}: {LIGHT} {CURVES}, {SPECTRA}, {AND} {SHOCK} {BREAKOUT}},
	volume = {734},
	issn = {0004-637X, 1538-4357},
	shorttitle = {{PAIR} {INSTABILITY} {SUPERNOVAE}},
	url = {https://iopscience.iop.org/article/10.1088/0004-637X/734/2/102},
	doi = {10.1088/0004-637X/734/2/102},
	
	number = {2},
	urldate = {2022-08-16},
	journal = {The Astrophysical Journal},
	author = {Kasen, Daniel and Woosley, S. E. and Heger, Alexander},
	month = jun,
	year = {2011},
	pages = {102},
}

@article{munoz_robust_2019,
	title = {Robust velocity-induced acoustic oscillations at cosmic dawn},
	volume = {100},
	issn = {2470-0010, 2470-0029},
	url = {https://link.aps.org/doi/10.1103/PhysRevD.100.063538},
	doi = {10.1103/PhysRevD.100.063538},
	
	number = {6},
	urldate = {2022-07-26},
	journal = {Physical Review D},
	author = {Mu\~{n}oz, Julian B.},
	month = sep,
	year = {2019},
	pages = {063538},
}

@article{mckee_formation_2008,
	title = {The {Formation} of the {First} {Stars}. {II}. {Radiative} {Feedback} {Processes} and {Implications} for the {Initial} {Mass} {Function}},
	volume = {681},
	issn = {0004-637X, 1538-4357},
	url = {https://iopscience.iop.org/article/10.1086/587434},
	doi = {10.1086/587434},
	
	number = {2},
	urldate = {2022-07-12},
	journal = {The Astrophysical Journal},
	author = {McKee, Christopher F. and Tan, Jonathan C.},
	month = jul,
	year = {2008},
	pages = {771--797},
}

@article{lazar_probing_2022,
	title = {Probing the initial mass function of the first stars with transients},
	volume = {511},
	issn = {0035-8711, 1365-2966},
	url = {https://academic.oup.com/mnras/article/511/2/2505/6516967},
	doi = {10.1093/mnras/stac176},
	
	number = {2},
	urldate = {2022-07-12},
	journal = {Monthly Notices of the Royal Astronomical Society},
	author = {Lazar, Alexandres and Bromm, Volker},
	month = feb,
	year = {2022},
	pages = {2505--2514},
}

@article{whalen_finding_2013,
	title = {{FINDING} {THE} {FIRST} {COSMIC} {EXPLOSIONS}. {I}. {PAIR}-{INSTABILITY} {SUPERNOVAE}},
	volume = {777},
	issn = {0004-637X, 1538-4357},
	url = {https://iopscience.iop.org/article/10.1088/0004-637X/777/2/110},
	doi = {10.1088/0004-637X/777/2/110},
	
	number = {2},
	urldate = {2022-07-12},
	journal = {The Astrophysical Journal},
	author = {Whalen, Daniel J. and Even, Wesley and Frey, Lucille H. and Smidt, Joseph and Johnson, Jarrett L. and Lovekin, C. C. and Fryer, Chris L. and Stiavelli, Massimo and Holz, Daniel E. and Heger, Alexander and Woosley, S. E. and Hungerford, Aimee L.},
	month = oct,
	year = {2013},
	pages = {110},
}

@article{munoz_impact_2022,
	title = {The {Impact} of the {First} {Galaxies} on {Cosmic} {Dawn} and {Reionization}},
	volume = {511},
	issn = {0035-8711, 1365-2966},
	url = {http://arxiv.org/abs/2110.13919},
	doi = {10.1093/mnras/stac185},
	number = {3},
	urldate = {2022-03-04},
	journal = {Monthly Notices of the Royal Astronomical Society},
	author = {Mu\~{n}oz, Julian B. and Qin, Yuxiang and Mesinger, Andrei and Murray, Steven G. and Greig, Bradley and Mason, Charlotte},
	month = feb,
	year = {2022},
	note = {arXiv: 2110.13919},
	pages = {3657--3681},
}

@article{naoz_simulations_2013,
	title = {Simulations of {Early} {Baryonic} {Structure} {Formation} with {Stream} {Velocity}: {II}. {The} {Gas} {Fraction}},
	volume = {763},
	issn = {0004-637X, 1538-4357},
	shorttitle = {Simulations of {Early} {Baryonic} {Structure} {Formation} with {Stream} {Velocity}},
	url = {http://arxiv.org/abs/1207.5515},
	doi = {10.1088/0004-637X/763/1/27},
	number = {1},
	urldate = {2022-03-04},
	journal = {The Astrophysical Journal},
	author = {Naoz, Smadar and Yoshida, Naoki and Gnedin, Nickolay Y.},
	month = jan,
	year = {2013},
	note = {arXiv: 1207.5515},
	pages = {27},
}

@article{tseliakhovich_relative_2010,
	title = {Relative velocity of dark matter and baryonic fluids and the formation of the first structures},
	volume = {82},
	issn = {1550-7998, 1550-2368},
	url = {http://arxiv.org/abs/1005.2416},
	doi = {10.1103/PhysRevD.82.083520},
	number = {8},
	urldate = {2022-03-04},
	journal = {Physical Review D},
	author = {Tseliakhovich, Dmitriy and Hirata, Christopher},
	month = oct,
	year = {2010},
	note = {arXiv: 1005.2416},
	pages = {083520},
}

@article{kulkarni_critical_2021,
	title = {The critical dark matter halo mass for {Population} {III} star formation: dependence on {Lyman}-{Werner} radiation, baryon-dark matter streaming velocity, and redshift},
	volume = {917},
	issn = {0004-637X, 1538-4357},
	shorttitle = {The critical dark matter halo mass for {Population} {III} star formation},
	url = {http://arxiv.org/abs/2010.04169},
	doi = {10.3847/1538-4357/ac08a3},
	number = {1},
	urldate = {2022-02-24},
	journal = {The Astrophysical Journal},
	author = {Kulkarni, Mihir and Visbal, Eli and Bryan, Greg L.},
	month = aug,
	year = {2021},
	note = {arXiv: 2010.04169},
	pages = {40},
}

@article{schauer_influence_2021,
	title = {The influence of streaming velocities and {Lyman}-{Werner} radiation on the formation of the first stars},
	volume = {507},
	issn = {0035-8711, 1365-2966},
	url = {http://arxiv.org/abs/2008.05663},
	doi = {10.1093/mnras/stab1953},
	number = {2},
	urldate = {2022-02-24},
	journal = {Monthly Notices of the Royal Astronomical Society},
	author = {Schauer, Anna T. P. and Glover, Simon C. O. and Klessen, Ralf S. and Clark, Paul},
	month = aug,
	year = {2021},
	note = {arXiv: 2008.05663},
	pages = {1775--1787},
}

@ARTICLE{pahl_spectroscopic_2025,
       author = {{Pahl}, Anthony and {Topping}, Michael W. and {Shapley}, Alice and {Sanders}, Ryan and {Reddy}, Naveen A. and {Clarke}, Leonardo and {Kehoe}, Emily and {Bento}, Trinity and {Brammer}, Gabe},
        title = "{A Spectroscopic Analysis of the Ionizing Photon Production Efficiency in JADES and CEERS: Implications for the Ionizing Photon Budget}",
      journal = {\apj},
         year = 2025,
        month = mar,
       volume = {981},
       number = {2},
          eid = {134},
        pages = {134},
          doi = {10.3847/1538-4357/adb1ab},
archivePrefix = {arXiv},
       eprint = {2407.03399},
 primaryClass = {astro-ph.GA},
       adsurl = {https://ui.adsabs.harvard.edu/abs/2025ApJ...981..134P}
}

@ARTICLE{madau_cosmic_2014,
       author = {{Madau}, Piero and {Dickinson}, Mark},
        title = "{Cosmic Star-Formation History}",
      journal = {\araa},
         year = 2014,
        month = aug,
       volume = {52},
        pages = {415-486},
          doi = {10.1146/annurev-astro-081811-125615},
archivePrefix = {arXiv},
       eprint = {1403.0007},
 primaryClass = {astro-ph.CO},
       adsurl = {https://ui.adsabs.harvard.edu/abs/2014ARA&A..52..415M}
}

@ARTICLE{raiter_predicted_2010,
       author = {{Raiter}, A. and {Schaerer}, D. and {Fosbury}, R.~A.~E.},
        title = "{Predicted UV properties of very metal-poor starburst galaxies}",
      journal = {\aap},
         year = 2010,
        month = nov,
       volume = {523},
          eid = {A64},
        pages = {A64},
          doi = {10.1051/0004-6361/201015236},
archivePrefix = {arXiv},
       eprint = {1008.2114},
 primaryClass = {astro-ph.CO},
       adsurl = {https://ui.adsabs.harvard.edu/abs/2010A&A...523A..64R}
}

@ARTICLE{williams_supersonic_2023,
       author = {{Williams}, Claire E. and {Naoz}, Smadar and {Lake}, William and {Chiou}, Yeou S. and {Burkhart}, Blakesley and {Marinacci}, Federico and {Vogelsberger}, Mark and {Chiaki}, Gen and {Nakazato}, Yurina and {Yoshida}, Naoki},
        title = "{The Supersonic Project: The Eccentricity and Rotational Support of SIGOs and DM GHOSts}",
      journal = {\apj},
         year = 2023,
        month = mar,
       volume = {945},
       number = {1},
          eid = {6},
        pages = {6},
          doi = {10.3847/1538-4357/acb820},
archivePrefix = {arXiv},
       eprint = {2211.02066},
 primaryClass = {astro-ph.GA},
       adsurl = {https://ui.adsabs.harvard.edu/abs/2023ApJ...945....6W}
}

@ARTICLE{hirano_primordial_2015,
       author = {{Hirano}, S. and {Hosokawa}, T. and {Yoshida}, N. and {Omukai}, K. and {Yorke}, H.~W.},
        title = "{Primordial star formation under the influence of far ultraviolet radiation: 1540 cosmological haloes and the stellar mass distribution}",
      journal = {\mnras},
         year = 2015,
        month = mar,
       volume = {448},
       number = {1},
        pages = {568-587},
          doi = {10.1093/mnras/stv044},
archivePrefix = {arXiv},
       eprint = {1501.01630},
 primaryClass = {astro-ph.GA},
       adsurl = {https://ui.adsabs.harvard.edu/abs/2015MNRAS.448..568H}
}

@ARTICLE{hirano_one_2014,
       author = {{Hirano}, Shingo and {Hosokawa}, Takashi and {Yoshida}, Naoki and {Umeda}, Hideyuki and {Omukai}, Kazuyuki and {Chiaki}, Gen and {Yorke}, Harold W.},
        title = "{One Hundred First Stars: Protostellar Evolution and the Final Masses}",
      journal = {\apj},
         year = 2014,
        month = feb,
       volume = {781},
       number = {2},
          eid = {60},
        pages = {60},
          doi = {10.1088/0004-637X/781/2/60},
archivePrefix = {arXiv},
       eprint = {1308.4456},
 primaryClass = {astro-ph.CO},
       adsurl = {https://ui.adsabs.harvard.edu/abs/2014ApJ...781...60H}
}

@ARTICLE{stacy_building_2016,
       author = {{Stacy}, Athena and {Bromm}, Volker and {Lee}, Aaron T.},
        title = "{Building up the Population III initial mass function from cosmological initial conditions}",
      journal = {\mnras},
         year = 2016,
        month = oct,
       volume = {462},
       number = {2},
        pages = {1307-1328},
          doi = {10.1093/mnras/stw1728},
archivePrefix = {arXiv},
       eprint = {1603.09475},
 primaryClass = {astro-ph.GA},
       adsurl = {https://ui.adsabs.harvard.edu/abs/2016MNRAS.462.1307S}
}

@ARTICLE{ventura_semi_2025,
       author = {{Ventura}, Emanuele M. and {Qin}, Yuxiang and {Balu}, Sreedhar and {Wyithe}, J. Stuart B.},
        title = "{Semi-analytical modelling of Pop. III star formation and metallicity evolution {\textendash} II. Impact on 21 cm power spectrum}",
      journal = {\mnras},
         year = 2025,
        month = jun,
       volume = {540},
       number = {1},
        pages = {483-497},
          doi = {10.1093/mnras/staf699},
       adsurl = {https://ui.adsabs.harvard.edu/abs/2025MNRAS.540..483V}
}

@ARTICLE{abel_formation_2002,
       author = {{Abel}, Tom and {Bryan}, Greg L. and {Norman}, Michael L.},
        title = "{The Formation of the First Star in the Universe}",
      journal = {Science},
         year = 2002,
        month = jan,
       volume = {295},
       number = {5552},
        pages = {93-98},
          doi = {10.1126/science.1063991},
archivePrefix = {arXiv},
       eprint = {astro-ph/0112088},
 primaryClass = {astro-ph},
       adsurl = {https://ui.adsabs.harvard.edu/abs/2002Sci...295...93A}
}

@ARTICLE{heger_nucleosynthetic_2002,
       author = {{Heger}, A. and {Woosley}, S.~E.},
        title = "{The Nucleosynthetic Signature of Population III}",
      journal = {\apj},
         year = 2002,
        month = mar,
       volume = {567},
       number = {1},
        pages = {532-543},
          doi = {10.1086/338487},
archivePrefix = {arXiv},
       eprint = {astro-ph/0107037},
 primaryClass = {astro-ph},
       adsurl = {https://ui.adsabs.harvard.edu/abs/2002ApJ...567..532H}
}

@ARTICLE{barkat_dynamics_1967,
       author = {{Barkat}, Z. and {Rakavy}, G. and {Sack}, N.},
        title = "{Dynamics of Supernova Explosion Resulting from Pair Formation}",
      journal = {\prl},
         year = 1967,
        month = mar,
       volume = {18},
       number = {10},
        pages = {379-381},
          doi = {10.1103/PhysRevLett.18.379},
       adsurl = {https://ui.adsabs.harvard.edu/abs/1967PhRvL..18..379B}
}

@ARTICLE{fryer_pair_2001,
       author = {{Fryer}, C.~L. and {Woosley}, S.~E. and {Heger}, A.},
        title = "{Pair-Instability Supernovae, Gravity Waves, and Gamma-Ray Transients}",
      journal = {\apj},
         year = 2001,
        month = mar,
       volume = {550},
       number = {1},
        pages = {372-382},
          doi = {10.1086/319719},
archivePrefix = {arXiv},
       eprint = {astro-ph/0007176},
 primaryClass = {astro-ph},
       adsurl = {https://ui.adsabs.harvard.edu/abs/2001ApJ...550..372F}
}

@Article{Matplotlib,
  Author    = {Hunter, J. D.},
  Title     = {Matplotlib: A 2D graphics environment},
  Journal   = {Computing in Science \& Engineering},
  Volume    = {9},
  Number    = {3},
  Pages     = {90--95},
  publisher = {IEEE COMPUTER SOC},
  doi       = {10.1109/MCSE.2007.55},
  year      = 2007
}

@ARTICLE{Astropy,
       author = {{Astropy Collaboration} and {Robitaille}, Thomas P. and {Tollerud}, Erik J. and {Greenfield}, Perry and {Droettboom}, Michael and {Bray}, Erik and {Aldcroft}, Tom and {Davis}, Matt and {Ginsburg}, Adam and {Price-Whelan}, Adrian M. and {Kerzendorf}, Wolfgang E. and {Conley}, Alexander and {Crighton}, Neil and {Barbary}, Kyle and {Muna}, Demitri and {Ferguson}, Henry and {Grollier}, Fr{\'e}d{\'e}ric and {Parikh}, Madhura M. and {Nair}, Prasanth H. and {Unther}, Hans M. and {Deil}, Christoph and {Woillez}, Julien and {Conseil}, Simon and {Kramer}, Roban and {Turner}, James E.~H. and {Singer}, Leo and {Fox}, Ryan and {Weaver}, Benjamin A. and {Zabalza}, Victor and {Edwards}, Zachary I. and {Azalee Bostroem}, K. and {Burke}, D.~J. and {Casey}, Andrew R. and {Crawford}, Steven M. and {Dencheva}, Nadia and {Ely}, Justin and {Jenness}, Tim and {Labrie}, Kathleen and {Lim}, Pey Lian and {Pierfederici}, Francesco and {Pontzen}, Andrew and {Ptak}, Andy and {Refsdal}, Brian and {Servillat}, Mathieu and {Streicher}, Ole},
        title = "{Astropy: A community Python package for astronomy}",
      journal = {\aap},
         year = 2013,
        month = oct,
       volume = {558},
          eid = {A33},
        pages = {A33},
          doi = {10.1051/0004-6361/201322068},
archivePrefix = {arXiv},
       eprint = {1307.6212},
 primaryClass = {astro-ph.IM},
       adsurl = {https://ui.adsabs.harvard.edu/abs/2013A&A...558A..33A}
}

@ARTICLE{numpy,
       author = {{van der Walt}, St{\'e}fan and {Colbert}, S. Chris and {Varoquaux}, Ga{\"e}l},
        title = "{The NumPy Array: A Structure for Efficient Numerical Computation}",
      journal = {Computing in Science and Engineering},
         year = 2011,
        month = mar,
       volume = {13},
       number = {2},
        pages = {22-30},
          doi = {10.1109/MCSE.2011.37},
archivePrefix = {arXiv},
       eprint = {1102.1523},
 primaryClass = {cs.MS},
       adsurl = {https://ui.adsabs.harvard.edu/abs/2011CSE....13b..22V}
}

@ARTICLE{Scipy,
  author  = {Virtanen, Pauli and Gommers, Ralf and Oliphant, Travis E. and
            Haberland, Matt and Reddy, Tyler and Cournapeau, David and
            Burovski, Evgeni and Peterson, Pearu and Weckesser, Warren and
            Bright, Jonathan and {van der Walt}, St{\'e}fan J. and
            Brett, Matthew and Wilson, Joshua and Millman, K. Jarrod and
            Mayorov, Nikolay and Nelson, Andrew R. J. and Jones, Eric and
            Kern, Robert and Larson, Eric and Carey, C J and
            Polat, {\.I}lhan and Feng, Yu and Moore, Eric W. and
            {VanderPlas}, Jake and Laxalde, Denis and Perktold, Josef and
            Cimrman, Robert and Henriksen, Ian and Quintero, E. A. and
            Harris, Charles R. and Archibald, Anne M. and
            Ribeiro, Ant{\^o}nio H. and Pedregosa, Fabian and
            {van Mulbregt}, Paul and {SciPy 1.0 Contributors}},
  title   = {{{SciPy} 1.0: Fundamental Algorithms for Scientific
            Computing in Python}},
  journal = {Nature Methods},
  year    = {2020},
  volume  = {17},
  pages   = {261--272},
  adsurl  = {https://rdcu.be/b08Wh},
  doi     = {10.1038/s41592-019-0686-2},
}

@ARTICLE{prole_fragmentation_2022,
       author = {{Prole}, Lewis R. and {Clark}, Paul C. and {Klessen}, Ralf S. and {Glover}, Simon C.~O.},
        title = "{Fragmentation-induced starvation in Population III star formation: a resolution study}",
      journal = {\mnras},
         year = 2022,
        month = mar,
       volume = {510},
       number = {3},
        pages = {4019-4030},
          doi = {10.1093/mnras/stab3697},
archivePrefix = {arXiv},
       eprint = {2112.10800},
 primaryClass = {astro-ph.GA},
       adsurl = {https://ui.adsabs.harvard.edu/abs/2022MNRAS.510.4019P}
}

@ARTICLE{stacy_constraining_2013,
       author = {{Stacy}, Athena and {Bromm}, Volker},
        title = "{Constraining the statistics of Population III binaries}",
      journal = {\mnras},
         year = 2013,
        month = aug,
       volume = {433},
       number = {2},
        pages = {1094-1107},
          doi = {10.1093/mnras/stt789},
archivePrefix = {arXiv},
       eprint = {1211.1889},
 primaryClass = {astro-ph.CO},
       adsurl = {https://ui.adsabs.harvard.edu/abs/2013MNRAS.433.1094S}
}

@ARTICLE{jaura_trapping_2022,
       author = {{Jaura}, Ondrej and {Glover}, Simon C.~O. and {Wollenberg}, Katharina M.~J. and {Klessen}, Ralf S. and {Geen}, Sam and {Haemmerl{\'e}}, Lionel},
        title = "{Trapping of H II regions in Population III star formation}",
      journal = {\mnras},
         year = 2022,
        month = may,
       volume = {512},
       number = {1},
        pages = {116-136},
          doi = {10.1093/mnras/stac487},
archivePrefix = {arXiv},
       eprint = {2202.09803},
 primaryClass = {astro-ph.SR},
       adsurl = {https://ui.adsabs.harvard.edu/abs/2022MNRAS.512..116J}
}

@ARTICLE{lake_stellar_2025,
       author = {{Lake}, William and {Grudi{\'c}}, Michael Y. and {Naoz}, Smadar and {Yoshida}, Naoki and {Williams}, Claire E. and {Burkhart}, Blakesley and {Marinacci}, Federico and {Vogelsberger}, Mark and {Chen}, Avi},
        title = "{The Stellar Initial Mass Function of Early Dark Matter{\textendash}free Gas Objects}",
      journal = {\apjl},
         year = 2025,
        month = may,
       volume = {985},
       number = {1},
          eid = {L6},
        pages = {L6},
          doi = {10.3847/2041-8213/add347},
archivePrefix = {arXiv},
       eprint = {2410.02868},
 primaryClass = {astro-ph.GA},
       adsurl = {https://ui.adsabs.harvard.edu/abs/2025ApJ...985L...6L}
}

@ARTICLE{sharda_population_2025,
       author = {{Sharda}, Piyush and {Menon}, Shyam H.},
        title = "{Population III star formation in the presence of turbulence, magnetic fields and ionizing radiation feedback}",
      journal = {\mnras},
         year = 2025,
        month = may,
          doi = {10.1093/mnras/staf803},
archivePrefix = {arXiv},
       eprint = {2405.18265},
 primaryClass = {astro-ph.GA},
       adsurl = {https://ui.adsabs.harvard.edu/abs/2025MNRAS.tmp..770S}
}

@ARTICLE{chiti_enrichment_2024,
       author = {{Chiti}, Anirudh and {Mardini}, Mohammad and {Limberg}, Guilherme and {Frebel}, Anna and {Ji}, Alexander P. and {Reggiani}, Henrique and {Ferguson}, Peter and {Andales}, Hillary Diane and {Brauer}, Kaley and {Li}, Ting S. and {Simon}, Joshua D.},
        title = "{Enrichment by extragalactic first stars in the Large Magellanic Cloud}",
      journal = {Nature Astronomy},
         year = 2024,
        month = may,
       volume = {8},
        pages = {637-647},
          doi = {10.1038/s41550-024-02223-w},
archivePrefix = {arXiv},
       eprint = {2401.11307},
 primaryClass = {astro-ph.GA},
       adsurl = {https://ui.adsabs.harvard.edu/abs/2024NatAs...8..637C}
}

@ARTICLE{lazare_when_2026,
       author = {{Lazare}, Hovav and {Libanore}, Sarah and {Vanzan}, Eleonora and {Mu{\~n}oz}, Julian B. and {Kovetz}, Ely D.},
        title = "{When galaxies burst II. Implications of enhanced burstiness for the 21-cm Cosmic Dawn signal}",
      journal = {arXiv e-prints},
         year = 2026,
        month = jul,
          eid = {arXiv:2607.20651},
        pages = {arXiv:2607.20651},
          doi = {10.48550/arXiv.2607.20651},
archivePrefix = {arXiv},
       eprint = {2607.20651},
 primaryClass = {astro-ph.CO},
       adsurl = {https://ui.adsabs.harvard.edu/abs/2026arXiv260720651L}
}

@ARTICLE{chiti_detailed_2023,
       author = {{Chiti}, Anirudh and {Frebel}, Anna and {Ji}, Alexander P. and {Mardini}, Mohammad K. and {Ou}, Xiaowei and {Simon}, Joshua D. and {Jerjen}, Helmut and {Kim}, Dongwon and {Norris}, John E.},
        title = "{Detailed Chemical Abundances of Stars in the Outskirts of the Tucana II Ultrafaint Dwarf Galaxy}",
      journal = {\aj},
         year = 2023,
        month = feb,
       volume = {165},
       number = {2},
          eid = {55},
        pages = {55},
          doi = {10.3847/1538-3881/aca416},
archivePrefix = {arXiv},
       eprint = {2205.01740},
 primaryClass = {astro-ph.GA},
       adsurl = {https://ui.adsabs.harvard.edu/abs/2023AJ....165...55C}
}

@misc{cai_metal_2025,
      title={A Metal-Free Galaxy at $z = 3.19$? Evidence of Late Population III Star Formation at Cosmic Noon}, 
      author={Sijia Cai and Mingyu Li and Zheng Cai and Yunjing Wu and Fujiang Yu and Mark Dickinson and Fengwu Sun and Xiaohui Fan and Ben Wang and Fergus Cullen and Fuyan Bian and Xiaojing Lin and Jiaqi Zou},
      year={2025},
      eprint={2507.17820},
      archivePrefix={arXiv},
      primaryClass={astro-ph.GA},
      url={https://arxiv.org/abs/2507.17820}, 
}

@ARTICLE{jeon_hunting_2026,
       author = {{Jeon}, Junehyoung and {Bromm}, Volker and {Venditti}, Alessandra and {Finkelstein}, Steven L. and {Hsiao}, Tiger Yu-Yang},
        title = "{Hunting for the First Explosions at the High-Redshift Frontier}",
      journal = {arXiv e-prints},
         year = 2026,
        month = jan,
          eid = {arXiv:2601.02469},
        pages = {arXiv:2601.02469},
          doi = {10.48550/arXiv.2601.02469},
archivePrefix = {arXiv},
       eprint = {2601.02469},
 primaryClass = {astro-ph.GA},
       adsurl = {https://ui.adsabs.harvard.edu/abs/2026arXiv260102469J}
}

@ARTICLE{vanzella_pristine_2025,
       author = {{Vanzella}, E. and {Messa}, M. and {Zanella}, A. and {Bolamperti}, A. and {Castellano}, M. and {Loiacono}, F. and {Bergamini}, P. and {Roberts-Borsani}, G. and {Adamo}, A. and {Fontana}, A. and {Treu}, T. and {Calura}, F. and {Grillo}, C. and {Lombardi}, M. and {Rosati}, P. and {Gilli}, R. and {Meneghetti}, M.},
        title = "{A Pristine Star-Forming Complex at z=4.19}",
      journal = {arXiv e-prints},
         year = 2025,
        month = sep,
          eid = {arXiv:2509.07073},
        pages = {arXiv:2509.07073},
          doi = {10.48550/arXiv.2509.07073},
archivePrefix = {arXiv},
       eprint = {2509.07073},
 primaryClass = {astro-ph.GA},
       adsurl = {https://ui.adsabs.harvard.edu/abs/2025arXiv250907073V}
}

@ARTICLE{schaerer_observable_2025,
       author = {{Schaerer}, D. and {Guibert}, J. and {Marques-Chaves}, R. and {Martins}, F.},
        title = "{Observable and ionizing properties of star-forming galaxies with very massive stars and different initial mass functions}",
      journal = {\aap},
         year = 2025,
        month = jan,
       volume = {693},
          eid = {A271},
        pages = {A271},
          doi = {10.1051/0004-6361/202451454},
archivePrefix = {arXiv},
       eprint = {2407.12122},
 primaryClass = {astro-ph.GA},
       adsurl = {https://ui.adsabs.harvard.edu/abs/2025A&A...693A.271S}
}

@ARTICLE{trussler_out_2026,
       author = {{Trussler}, James A.~A. and {Eisenstein}, Daniel J. and {Bunker}, Andrew J. and {Cameron}, Alex J. and {Carniani}, Stefano and {Charlot}, St{\'e}phane and {Chevallard}, Jacopo and {Conselice}, Christopher J. and {Curti}, Mirko and {Curtis-Lake}, Emma and {D'Eugenio}, Francesco and {Egami}, Eiichi and {Hainline}, Kevin and {Hausen}, Ryan and {Helton}, Jakob M. and {Hsiao}, Tiger Yu-Yang and {Ji}, Zhiyuan and {Johnson}, Benjamin D. and {Looser}, Tobias J. and {Maiolino}, Roberto and {Pusk{\'a}s}, D{\'a}vid and {Rinaldi}, Pierluigi and {Robertson}, Brant and {Sun}, Fengwu and {Tacchella}, Sandro and {{\"U}bler}, Hannah and {Williams}, Christina C. and {Willmer}, Christopher N.~A. and {Witstok}, Joris and {Wu}, Zihao},
        title = "{Out of oxygen: Extremely metal-poor galaxy candidates identified at $2.5 < z < 6.5$ with deep JADES medium-band imaging}",
      journal = {arXiv e-prints},
         year = 2026,
        month = mar,
          eid = {arXiv:2603.15761},
        pages = {arXiv:2603.15761},
          doi = {10.48550/arXiv.2603.15761},
archivePrefix = {arXiv},
       eprint = {2603.15761},
 primaryClass = {astro-ph.GA},
       adsurl = {https://ui.adsabs.harvard.edu/abs/2026arXiv260315761T}
}

@ARTICLE{maiolino_search_2026,
       author = {{Maiolino}, Roberto and {{\"U}bler}, Hannah and {Perna}, Michele and {Witstok}, Joris and {Jones}, Gareth C. and {Perez-Gonzalez}, Pablo G. and {Nakajima}, Kimihiko and {Rusta}, Elka and {Salvadori}, Stefania and {Tacchella}, Sandro and {Madau}, Piero and {Trussler}, James A.~A. and {D'Eugenio}, Francesco and {Ji}, Xihan and {Scholtz}, Jan and {Carniani}, Stefano and {Isobe}, Yuki and {Arribas}, Santiago and {Baker}, William M. and {B{\"o}ker}, Torsten and {Bromm}, Volker and {Bunker}, Andrew J. and {Charlot}, Stephane and {Chevallard}, Jacopo and {Curti}, Mirko and {Curtis-Lake}, Emma and {Eisenstein}, Daniel and {Egami}, Eiichi and {Ferrara}, Andrea and {Graziani}, Luca and {Hainline}, Kevin and {Helton}, Jakob M. and {Ivey}, Lucy and {Jonson}, Benjamin and {Koller}, Maria and {Kumari}, Nimisha and {Marconi}, Alessandro and {Mazzolari}, Giovanni and {Laporte}, Nicolas and {Parlanti}, Eleonora and {Pascalau}, Robert and {Rinaldi}, Pierluigi and {Robertson}, Brant and {Rodr{\'\i}guez Del Pino}, Bruno and {Schneider}, Raffaella and {Venditti}, Alessandra and {Venturi}, Giacomo and {Willmer}, Christopher N.~A. and {Witten}, Callum and {Pentericci}, Laura and {Zamora}, Sandra},
        title = "{The search for Population III: Confirmation of a HeII emitter with no metal lines at z=10.6}",
      journal = {arXiv e-prints},
         year = 2026,
        month = mar,
          eid = {arXiv:2603.20362},
        pages = {arXiv:2603.20362},
archivePrefix = {arXiv},
       eprint = {2603.20362},
 primaryClass = {astro-ph.GA},
       adsurl = {https://ui.adsabs.harvard.edu/abs/2026arXiv260320362M}
}

@ARTICLE{trapp_flexible_2020,
       author = {{Trapp}, A.~C. and {Furlanetto}, Steven R.},
        title = "{A flexible analytic model of cosmic variance in the first billion years}",
      journal = {\mnras},
         year = 2020,
        month = dec,
       volume = {499},
       number = {2},
        pages = {2401-2415},
          doi = {10.1093/mnras/staa2828},
archivePrefix = {arXiv},
       eprint = {2009.05059},
 primaryClass = {astro-ph.GA},
       adsurl = {https://ui.adsabs.harvard.edu/abs/2020MNRAS.499.2401T}
}
\end{document}